\documentclass[11pt]{article}
\pdfoutput=1
\usepackage{jcappub,natbib}
\input{colordvi.tex}

\def\la{\mathrel{\raise.3ex\hbox{$<$\kern-.75em\lower1ex\hbox{$\sim$}}}}

\title{Gravitational wave signatures from warm dark energy}

\author[a]{Alexandros Papageorgiou}

\emailAdd{papag008@umn.edu}

\affiliation[a]{School of Physics and Astronomy, University of Minnesota, Minneapolis, 55455, USA}

\abstract{Motivated by some of the recent swampland conjectures, we study a model of dark energy, in which a quintessence axion slowly rolls in a steep potential due to its interactions with a U(1) or an SU(2) gauge field. 
The gauge fields produced by this interaction can generate a stochastic gravitational wave background with frequencies in the range $f \sim \left(10^{-16} - 10^{-14}\right)\;  {\rm Hz}$. Gravitational waves in this range can in principle be probed by CMB spectral distortions. We show that the amplitude of the signal produced in this model is above the level for detectability only under very favourable choices of parameters. }

\begin{document}

\maketitle
\flushbottom

\section{Introduction}
\label{sec:intro} 

{\hskip 2em}After over two decades since its discovery, the nature of dark energy remains largely a mystery today \cite{Riess:1998cb}. Of the various mechanisms that have been put forward, one of the most compelling is the possibility that the dark energy component is a scalar field which is undergoing slow-roll. Such a scenario is referred to as "quintessence" and it recently rose to prominence as a result of the various conjectures that arise from string theory and quantum gravity. The de Sitter conjecture \cite{Andriot:2018wzk,Dvali:2018fqu,Andriot:2018mav,Garg:2018reu,Ooguri:2018wrx,Rudelius:2019cfh,Obied:2018sgi}, in the context of the swampland programme \cite{Palti:2019pca} and the Trans-Planckian-Censorship conjecture \cite{Bedroya:2019snp,Bedroya:2019tba} constrain the slope of the quintessence potential $V(\phi)$ to satisfy

\begin{equation}
|V'|\geq c\, V/M_p
\end{equation}

for some ${\cal O}(1)$ constant, where prime denotes the derivative with respect to the quintessence field. 

The revived interest of the quintessence scenario soon led to several phenomenological studies which constrained the slope of the quintessence field to be $c\leq {\cal O}(0.1)$ \cite{Akrami:2018ylq}. This creates a small tension between the requirements of the swampland conjectures and cosmological observations. 

The recent work  \cite{DallAgata:2019yrr} developed a model called "warm dark energy" which attempts to reconcile this tension. In this model, the role of the quintessence field is played by an axion field coupled to some U(1) gauge field via the Chern-Simons coupling. The slope of the potential is chosen to be large enough to be compatible with the swampland conjectures while slow-roll is maintained as the axion field produces the gauge field at the expense of its own kinetic energy. The copious gauge field production in this model results in an effective friction that is greater than the standard Hubble friction due to the background expansion. The idea of coupling an axion field with a high slope to a U(1) gauge field as a means of inducing additional friction has been introduced by Anber and Sorbo \cite{Anber:2009ua} in the context of primordial inflation, and it has then been applied in \cite{DallAgata:2019yrr} as a mechanism for late time quintessence.

\begin{figure}[ht!]
\centerline{
\includegraphics[width=0.70\textwidth,angle=0]{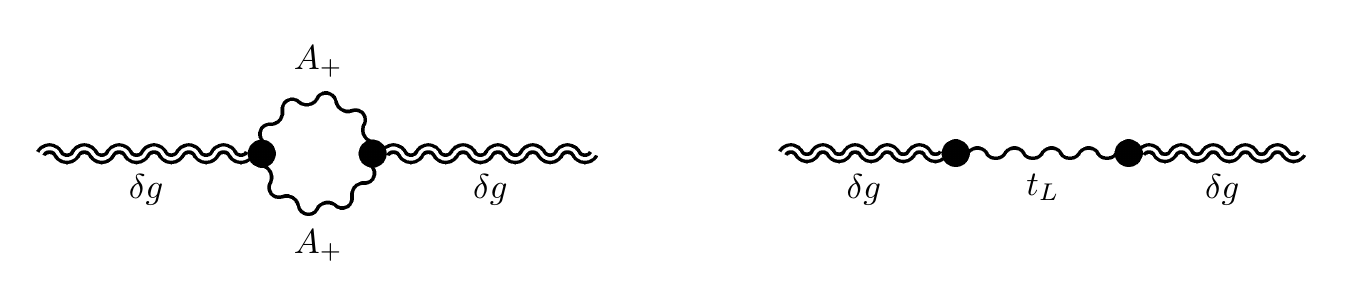}
}
\caption{\textit{Left Panel}: diagrammatic representation of the nonlinear sourcing of gravitational waves by gauge quanta of vector fields. \textit{Right Panel}: diagrammatic representation of the linear sourcing of gravitational waves by "tensor" perturbations of a massless non-Abelian SU(2) gauge field.
}
\label{fig:diagrams}
\end{figure}

It is well known, in the context of natural inflation, that the copious amount of gauge field amplification that arises from a Chern-Simons coupling can source substantial amounts of gravitational waves \cite{Barnaby:2010vf,Sorbo:2011rz}. Gravitational waves in this case are sourced nonlinearly by a process that is diagrammatically depicted on the left panel of figure \ref{fig:diagrams}. We study in this work the sourcing of gravitational waves in the context of the warm dark energy model. These are gravitational waves that are produced during matter domination and the typical frequencies that we find are in the ball park of $f\sim {\cal O}(10^{-15})\,{\rm Hz}$. As recently studied in \cite{Kite:2020uix}, gravitational waves of such frequencies can in principle be probed by measurements of spectral distortions in the CMB. This offers an exciting new opportunity for the previously inaccessible by experiments $f\sim \left(10^{-15} - 10^{-9}\right)\,{\rm Hz}$ frequency range. This also provides solid motivation for a careful numerical analysis of the gravitational wave power spectrum produced in the warm dark energy model which we undertake in subsection \ref{sec:modelU1-gw}.

Continuing to draw inspiration from inflation we also explore in this work the possibility that a Chern-Simons coupling between an axion field and a massless non-Abelian SU(2) gauge field with a classical isotropic vaccum expectation value (vev) can produce additional friction to slow down the axion field on a steep potential, in the context of dark energy\footnote{See also \cite{Alexander:2016nrg,Alexander:2016mrw} for similar quintessence models in which an axion field coupled to massive non-Abelian SU(2) gauge fields are considered.}. This idea is inspired by the Chromo-Natural Inflation model proposed in \cite{Adshead:2012kp} and further studied in \cite{Dimastrogiovanni:2012ew,Adshead:2013nka,SheikhJabbari:2012qf} (see also \cite{DallAgata:2018ybl} for a Supergravity realization of the model). The mechanism we explore is fundamentally different from the one used for the Abelian U(1) field in that all the dynamics take place at the background level using only the classical equations of motion for the axion and the gauge field vev. We explore the early time dynamics of the model and we specify the region of parameter space that allows for compatibility with observational data. 

At the level of perturbations, the presence of the isotropic vev, induces a linear coupling between tensor perturbations of the metric and of the gauge field\footnote{Following standard notation for these models, we denote as tensor perturbations of the metric those modes that transform as a rank 2 tensor under spatial notations; we denote as tensor perturbations of the gauge field the modes that, due to the presence of gauge vevs, couple linearly to the tensor metric perturbations"}.  These "tensor" perturbations of the gauge field, undergo an instability and then linearly source gravitational waves in a way that is diagrammatically depicted on the right panel of figure \ref{fig:diagrams}. This is well understood in the context of inflation \cite{Adshead:2013nka,Dimastrogiovanni:2012ew}. We explore this mechanism in the context of dark energy and characterize the frequencies and magnitude of the gravitational waves produced during matter domination.

In summary, the aim of this paper is two-fold. We want to give a homogeneous presentation of axion-gauge quintessence with a coupling to a U(1) or SU(2) gauge field described by the action 

\begin{equation}
S = \int d^4 x \sqrt{-g} \left[ - \frac{1}{2} \left( \partial \phi \right)^2 - V \left( \phi \right) - \frac{1}{4} F_{\mu \nu} F^{\mu \nu} - \frac{\phi}{4 f} F_{\mu \nu} {\tilde F}^{\mu \nu} \right] \;, 
\label{eq:act}
\end{equation} 

where $F^{\mu \nu}$ is the strength tensor of either the Abelian or non-Abelian (with suppressed gauge indexes) gauge field. In order to avoid any ambiguity we refer to the first case as "axion-U(1)" while the second case will be referred to as "axion-SU(2)" case. Additionally we study the gravitational wave production for both of these cases and compute the power spectrum.

This paper is separated into two main sections each of which is separated into two subsections. In subsection \ref{sec:modelU1-M} we provide a review of the axion-U(1) model, which was studied in \cite{DallAgata:2019yrr}, for the sake of completeness. Afterwards we compute the gravitational wave power spectrum in subsection \ref{sec:modelU1-gw}. In the second part of the paper we study the axion-SU(2) case. In subsection \ref{sec:modelSU2-M} we analyze the model at the level of the background and we find the parameter space for which the model is compatible with observations, for a steep potential. Finally in subsection \ref{sec:modelSU2-gw} the axion-SU(2) model is studied at the level of perturbations and the possibility that gravitational waves could be probed by CMB spectral distortions is explored. We make our concluding remarks in section \ref{sec:conclusions} and we present some mathematical details, which were omitted from the main text, in Appendix \ref{app:twopoint}.

\section{Accelerated expansion from dissipation with an axion-U(1) coupling} 
\label{sec:modelU1} 

{\hskip 2em}We initiate the main body of this paper by providing a short review of the results found for the axion-U(1) coupling case presented in \cite{DallAgata:2019yrr}. Our purpose is to study the phenomenological consequences of the low-energy effective theory whose bosonic sector is given by (\ref{eq:act}) and to this end we remain agnostic about the mechanism which leads to our effective theory (see \cite{DallAgata:2019yrr} for a supergravity realization). In addition to reviewing previous results we, extend the analysis of the axion-U(1) quintessence model by computing the gravitational wave power spectrum produced by this mechanism. We find that the large amplification of vector fields can lead to substantial production of gravitational waves whose frequency can be as high as $f\simeq{\cal O}(10^{-15})\;{\rm Hz}$. These gravitational waves could in principle be detected by measurements of $\mu$-distortions as was recently pointed out in \cite{Kite:2020uix}.

This section is divided into two subsections. In subsection \ref{sec:modelU1-M} we establish our notation and summarize the analytic and numerical results of \cite{DallAgata:2019yrr}. In \ref{sec:modelU1-gw} we provide approximate analytic formulas that can serve in diagnosing the frequencies of the gravitational waves produced and finally we display the full gravitational wave spectrum for two distinct values of the coupling parameter $f$.

\subsection{axion-U(1) Quintessence model} 
\label{sec:modelU1-M} 

{\hskip 2em}Our action takes the form
\begin{equation}
S = \int d^4 x \sqrt{-g} \left[ - \frac{1}{2} \left( \partial \phi \right)^2 - V \left( \phi \right) - \frac{1}{4} F_{\mu \nu} F^{\mu \nu} - \frac{\phi}{4 f} F_{\mu \nu} {\tilde F}^{\mu \nu} \right] \;, 
\label{eq:Lag}
\end{equation} 

where $\phi$ is an axion-like field (from this point on we refer to it simply as "axion") which enjoys a shift symmetry $\phi\rightarrow \phi+C$. This symmetry is respected by the interaction term but violated by the potential term $V(\phi)$. The gauge field strength tensor is denoted as $F_{\mu\nu}$ and $\tilde{F}^{\mu\nu} \equiv \frac{\epsilon^{\mu\nu\alpha\beta}}{2\sqrt{-g}} F_{\alpha \beta}$ represents its dual.

The background geometry of the universe is assumed to be the standard FLRW whose line element is given by $ds^2=a(\tau)^2\left(-d\tau^2+\delta_{ij}dx^i dx^j\right)$. Where $a(\tau)$ is the scale factor and $\tau$ is the conformal time. We assume that the field $\phi$ enjoys high occupation numbers and is cosmologically important. As a consequence, in order to have a cosmology that is compatible with the FLRW background, its value is only dependent on time $\phi=\phi\left(\tau\right)$ and its motion can be modeled by solving the classical equations of motion that result from (\ref{eq:Lag}).

The relevant dynamical equations that arise from our action are the equation of motion of of the axion field, the equations of motion of the gauge field right and left handed polarizations ($A_+$ and $A_-$ respectively) and the Friedman equation given by the 00 component of the Einstein field equations

\begin{eqnarray} 
&&\frac{\partial^2 \phi}{\partial \tau^2} + \frac{2}{a} \, \frac{\partial a}{\partial \tau} \, \frac{\partial \phi}{\partial \tau} + a^2 
\, \frac{\partial V}{\partial \phi} = \frac{a^2}{f} \, \left\langle\vec{E} \cdot \vec{B}\right\rangle \;\;,
\label{eq:eomphi}\\
&&\left( \partial_\tau^2 + k^2 \mp 2 a H k \xi \right) A_\pm \left( \tau ,\, k \right)=0 \;\;,\;\; \xi \equiv \frac{\partial_\tau \phi}{2 a f H} \;\;, 
\label{A-eom}\\
&&\left( \frac{1}{a} \, \frac{\partial a}{\partial \tau} \right)^2 = \frac{a^2}{3 M_p^2} \left[ \frac{1}{2 a^2} \left( \frac{\partial \phi}{\partial \tau} \right)^2 + V + \frac{\rho_{m,{\rm in}} \, a_{\rm in}^3}{a^3} + \left\langle\frac{E^2+B^2}{2}\right\rangle \right] \;\; , 
\label{eom-00}
\end{eqnarray} 

where the Hubble rate is defined as usual $H\equiv \frac{\dot{a}}{a}=\frac{\partial_\tau a}{a^2}$ and the two gauge field backreaction terms are given by

\begin{eqnarray}
\left\langle\vec{E} \cdot \vec{B}\right\rangle &=& -\frac{1}{4 \pi^2 a^4} \sum_{i=\pm}\int d k \, k^3 \, \frac{\partial}{\partial \tau} \left\vert A_i \right\vert^2\,,\nonumber\\
\left\langle\frac{E^2+B^2}{2}\right\rangle &=& \frac{1}{4 \pi^2 a^4} \sum_{i=\pm}\int d k \, k^2 \, \left[ \left\vert \frac{\partial A_i}{\partial \tau} \right\vert^2 
 + k^2 \, \left\vert A_i \right\vert^2 \right]\;, 
\end{eqnarray}

note that we have added the term $\rho_{m}\left(\tau\right)=\frac{\rho_{m,{\rm in}} \, a_{\rm in}^3}{a^3}$ on the right hand side of the Friedman equation which represents the total energy density of non-relativistic matter as a function of time. The non-relativistic matter is assumed not to have any direct interaction with either the axion or the gauge field.

Without loss of generality we can assume that the speed of the axion is positive $\frac{d\phi}{d\tau}> 0$. One consequence of this choice is that, assuming the speed is sufficiently high, the right handed polarization of the gauge field becomes unstable and grows tachyonically. These tachyonically enhanced modes can then grow enough to cause the backreaction term in equation (\ref{eq:eomphi}) to become dominant, therefore slowing down the evolution of the axion field. To put it in a difference perspective, as the axion field rolls down its potential, it is producing gauge modes via the Chern-Simons coupling at the expense of its own kinetic energy. In this way, the kinetic energy remains small and the axion field maintains a slow-roll configuration through the transition from matter domination to axion field domination. This is the same mechanism as the one used by Anber and Sorbo in the context of primordial inflation \cite{Anber:2009ua}. Throughout this process, the left handed polarization of the gauge field remains stable and therefore we can ignore it in the following study.

In order to facilitate our analysis we choose a simple potential of the form 

\begin{equation}
V\left(\phi\right)=V_0{\rm e}^{-\frac{\lambda \phi}{M_p}}
\end{equation}

This choice of potential is convenient because the late time behavior of the equation of state of the scalar field can be understood analytically in the absence of the Chern-Simons coupling. In the late time limit the equation of state parameter is given by the simple formula \cite{Ferreira:1997hj}

\begin{equation}
w_\phi\equiv\frac{p_\phi}{\rho_\phi}=\frac{\lambda^2}{3}-1
\label{eq:asympt}
\end{equation}

Which is valid for $\lambda<\sqrt{3}$. 

For the exponential potential the generic expectation that arises from string theory is $\lambda\gtrsim\sqrt{6}$ \cite{Obied:2018sgi,Bedroya:2019snp}. On the other hand, observations seem to suggest that a value of $\lambda \simeq 1$ is only acceptable at the $3\sigma$ level\footnote{the study of   \cite{Akrami:2018ylq} obtained $\lambda \la 0.49,\, 0.80 ,\, {\rm and} \, 1.02$ at $68\%$, $95\%$, and $99.7\%$ while \cite{Raveri:2018ddi} , obtains the limits 
$\lambda \la 0.28,\, 0.51$  at $68\%$,  and $95\%$ (See section  ``Dark energy and the cosmological data'' of \cite{Akrami:2018ylq} for a detailed discussion on the data sets used).}. For definiteness we fix $\lambda=1$ as done in \cite{DallAgata:2019yrr}. This choice is at the interface of compatibility between the swampland conjectures and observational data. 

\subsubsection{Early time analytic solutions} 
\label{sec:modelU1-MAna} 

{\hskip 2em}At very early times, $\rho_m\gg V\left(\phi\right)$ and the speed of the axion is negligibly small. This has two important consequences, namely the background evolves as in standard matter domination, with no backreaction from the gauge field modes, that remain in their vacuum configuration. In this regime, reference \cite{DallAgata:2019yrr} gives the following solutions for the dynamically relevant fields and the background

\begin{eqnarray}
&& a_{\rm early}\left(\tau\right)=\left(\frac{\tau}{\tau_{\rm in}}\right)^2\nonumber\\
&& \phi_{\rm early}\left(\tau\right) = \frac{2}{9}\frac{\lambda}{\bar{\rho}_m}M_p\left[\left(\frac{\tau}{\tau_{\rm in}}\right)^6-1\right]\nonumber\\
&& A_{+,{\rm early}}\left(\tau,k\right)=\frac{1}{\sqrt{2 k}}{\rm e}^{\frac{2\tau\sqrt{f k M_p\left(V_0 \lambda \tau^5-9 f k M_p \tau_{\rm in}^4\right)}}{21 f M_p \tau_{\rm in}^2}+\frac{5 i 3^{2/5}\sqrt{\pi}\left(\frac{f k^6 M_p \tau_{\rm in}^4}{V_0 \lambda}\right)^{1/5}\Gamma\left(\frac{6}{5}\right)}{7 \Gamma\left(\frac{7}{10}\right)}-\frac{5}{7}i k \tau \,{}_2F_1\left[\frac{1}{5},\frac{1}{2},\frac{6}{5},\frac{V_0 \lambda \tau^5}{9 f k M_p \tau_{\rm in}^4}\right]}\nonumber\\
\label{eq:early}
\end{eqnarray}

where $\tau_{\rm in}=\frac{2\sqrt{3}M_p}{V_0^{1/2}} \frac{1}{\bar{\rho}_m^{1/2}}$ and $\bar{\rho}_m$ is the initial ratio of the energy of matter to that of the potential of the axion field $\bar{\rho}_m\equiv\frac{\rho_{m,{\rm in}}}{V_0}$. For a more detailed discussion on the derivation of the early time solution of the axion field see subsection 3.3 and Appendix A of \cite{DallAgata:2019yrr}. These expressions are useful for the purpose of setting the initial conditions of a numerical evolution which evolves the set of equations (\ref{eq:eomphi}), (\ref{A-eom}) and (\ref{eom-00}) beyond the period of matter domination and also beyond the regime of negligible backreaction. The early time solutions can also be used as a diagnostic tool to estimate the time in which the backreaction becomes important and also to estimate the range of gauge field modes in k-space which become unstable and therefore have to be included in the full numerical evolution. Generally speaking, the behavior of the various fields will deviate from their corresponding early time solutions because either the backreaction term will become dominant in the equation of motion of the axion field, or because the energy of the axion field will start to dominate the universe, or both. The magnitude of the coupling parameter $f$ controls which approximation breaks first. For sufficiently strong coupling (i.e. sufficiently small $f$) backreaction becomes dominant during the matter dominated stage that preceded the current acceleration. For small coupling (high $f$), backreaction dominates later, possibly even after the present moment (we define the "present moment" as the instant in which the ratio of dark energy to matter is the observed one).

To conclude this subsection, we can estimate the time of transition from matter domination to dark energy domination and also the maximum wavenumber that becomes unstable at each point in time by making use of our analytic formulas. We have parametrically

\begin{equation}
a_{\rm transition}\simeq\bar{\rho}_m^{1/3}\;\;\;\; \Rightarrow\;\;\;\; \tau_{\rm transition}\simeq \frac{2\sqrt{3}M_p}{\bar{\rho}_m^{1/3}V_0^{1/2}}\,,\;\;\;\;k_{\rm max}\left(\tau\right)=\frac{1}{f}\frac{d\phi}{d\tau}\simeq \frac{2\lambda}{3\sqrt{3}}\frac{ V_0^{1/2}}{\bar{\rho}_m^{1/2} f}\left(\frac{\tau}{\tau_{\rm in}}\right)^5
\label{eq:limits}
\end{equation}

Another useful quantity that quantifies the level of backreaction is the ratio between the third and fourth terms of (\ref{eq:eomphi}).

\begin{equation}
\left( \frac{\partial V}{\partial \phi} \right)^{-1} \frac{\vec{E} \cdot \vec{B}}{f}  = - \left( \frac{\partial V}{\partial \phi} \right)^{-1} \frac{1}{4 \pi^2 a^4 \, f} \int d k \, k^3 \, \frac{\partial}{\partial \tau} \left\vert A \right\vert^2 \equiv \int d {\tilde k} \; {\cal B}_{EB} \;\;, 
\label{back}
\end{equation} 

\begin{figure}[ht!]
\centerline{
\includegraphics[width=0.32\textwidth,angle=0]{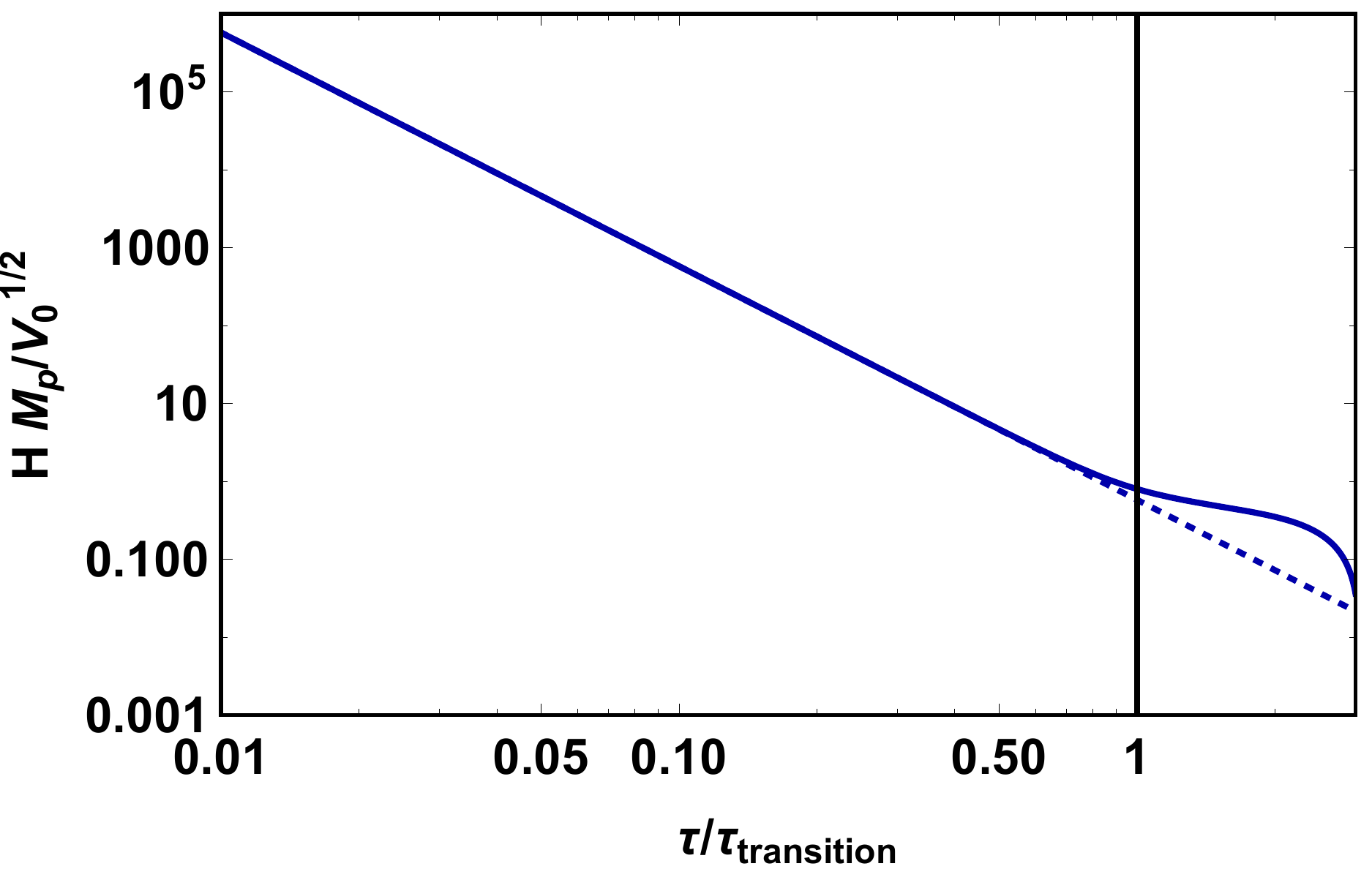}
\includegraphics[width=0.32\textwidth,angle=0]{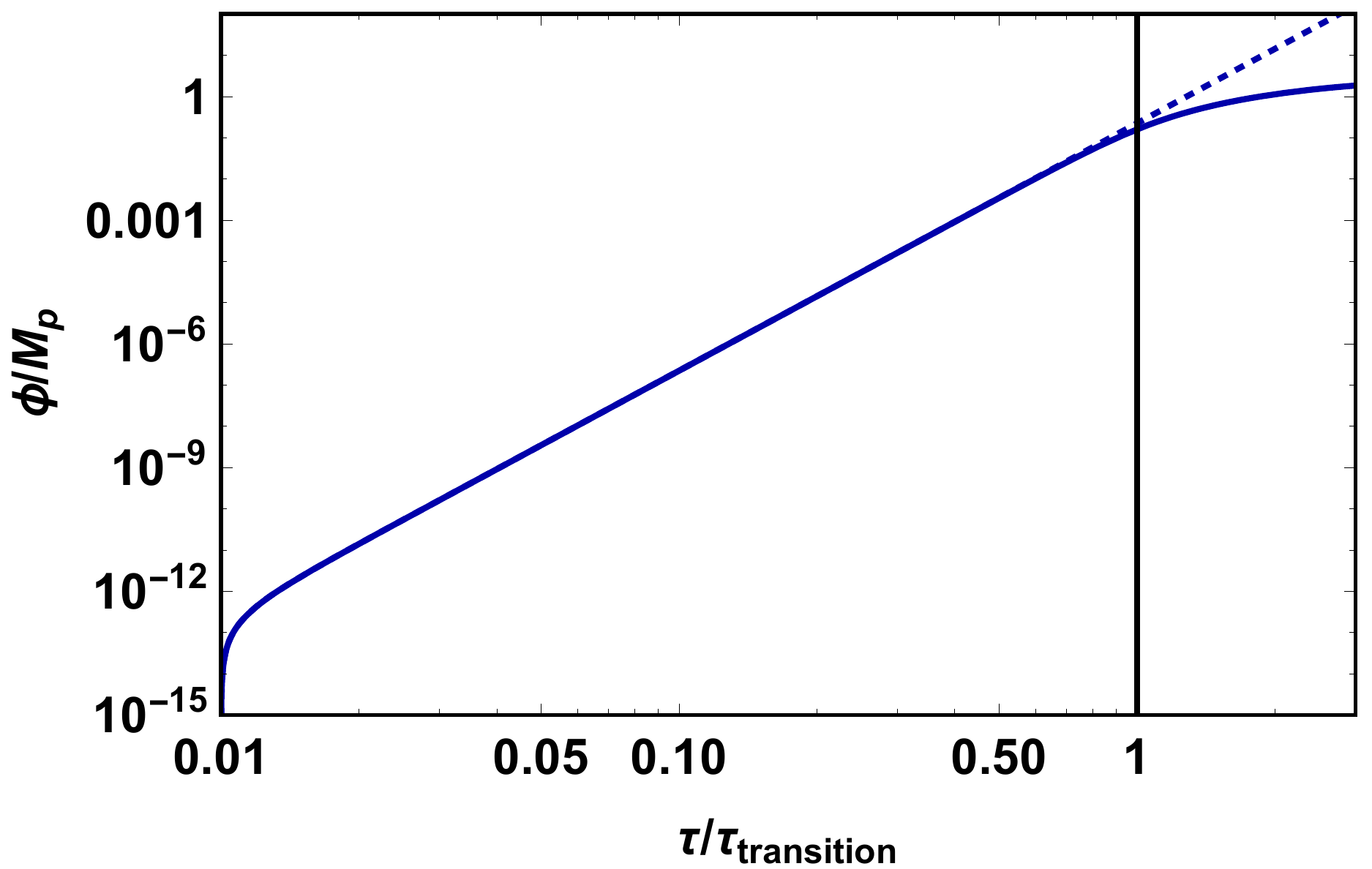}
\includegraphics[width=0.32\textwidth,angle=0]{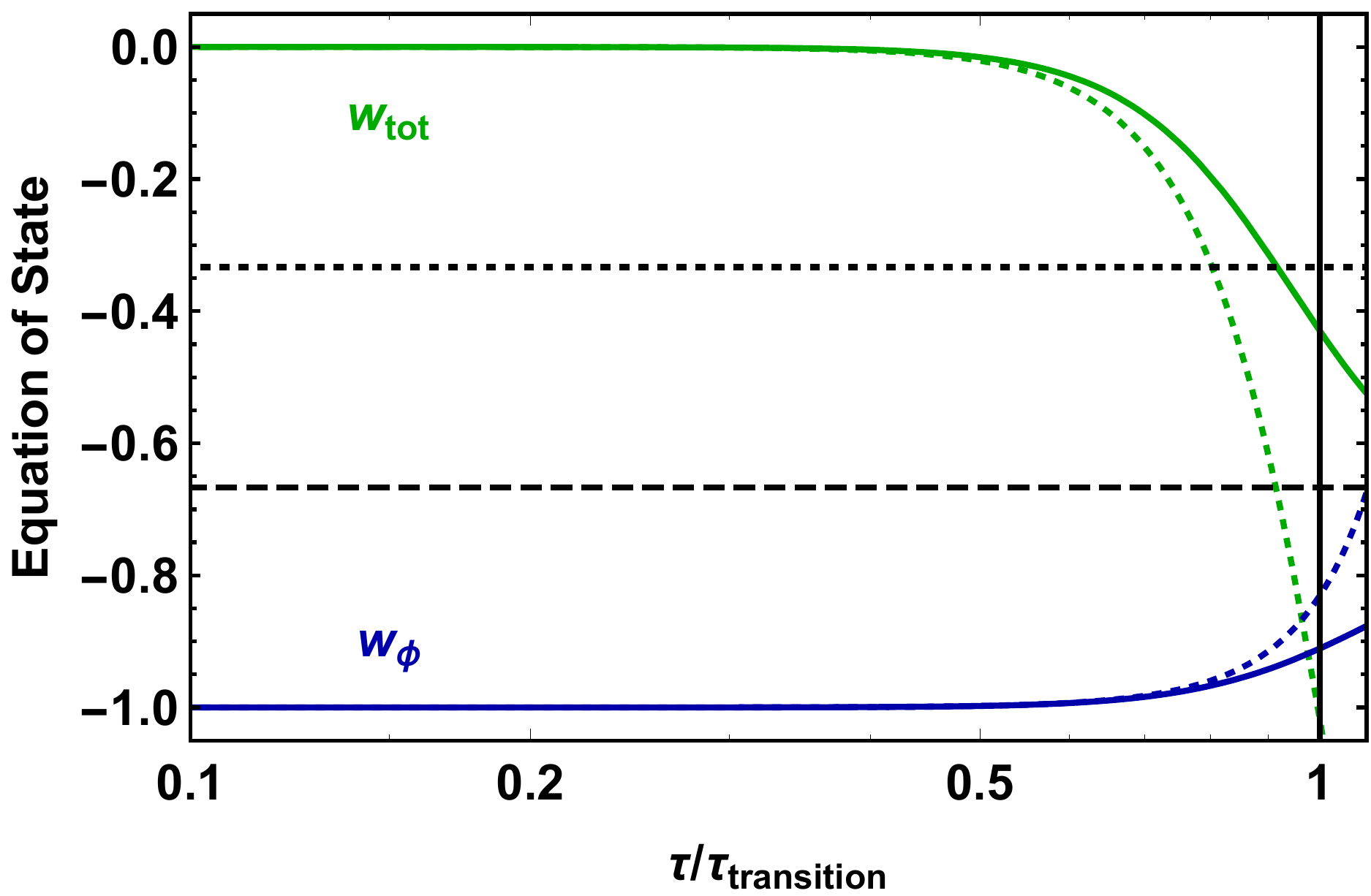}
}
\caption{The three panels display the comparison between a numerical evolution of the corresponding plotted quantities without backreaction (solid lines) and their early time approximations (\ref{eq:early}). \textit{Left panel}: Hubble rate.
\textit{Central panel}: Axion field value.
\textit{Right panel}: Total equation of state defined as $w_{\rm tot} \equiv \frac{p_{\rm tot}}{\rho_{\rm tot}} $. The black dotted line is the threshold beyond which the universe undergoes accelerated expansion and the black dashed line denotes the asymptotic value of the equation of state parameter defined in (\ref{eq:asympt})
}
\label{fig:system2}
\end{figure}

\begin{figure}[ht!]
\centerline{
\includegraphics[width=0.32\textwidth,angle=0]{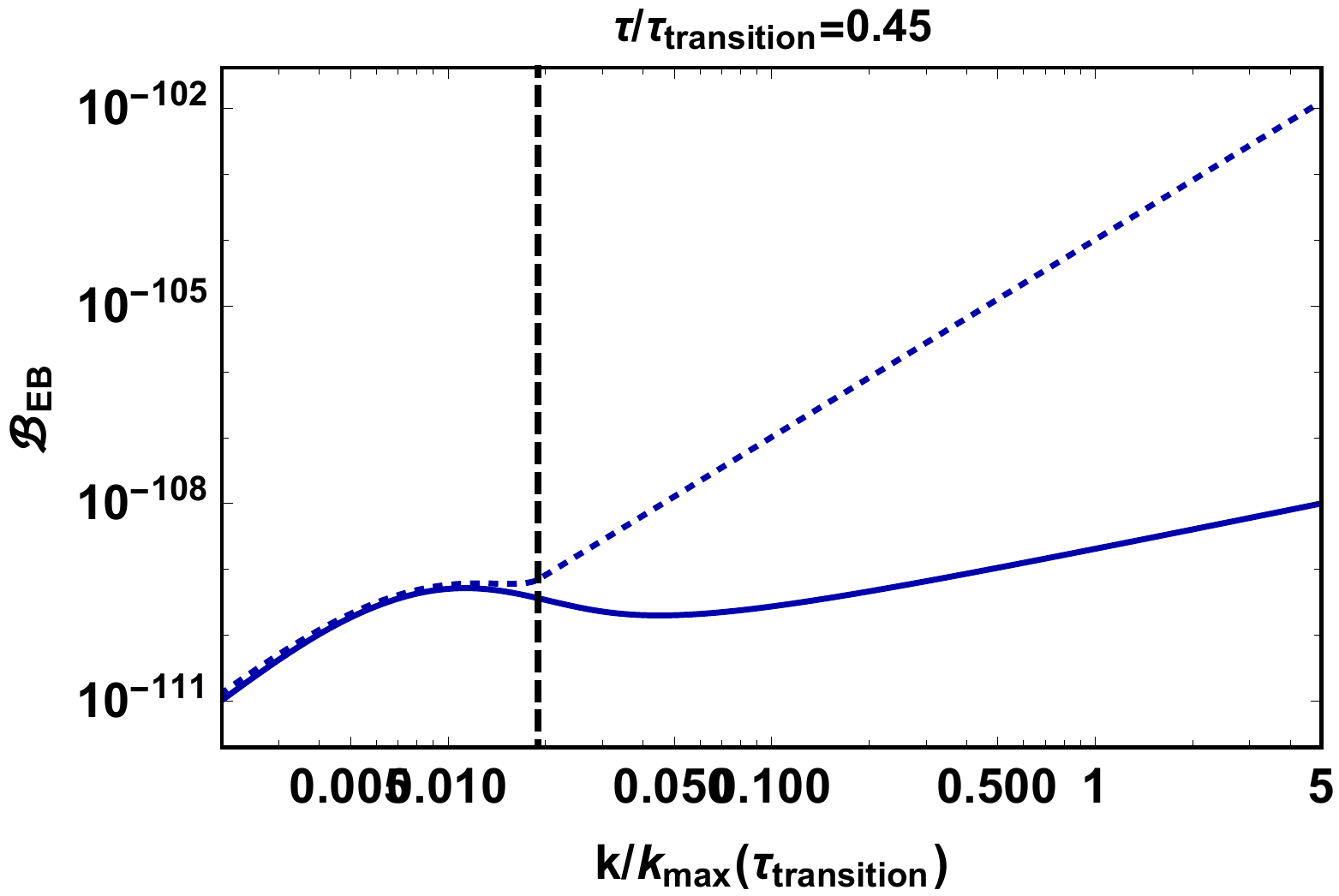}
\includegraphics[width=0.32\textwidth,angle=0]{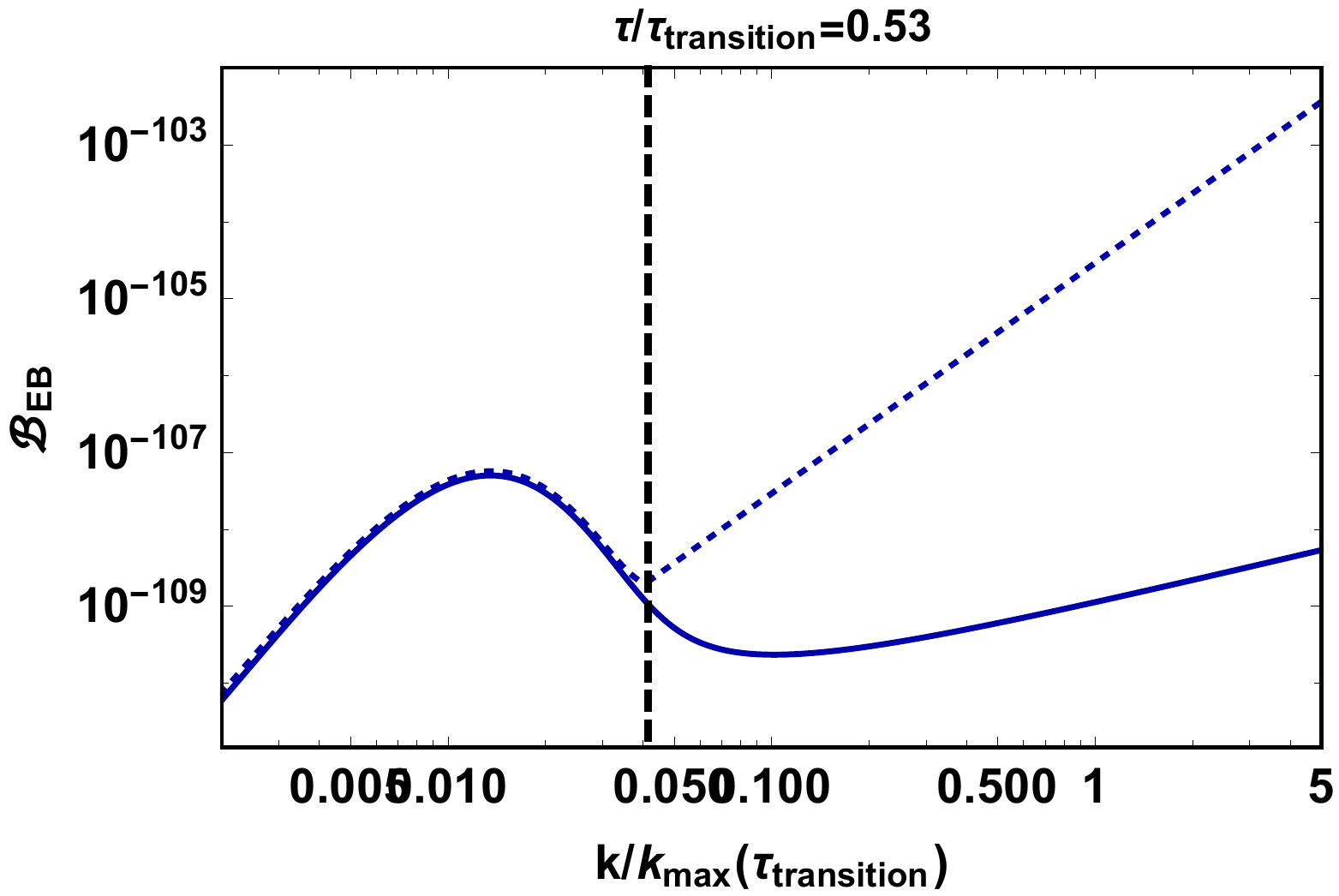}
\includegraphics[width=0.32\textwidth,angle=0]{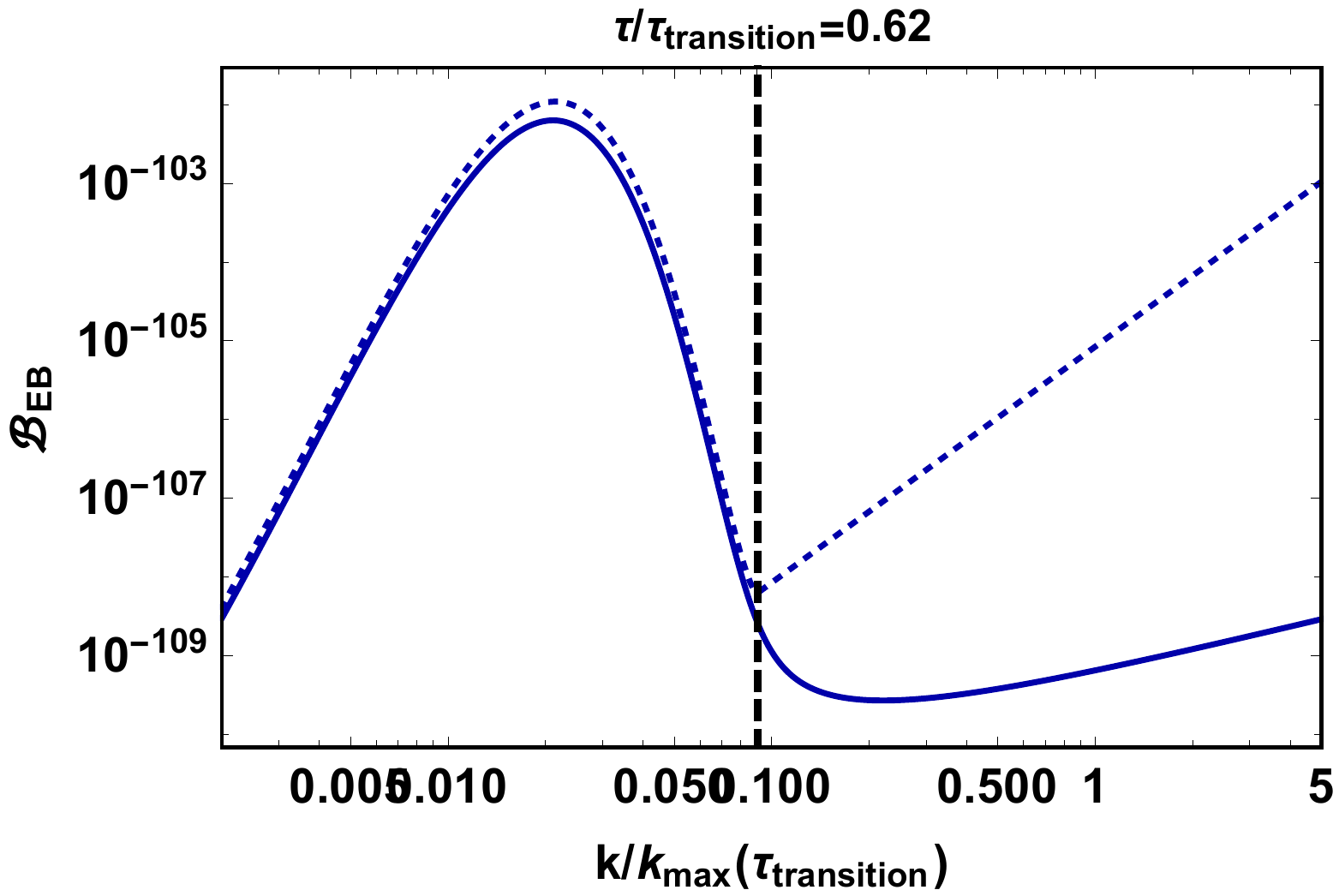}
}

\centerline{
\includegraphics[width=0.32\textwidth,angle=0]{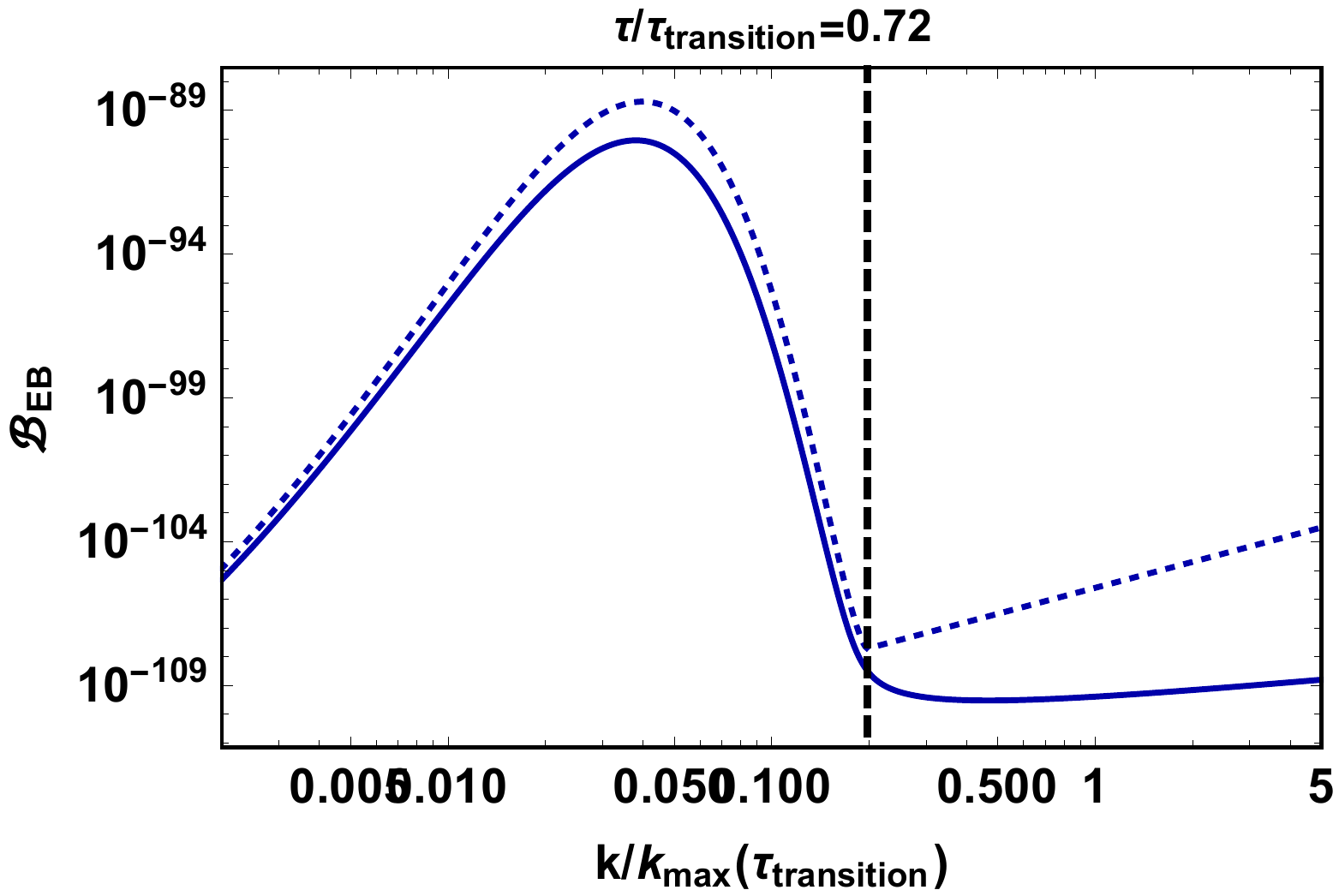}
\includegraphics[width=0.32\textwidth,angle=0]{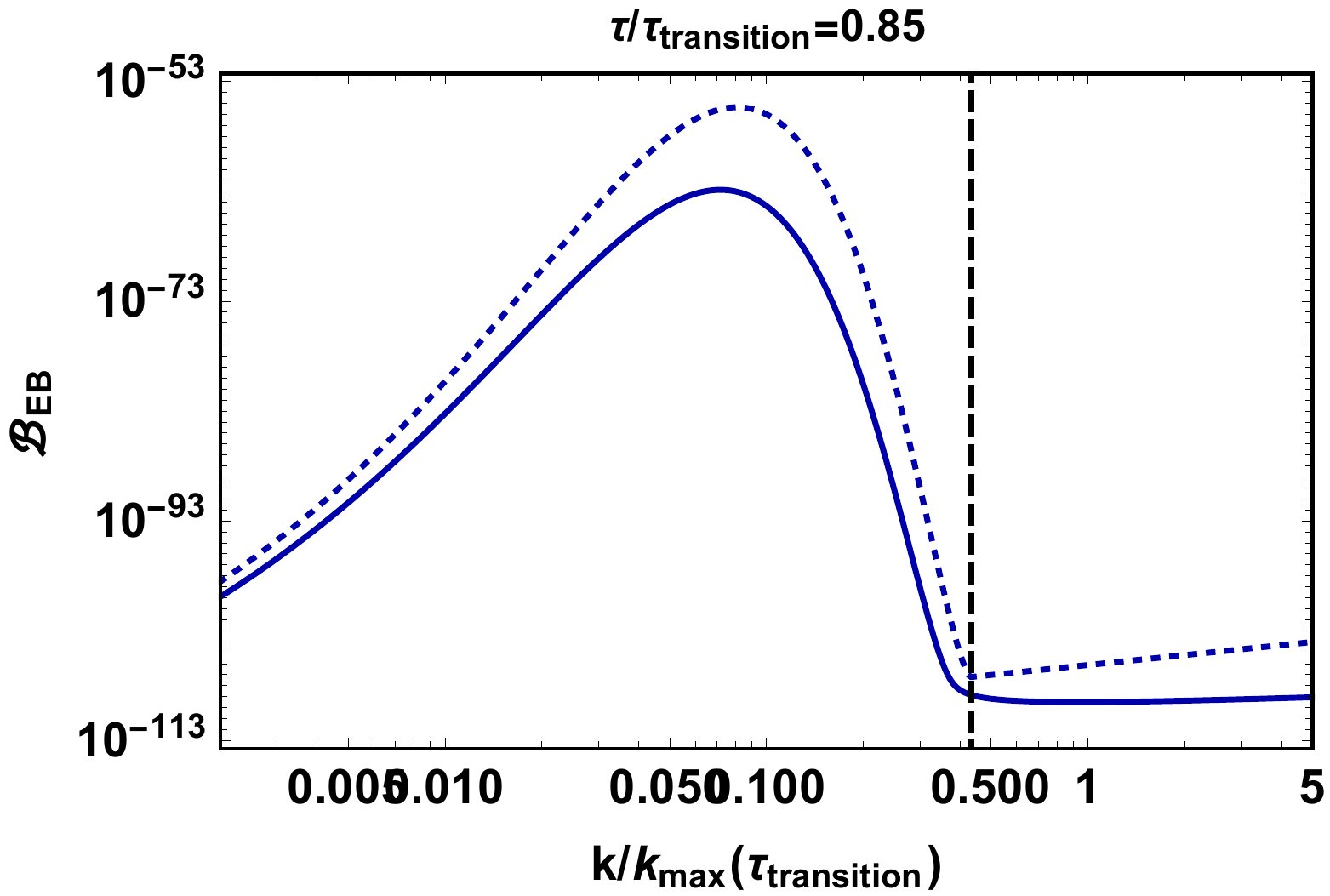}
\includegraphics[width=0.32\textwidth,angle=0]{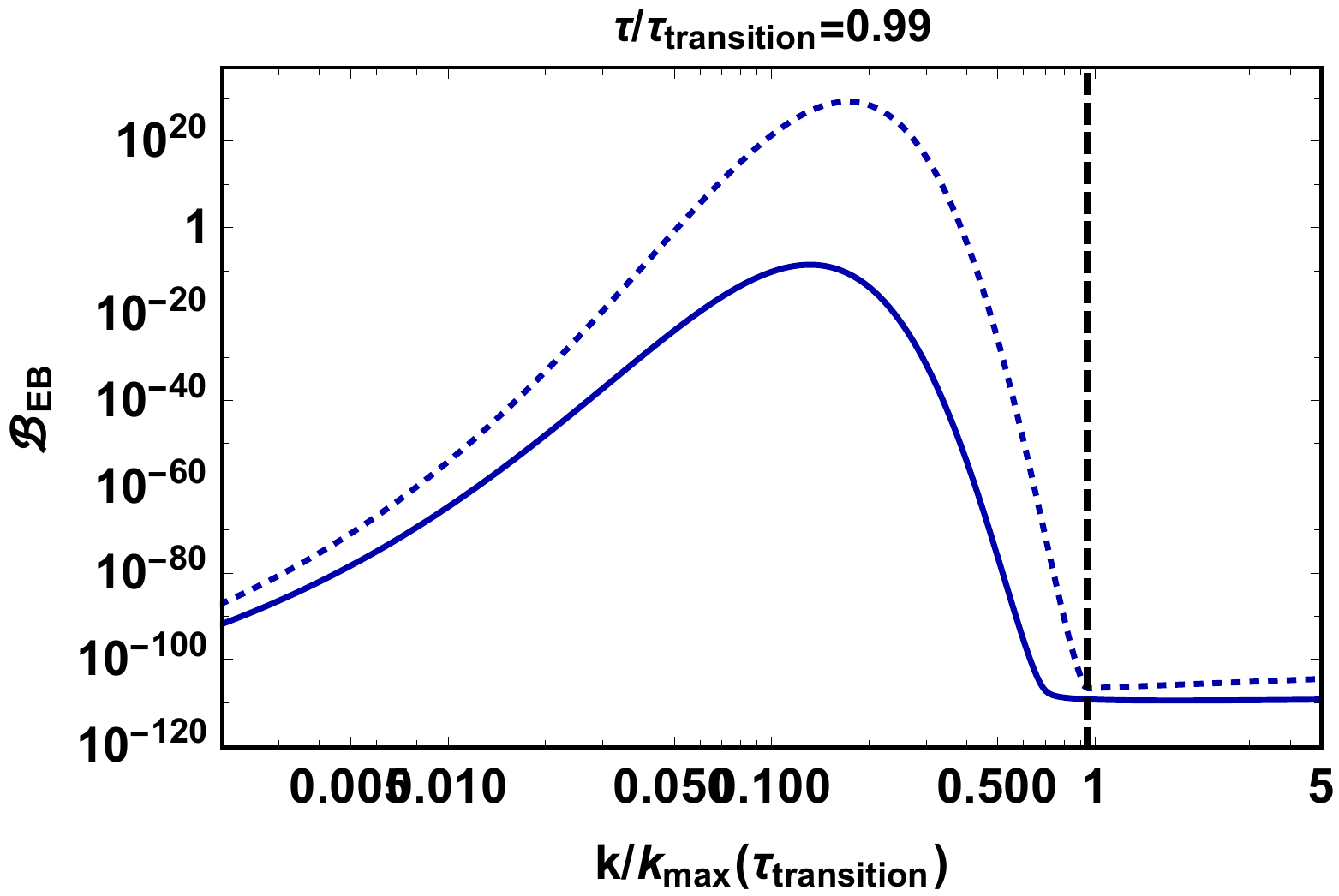}
}
\caption{Comparison of the integrand defined in (\ref{back}) using the analytic approximation given in eq (\ref{eq:early}) (dotted line) vs.
using the numerical evolution of the system of equations (\ref{A-eom}) and (\ref{eq:eomphi}) (solid line), for increasing values of the time.
The value of the coupling constant is $f=5 \cdot 10^{-4} M_p$ and we set $V_0=10^{-120} M_p^4$ for the amplitude of the scalar field potential.
The vertical line denotes the threshold between the stable (on the right) and unstable (on the left) modes and is given by (\ref{eq:limits}).
}
\label{fig:integrand}
\end{figure}

We fix $\bar{\rho}_m=10^{12}$ and present in figures \ref{fig:system2} and \ref{fig:integrand} the comparison between our analytic early time solutions and a numerical evolution that ignores contributions of backreaction, but takes into account the transition from matter to dark energy domination consistently. We see that our analytic approximations capture the dynamics very accurately deep in the matter domination regime and for negligible backreaction.

\subsubsection{Numerical results and comparison with observations} 
\label{sec:modelU1-MNume} 

\begin{figure}[ht!]
\centerline{
\includegraphics[width=0.32\textwidth,angle=0]{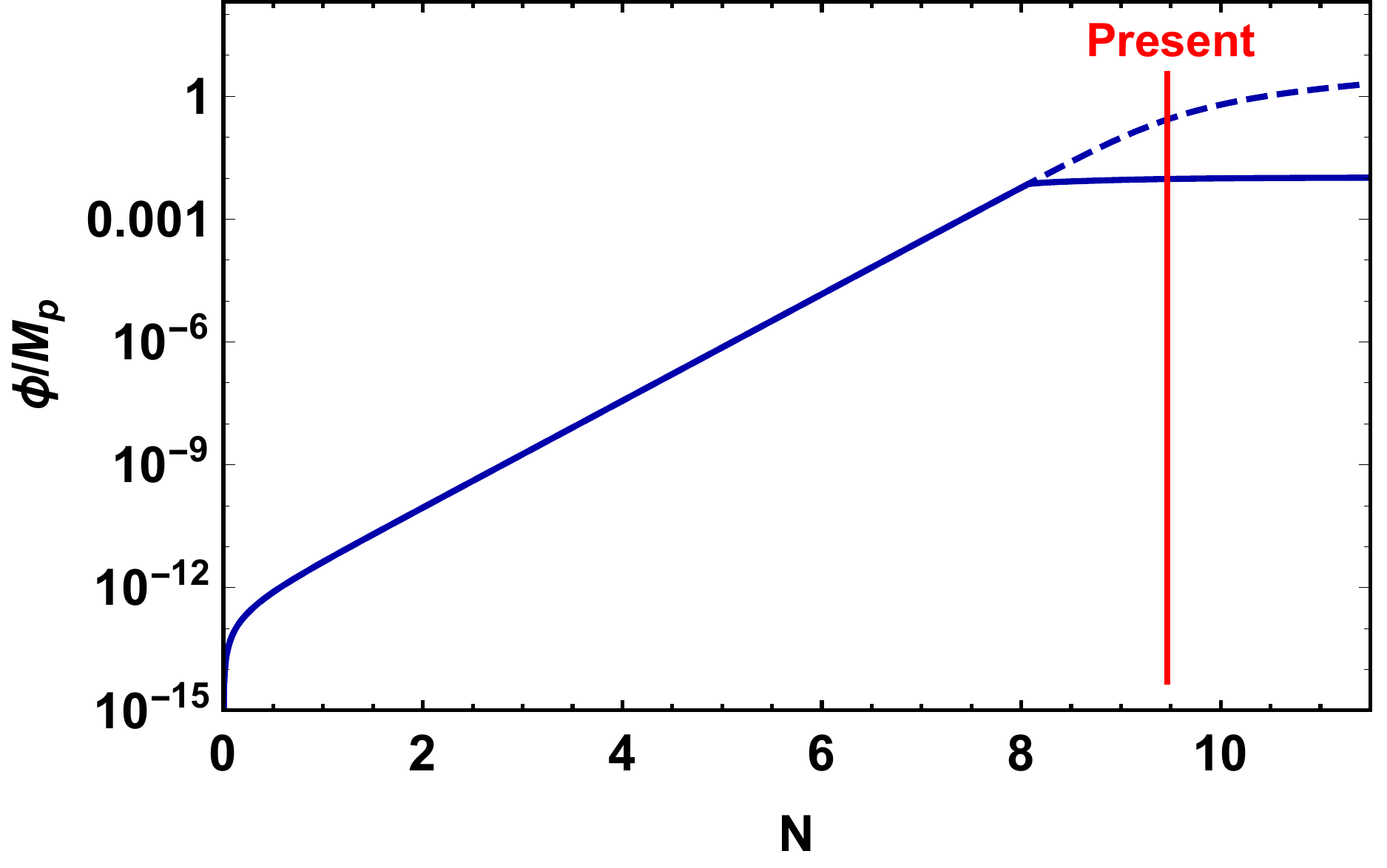}
\includegraphics[width=0.32\textwidth,angle=0]{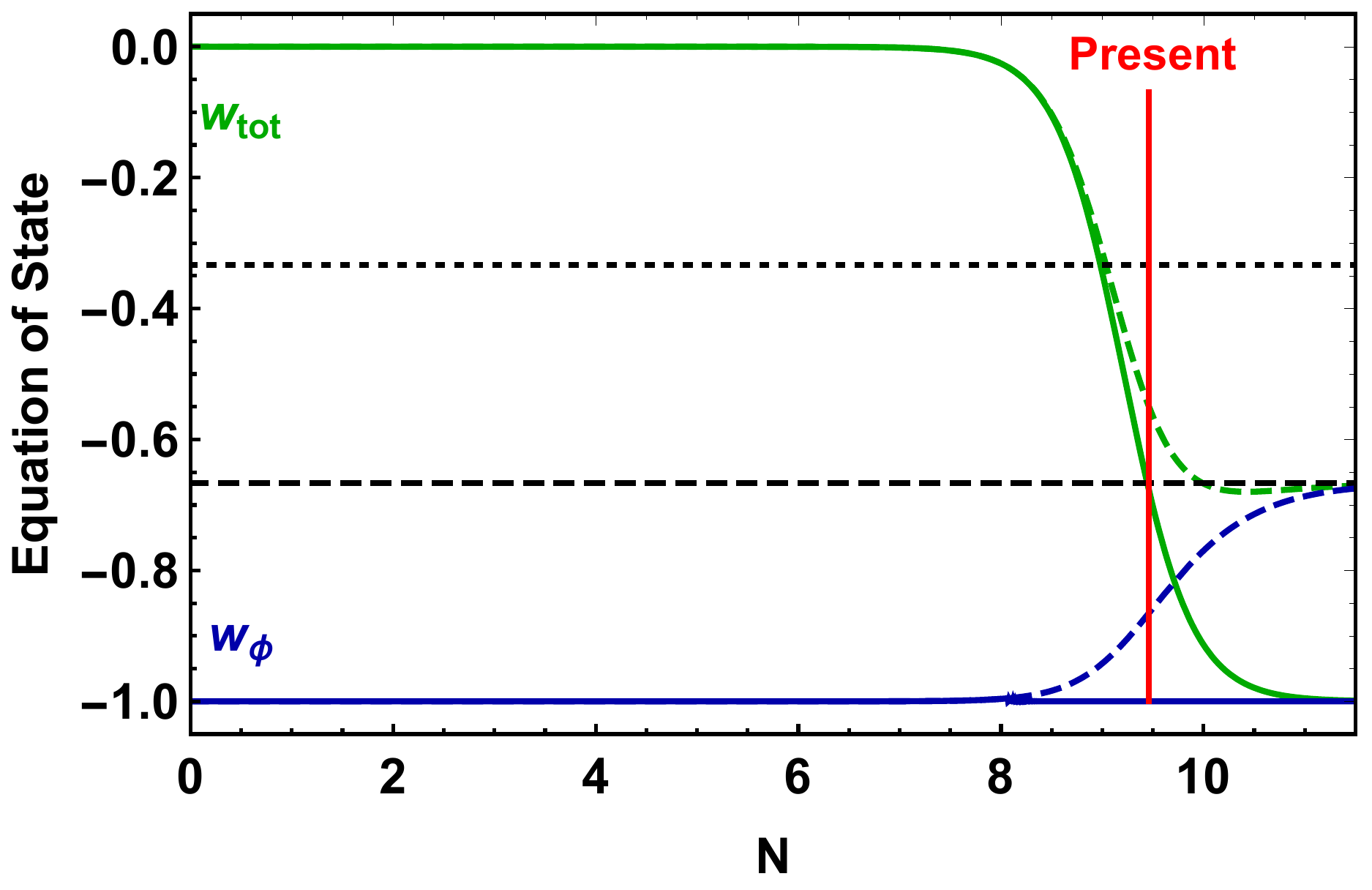}
\includegraphics[width=0.32\textwidth,angle=0]{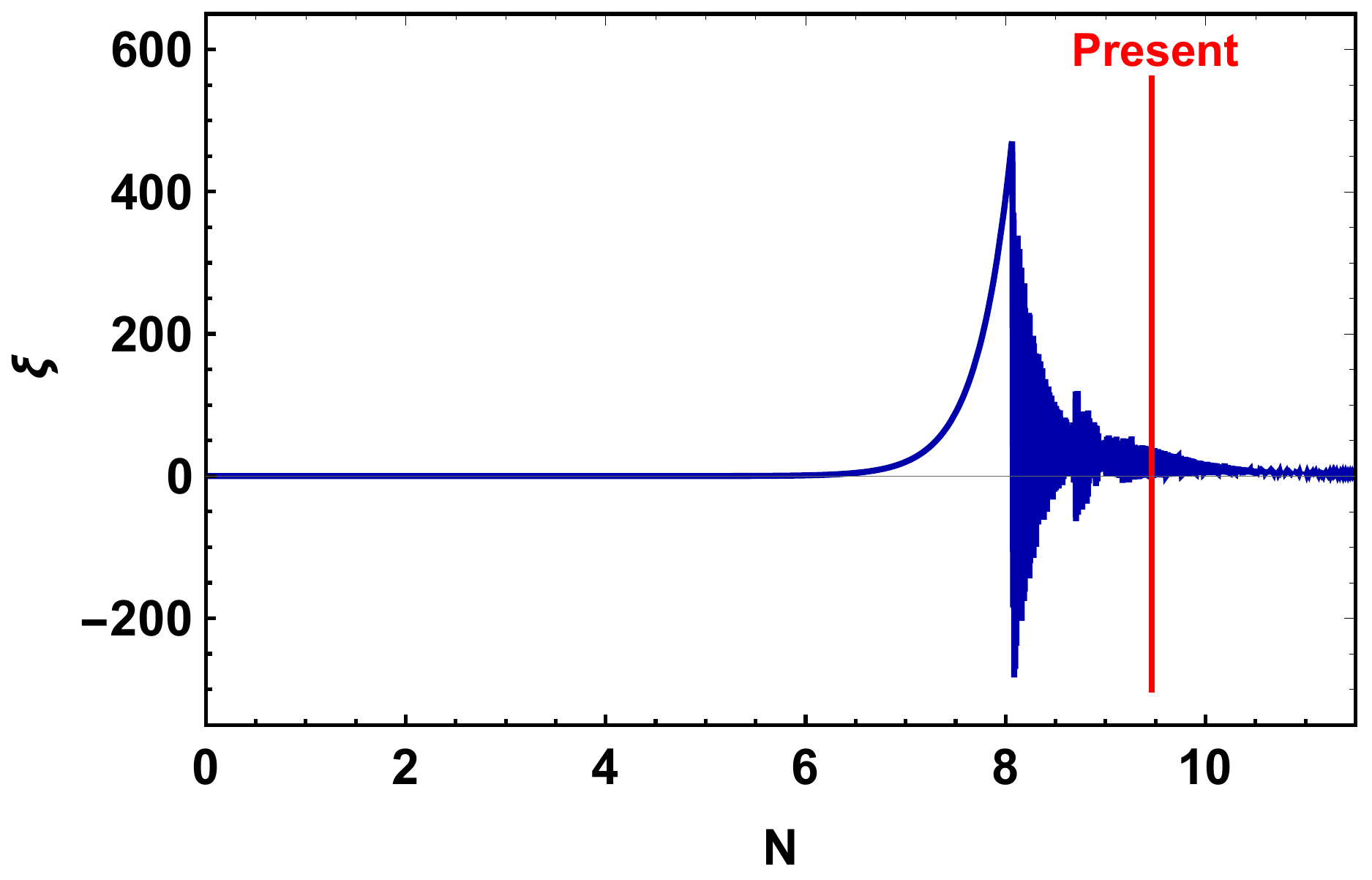}
}

\centerline{
\includegraphics[width=0.32\textwidth,angle=0]{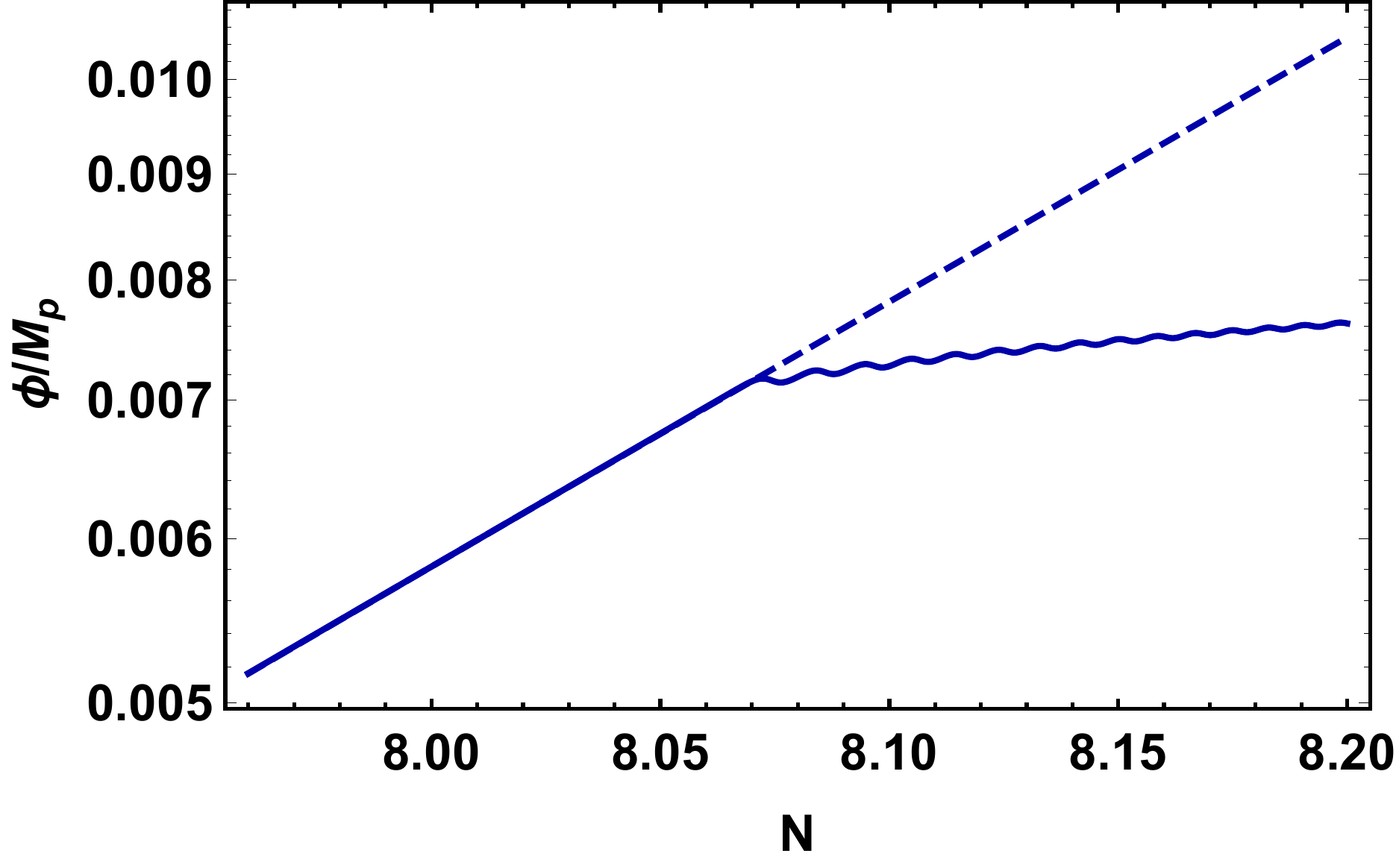}
\includegraphics[width=0.32\textwidth,angle=0]{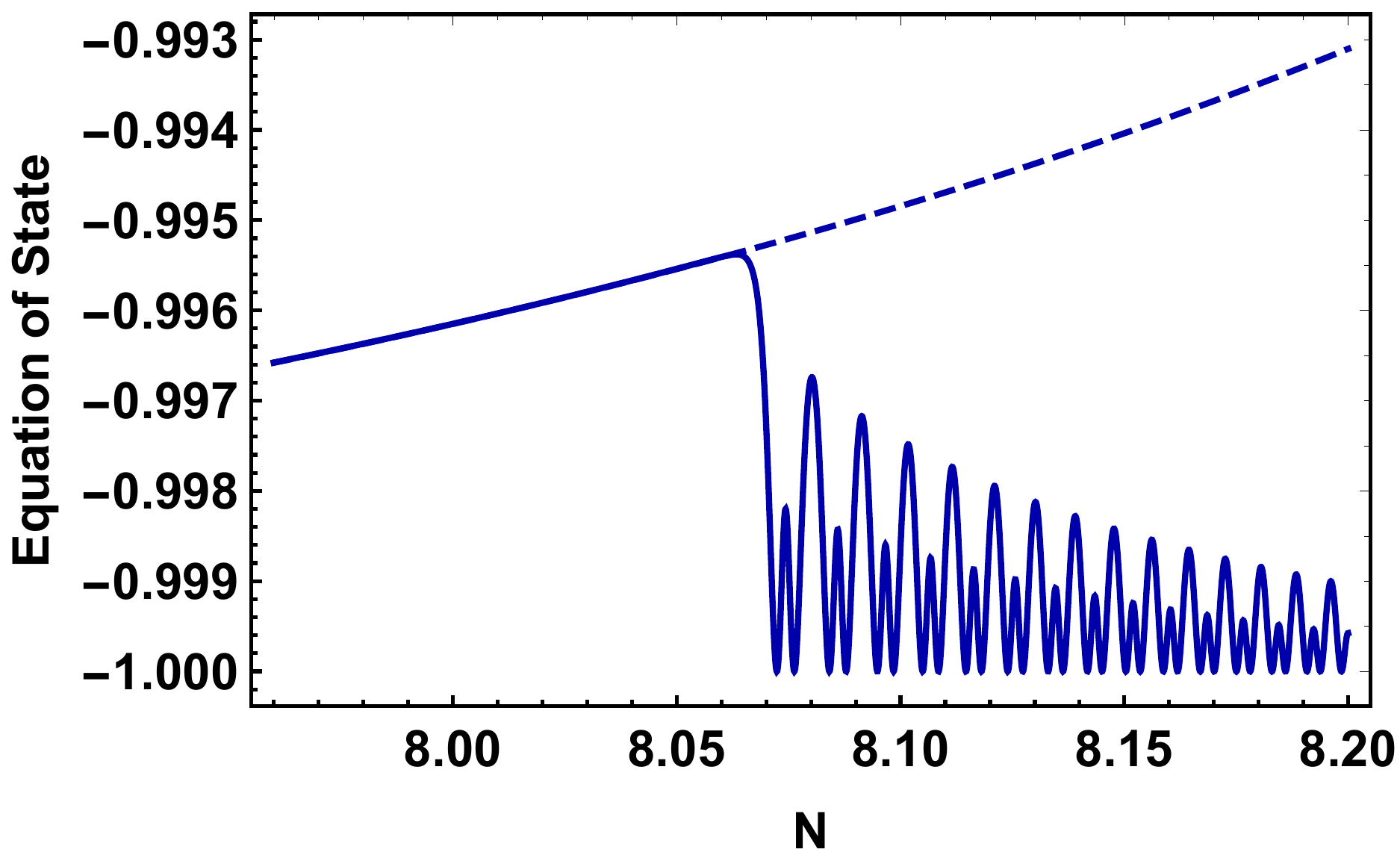}
\includegraphics[width=0.32\textwidth,angle=0]{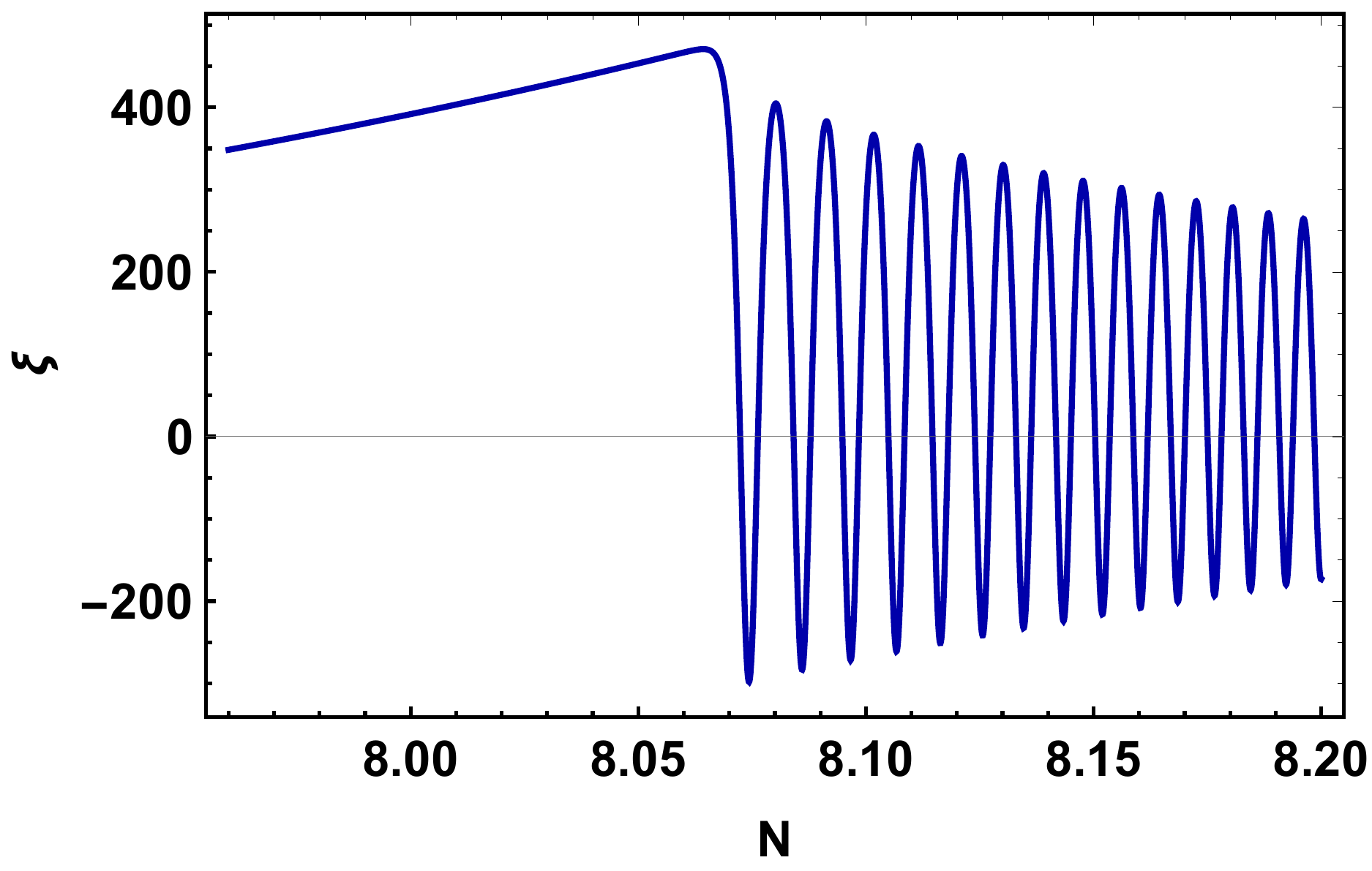}
}

\caption{The top three panels display the comparison between our full numerical solution including backreaction and the numerical solution that neglects backreaction. The lower three panels display the same quantity, but the "time" has been rescaled so that it is easier to see what happens to the various quantities at the instant of transition from negligible to dominant backreaction. The parameters used here are $\lambda=1$, $V_0=10^{-120} M_p^4$, $\bar{\rho}_m=10^{12}$ and $f=2.2\cdot 10^{-5} M_p$.
}
\label{fig:comparison}
\end{figure}

{\hskip 2em}Equipped with the insights gained in the previous section one can proceed to study the high backreaction regime numerically. At early times the backreaction is negligible but as the axion field speeds up the unstable modes grow exponentially rapidly until they eventually dominate the equation of motion of the axion field. We display the comparison between the full numerical solution including backreaction and a numerical solution neglecting backreaction in figure \ref{fig:comparison}. For this part we change the "time" variable from conformal time $\tau$ to the number of e-folds defined as $N\equiv \log({a})$. The exact equations that we evolve can be found in Appendix B of \cite{DallAgata:2019yrr} and the details of the numerical scheme used is found in Appendix C of the same reference. For other works that use similar discretization techniques on related models see \cite{Cheng:2015oqa,Cheng:2018yyr,Notari:2016npn,Ferreira:2017lnd,Domcke:2020zez}. Also for numerical analysis using lattice simulations see \cite{Adshead:2015pva,Adshead:2019lbr,Cuissa:2018oiw,Agrawal:2018vin}.

One can observe from the various panels, that the backreaction becomes dominant at a very sharp instant of time. After that point, the value of the axion field becomes nearly frozen while undergoing fast oscillations with very small amplitude. These fast oscillations decay relatively rapidly. Another important observation is that the presense of the backreaction term serves the purpose of keeping the value of the equation of state of the axion field $w_\phi$ close to the observationally required value of $-1$.

The stronger the value of the coupling (or equivalently, the smaller the value of parameter $f$) the more pronounced the effects of backreaction are. This implies that for every value of $\lambda$ there should be a value $f_{\rm thr} (\lambda)$ below which the effects of backreaction make the model compatible with observational data. For simplicity here, we will only explore the value $\lambda=1$. Finding the exact value of $f_{\rm thr}$ is challenging as a direct comparison with the data would require a complete data analysis that would use our numerical expressions as inputs. For example, in order to compare with observations of supernovae, one would have to compute the luminosity distance $d_L=\frac{a_0^2}{a}\int^{a_0}_{a}\frac{da'}{a'^2 H(a')}$ which requires knowledge of the time dependence of the scale factor $a(\tau)$ which, in turn, depends on the time evolution of the equation of state parameter of the axion field $w_\phi(\tau)$. Instead of performing a full numerical analysis \cite{DallAgata:2019yrr} finds two values labeled $f_1$ and $f_2$. Value $f_1$ denotes the upper limit for which the model could be compatible with observations since for that value backreaction becomes relevant exactly at the present moment. As a result the effects of backreaction of the model for any value $f > f_1$ is exactly the same as if there was no Chern-Simons coupling at all. At the other extreme, $f_2$ is defined as a value for which the model is guaranteed to be compatible with the data using a conservative criterion described in the next paragraph. This implies that for every value $f< f_2$ the model will be compatible with observations. 

In order to find $f_2$ we will increase the coupling sufficiently enough so that the backreaction becomes dominant at very early times. After backreaction becomes dominant, the fast oscillations observed in figure \ref{fig:comparison} decay sufficiently enough so that the evolution of $w_\phi$ can be considered linear. This allows us to use the standard CPL parametrization \cite{Chevallier:2000qy,Linder:2002et}

\begin{equation}
w_\phi(a)=w_0+\left(1-\frac{a}{a_0}\right)w_a
\end{equation}

We then use the results of \cite{Akrami:2018odb} which shows the allowed region of $(w_0,w_a)$ for different data sets. We focus here on the cyan contours which display the most stringent constrains. We set as our criterion for comparibility with the data that the values $\left(w_\phi,\frac{d w_\phi}{da}\right)$ have to be within the 1-$\sigma$ cyan contours for a full e-fold before the present moment. In this way, the late oscillations can be considered small while any large initial oscillations in the equation of state of the axion took place very early when the axion was a negligibly small contribution to the total energy of the universe. 

\begin{figure}[ht!]
\centerline{
\includegraphics[width=0.50\textwidth,angle=0]{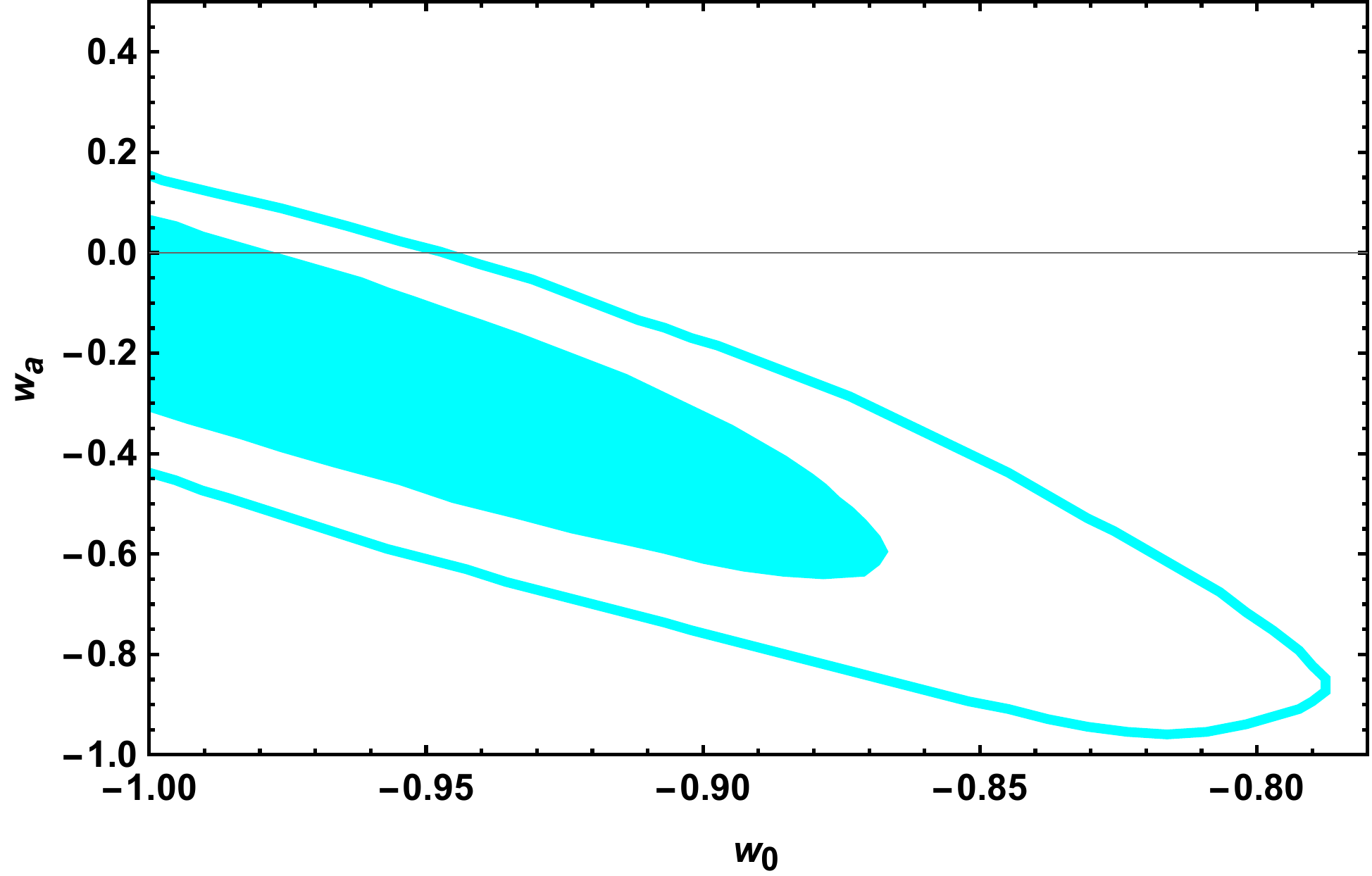}
\includegraphics[width=0.50\textwidth,angle=0]{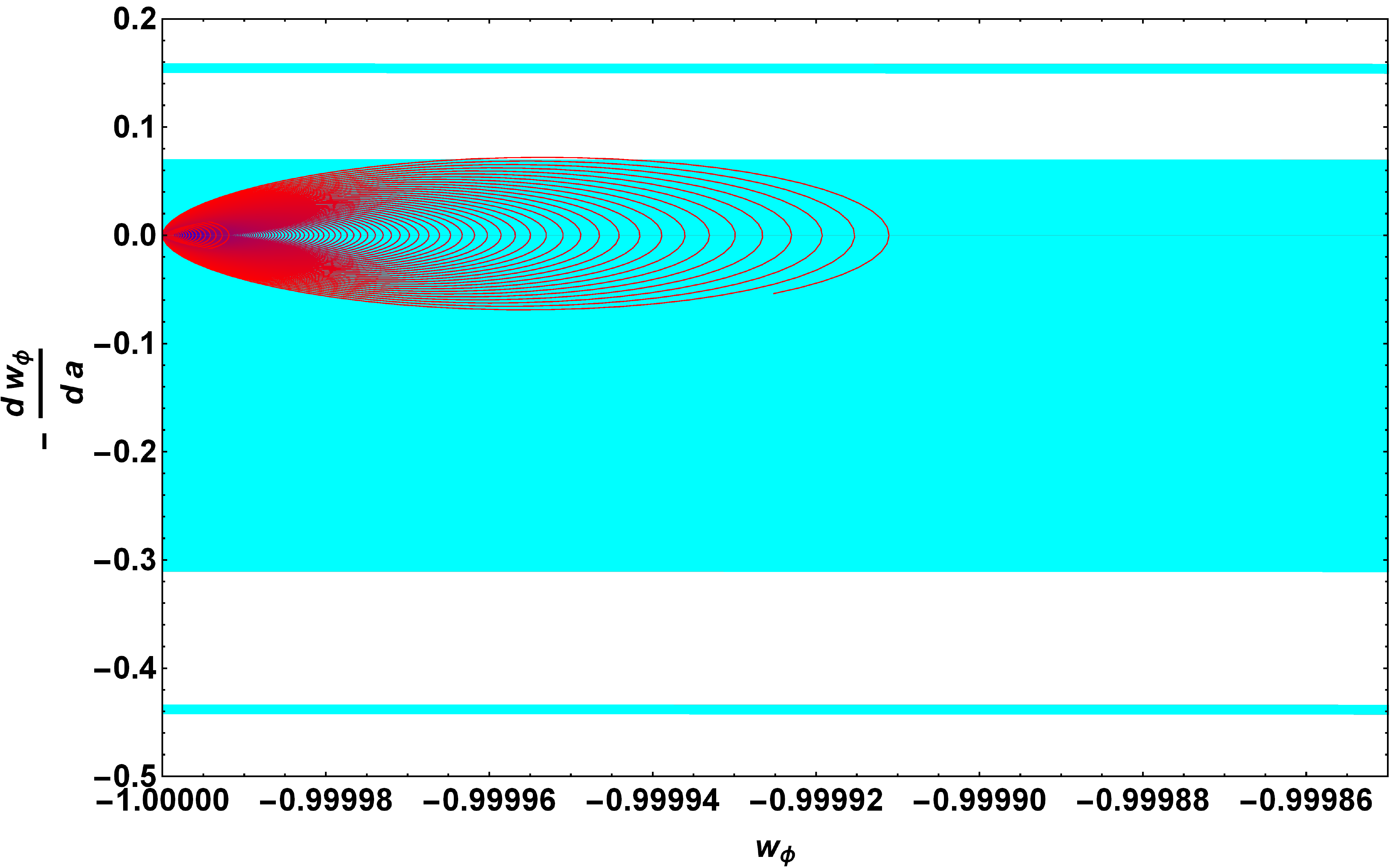}
}

\caption{\textit{Left Panel}: The allowed contours shown in figure 30 of \cite{Akrami:2018odb}. \textit{Right Panel}: evolution of the parameters $(w_\phi,\frac{d w_{\phi}}{da })$ for one e-fold before the present moment. The color evolves from red to blue (past to future). the parameters used are $f=2.2\cdot 10^{-5}$, $V_0=10^{-120} M_p$ and $\bar{\rho}_{m}=10^{12}$.
}
\label{fig:Planck}
\end{figure}

Figure \ref{fig:Planck} demonstrates the comparison between data and our numerical solution for the value $f_2=2.2\cdot 10^{-5} M_p$. Any value smaller than $f_2$ will yield an evolution that is even closer to the $(-1,0)$ value in the parameter space of $(w_\phi,\frac{d w_\phi}{da})$ and therefore will also be compatible with observations. We would also like to stress here that the comparison with the data is not affected by variation of the parameter $\bar{\rho}_m$ and our choices in the above figures are arbitrary as long as $\bar{\rho}_m \gg 1$

To conclude this subsection, instead of a rigorous data comparison, we have come up with criteria that allow us to identify the parameter space for which we can be confident that the model is compatible or incompatible with observations. We have demonstrated that compatibility of the model with observations is dependent on the value of the coupling $f$. For a typical value $\lambda=1$ the true threshold value of the coupling constant is in the range

\begin{equation}
2.2\cdot 10^{-5} < \frac{f_{\rm thr}}{M_p} < 9.25\cdot 10^{-4}
\end{equation}

\vspace*{0.5 cm}

\subsection{Gravitational wave production in the axion-U(1) case} 
\label{sec:modelU1-gw} 

{\hskip 2em}The copious amount of gauge field amplification that takes place in our model can be used to source gravitational waves. These gravitational waves could in principle impart their signature in the form of spectral distortions in the CMB \cite{Kite:2020uix}. In order for these gravitational waves to be detectable from spectral distortions, their frequencies should be at least $f_{\rm frequency}\sim 10^{-16} \;{\rm Hz}$ which is approximately the lowest frequency that can be probed from measurements of spectral distortions. In exploring the possibility of detectable gravitational waves, we will focus on values of the coupling constant that are $f< f_2=2.2 \cdot 10^{-5} M_p$ so that we can be confident that the model is compatible with observations. Another benefit of focusing on such strong couplings is that the backreaction becomes strong early enough that we can use the approximate early time analytic solutions (\ref{eq:early}) instead of relying on numerical solutions. We will compute the gravitational waves produced before the strong backreaction regime and neglect any gravitational waves produced in the strong backreaction regime. As we show below this approach captures the regime in which the greatest possible gravitational wave frequencies are produced and any contribution after that moment will be small. In this way we set a lower limit to the potential gravitational wave signal that could be produced by this mechanism while simultaneously relying on simple analytic formulas for the estimation of the gravitational wave spectrum.

We devote subsection \ref{sec:modelU1-gwApprox} to providing some simple analytic formulas that can assist in diagnosing the frequencies of the gravitational waves and then the full results of the power spectrum are presented in subsection \ref{sec:modelU1-gwPS}.

\subsubsection{Frequency diagnostics} 
\label{sec:modelU1-gwApprox} 

\begin{figure}[ht!]
\centerline{
\includegraphics[width=0.5\textwidth,angle=0]{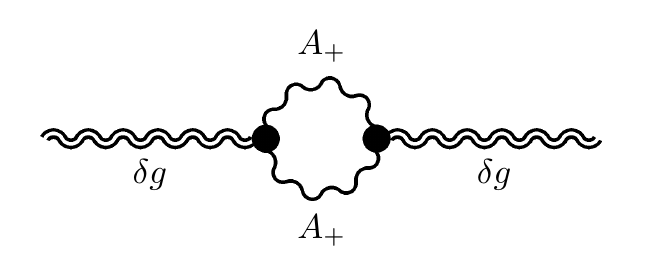}
}

\caption{Diagramatic representation of the nonlinear sourcing of gravitational waves by the unstable polarization of the gauge field $A_+$
}
\label{fig:FeynmanU(1)}
\end{figure}

{\hskip 2em}To begin our analysis, we will develop some diagnostic tools that allow us to estimate the frequencies of the gravitational waves that are produced as a function of the coupling strength $f$. The process that produces gravitational waves is graphically represented in figure \ref{fig:FeynmanU(1)}. The physics of this process is completely analogous to the gravitational wave production of the natural inflation case with a U(1) gauge field performed in \cite{Sorbo:2011rz,Barnaby:2011qe}. 

Even though the process is nonlinear, one expects that the frequencies of the gravitational waves will be in the same ball-park as the frequencies of the unstable gauge modes that source them. The value of the greatest excited comoving wavenumber at each instant in time is given simply by

\begin{equation}
k_{\rm max}\left(\tau\right)=2 a H \xi 
\end{equation}

Assuming that the production happens late during matter domination we can estimate the hubble rate by its present value. Converting the quantity above into a physical frequency we get

\begin{equation}
f_{\rm max}=\frac{k_{\rm max}}{2\,\pi\,a}=\frac{H_0 \xi}{\pi}
\end{equation}

The formula above provides a crude approximation of the frequencies that one should expect. Since $H_0\sim 2\cdot  10^{-18}\, {\rm Hz}$ the overall frequency can be as high as $f_{\rm max}=10^{-15}\, {\rm Hz}$ because $\xi$ can grow up to a value $\xi \sim {\cal O} \left(10^{2}\right)$ as can be seen by figure \ref{fig:comparison}.

We can in fact make a more precise estimation by making use of formula (\ref{eq:limits}).

\begin{equation}
k_{\rm max}\left(\tau\right)=\frac{1}{f}\frac{d\phi}{d\tau}\simeq \frac{2\lambda}{3\sqrt{3}}\frac{ V_0^{1/2}}{\bar{\rho}_m^{1/2} f}\left(\frac{\tau}{\tau_{\rm in}}\right)^5
\end{equation}

The dependence on the initial condition parameters $\bar{\rho}_m$ and $\tau_{\rm in}$ will vanish shortly when we convert the comoving wavenumber to the physical frequency. Since we only care about gravitational waves that are produced before backreaction becomes dominant, we evaluate the formula above at the instant when backreaction becomes important. We label that time as $\tau_{\rm backreaction}$. The maximum unstable wavenumber then becomes 

\begin{equation}
k_{\rm max}\left(\tau_{\rm backreaction}\right)\simeq \frac{2\lambda}{3\sqrt{3}}\frac{ V_0^{1/2}}{\bar{\rho}_m^{1/2} f}\left(\frac{\tau_{\rm backreaction}}{\tau_{\rm in}}\right)^5
\label{eq:kmax}
\end{equation}

If we focus our attention to the coupling constant $f$ for the time being, we see that there are two opposing effects that control the magnitude of $k_{\rm max}(\tau_{\rm backreaction})$. Since $\tau_{\rm in}$ is independent of $f$, a stronger coupling (smaller value of $f$) will yield a greater value of the greatest excited wavenumber because $f$ is present in the denominator on the right hand side of (\ref{eq:kmax}). On the other hand, a stronger coupling will make the backreaction dominant earlier therefore lowering the value of $\tau_{\rm backreaction}$ which results in a smaller value of $k_{\rm max}(\tau_{\rm backreaction})$. Since these are two opposing effects, at first sight it is difficult to discern whether a stronger coupling overall results in a greater wavenumber or not. In order to resolve this difficulty, we perform multiple numerical evolutions and record the time $\tau_{\rm backreaction}$ for various values of $f$. We then perform a parameter fitting and find the expression

\begin{equation}
\tau_{\rm backreaction}\left(f\right)=C_1 \left(\frac{f}{M_p}\right)^{C_2} \tau_{\rm in} \, \bar{\rho}_m^{1/6}
\label{eq:timeback}
\end{equation}

which is correct specifically for $\lambda=1$ and with the constants $C_1$ and $C_2$ given by 

\begin{equation}
C_1= 3.60,\;\;\;\;C_2=0.172
\end{equation}

This expression is accurate to a $1\%$ level and it has been checked for a range of values $\frac{f}{M_p}\rightarrow (10^{-8},10^{-5})$. We can now eliminate the variable $\tau_{\rm backreaction}\left(f\right)$ in (\ref{eq:kmax}) and see the true dependence of the greatest possible wavenumber on the various parameters.

\begin{equation}
k_{\rm max}\left(\tau_{\rm backreaction}\right)\simeq \frac{2 \, \bar{\rho}_m^{1/3}C_1^5}{3\sqrt{3}}\left(\frac{V_0^{1/2}}{f}\right)\left(\frac{f}{M_p}\right)^{5\,C_2}
\label{eq:kmaxtrue1}
\end{equation}

This can be converted to the physical wavenumber evaluated at the present moment by dividing with the value of the scale factor today $a_0=\left(\frac{\rho_{0,\rm dark\;energy}}{\rho_{0,\rm matter}}\right)^{1/3}\bar{\rho}_m^{1/3} \simeq 1.283 \;\bar{\rho}_m^{1/3}$ where this expression assumes that the total field displacement of the axion field is much smaller than $M_p$. This will always be the case in our model. The maximum physical frequency evaluated today becomes.

\begin{equation}
f_{\rm  max}\left(\tau_{\rm backreaction}\right)\simeq \frac{181}{2\pi} \frac{\sqrt{V_0}}{M_p}\left(\frac{M_p}{f}\right)^{0.139}
\label{eq:kmaxtrue2}
\end{equation}

\begin{figure}[ht!]
\centerline{
\includegraphics[width=1\textwidth,angle=0]{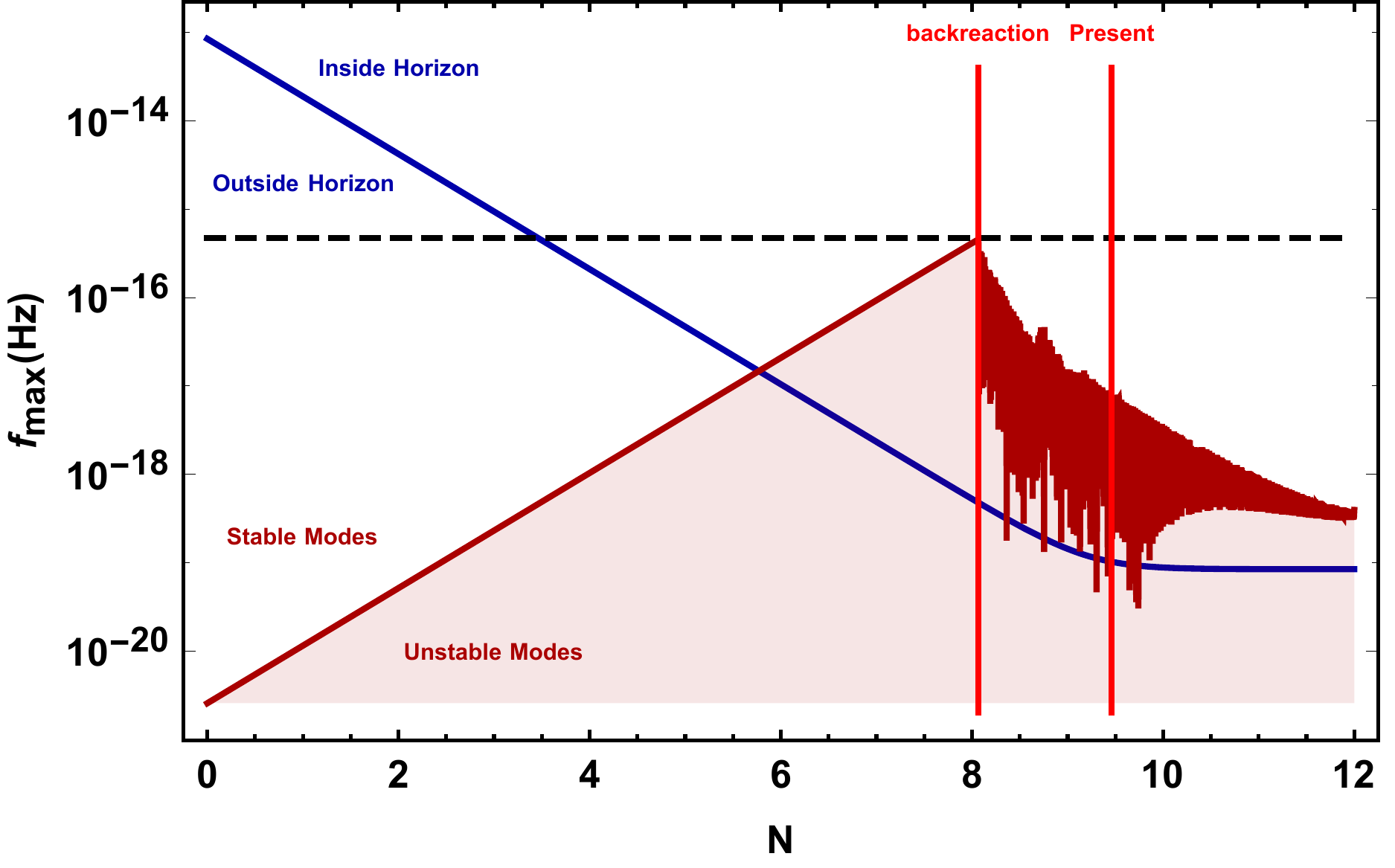}
}

\caption{Comparison between the maximum physical unstable frequency (maroon line), Hubble rate (blue line) and our analytic prediction (\ref{eq:kmaxtrue2}) (black dashed line). The first two quantities have been converted into frequencies and have been normalized by dividing them by the ratio of the value of the scale factor today over the value of the scale factor at $\tau_{\rm backreaction}$. The two red vertical lines point out the instants when the backreaction becomes important, and the present moment.
}
\label{fig:Comparisonfreq}
\end{figure}

This expression reveals the true dependence of the maximum physical frequency in terms of the coupling constant $f$. It is now apparent that the lower $f$ is, the greater the GW frequencies one should expect. However, since the power of the last factor is relatively small, one would need to decrease the value of $f$ by more than ten orders of magnitude in order to achieve frequencies of the order of PTA experiments. For this reason we limit our study to measurements of spectral distortions of the CMB which can probe gravitational waves in the range of $f\simeq{\cal O}(10^{-15})\;{\rm Hz}$. Finally, we display in figure \ref{fig:Comparisonfreq} the comparison between the maximum unstable frequency as a function of time as evaluated in an exact way using our numerical solution (maroon line), the prediction of our simple analytic estimation (\ref{eq:kmaxtrue2}) (black dashed line) and the value of the hubble rate (blue line). All of these quantities have been converted to a physical frequency and normalized by dividing them by the ratio of the value of the scale factor today over the value of the scale factor at $\tau_{\rm backreaction}$. One can observe that the maximum unstable frequency grows until backreaction becomes important, then the maximum frequency undergoes fast oscillations and decays away. This picture also illustrates why we will only attempt to compute the gravitational waves produced until the moment the backreaction becomes important. After the backreaction becomes important the maximum frequency stops increasing and since the particle production parameter also decreases (as seen in figure \ref{fig:comparison}) one expects any further gravitational waves produced after the $\tau_{\rm backreactiom}$ moment to be a small contribution to the power spectrum.

\subsubsection{Gravitational wave power spectrum} 
\label{sec:modelU1-gwPS} 

{\hskip 2em}We focus now on the computation of the gravitational wave power spectrum due to the burst of particle production that immediately precedes the strong backreaction regime, deep into matter domination. We define the line element as 

\begin{equation}
ds^2=a^2\left(\tau\right)\left[-d\tau^2+\left(\delta_{ij}+h_{ij}\right)dx^i dx^j\right]
\end{equation}

where $h_{ij}$ represents tensor perturbations of the metric. The energy density of gravitational waves in the TT gauge is given by the formula (1.136) of \cite{Maggiore:1900zz}.

\begin{equation}
\rho_{\rm GW}=\frac{M_p^2}{4 a^2}\left\langle h'_{ij} h'_{ij}\right\rangle
\end{equation}

expressed here with respect to conformal time. We decompose the tensor modes in fourier space and expand them in a chiral basis.

\begin{equation}
h_{ij}\left(\tau,\vec{x}\right)=\int\frac{d^3 k}{\left(2\pi\right)^{3/2}}{\rm e}^{i \vec{k}\cdot\vec{x}}\sum_{\lambda=+,-}\Pi^*_{ij,\lambda}\left(\hat{k}\right)h_\lambda\left(\tau,\vec{k}\right)
\end{equation}

The definitions used and more technical details can be found in Appendix \ref{app:twopoint}. The energy density then becomes 

\begin{equation}
\rho_{\rm GW}=\frac{M_p^2}{4 a^2} \int \frac{d^3 k\, d^3 k'}{\left(2\pi\right)^3}{\rm e}^{i \vec{x}\cdot\left(\vec{k}+\vec{k}'\right)}\left\langle\Pi^*_{ij,\lambda}\left(\vec{k}\right)\Pi^*_{ij,\sigma}\left(\vec{k'}\right)h'_\lambda\left(\tau,\vec{k}\right)h'_\sigma\left(\tau,\vec{k'}\right)\right\rangle
\end{equation}

At this point, for notational convenience, we define the quantity

\begin{equation}
\left\langle \partial_0^m h_\lambda\left(\tau,\vec{k}\right)\partial_0^n h_\sigma\left(\tau,\vec{k'}\right)\right\rangle=\delta_{\lambda\sigma}\delta^{(3)}\left(\vec{k}+\vec{k'}\right)\frac{2\pi^2}{k^3}P_\lambda^{(m,n)}\left(\tau,k\right)
\end{equation}

The expression for the energy density is simplified to

\begin{equation}
\rho_{\rm GW}=\frac{M_p^2}{4 a^2} \int \frac{d^3 k}{\left(2\pi\right)^3}\frac{2\pi^2}{k^3}\sum_\lambda\Pi^*_{ij,\lambda}\left(\vec{k}\right)\Pi^*_{ij,\lambda}\left(-\vec{k}\right)P^{(1,1)}_\lambda\left(\tau,k\right)
\end{equation}

Collecting all the various factors and simplifying we get

\begin{equation}
\rho_{\rm GW}=\frac{M_p^2}{4 a^2} \int \frac{dk}{k}\sum_\lambda P^{(1,1)}_\lambda\left(\tau,k\right)
\end{equation}

Using 

\begin{equation}
\rho_{\rm GW}=\int d\log k \,\frac{\partial \rho_{GW}}{\partial \log k}
\end{equation}

We define the gravitational wave power spectrum today as

\begin{equation}
\Omega_{\rm GW,0}\left(k\right)=\left(\frac{a}{a_0}\right)^4\frac{1}{3 M_p^2 H_0^2}\frac{\partial \rho_{GW}}{\partial \log k}= \left(\frac{a}{a_0}\right)^2\frac{1}{12 a_0^2 H_0^2}\sum_\lambda P_\lambda^{(1,1)}\left(\tau_{\rm backreaction},k\right)
\label{eq:power}
\end{equation}

where we have assumed that the energy density of GW decays as radiation after it has been produced. The power spectra in the right hand side of (\ref{eq:power}) are computed up to the instant $\tau_{\rm backreaction}$. We ignore further GW production after that instant. In that sense, our computation should be considered a lower limit for the power spectrum, but one does not expect much production to take place after that instant because the axion field slows down dramatically and therefore the instability in the gauge modes becomes less efficient. An added benefit to stopping our calculation at $\tau_{\rm backreaction}$ is that we can perform our calculation using the simple analytic formulas found in the previous subsection (\ref{eq:early}).

\begin{figure}[ht!]
\centerline{
\includegraphics[width=0.50\textwidth,angle=0]{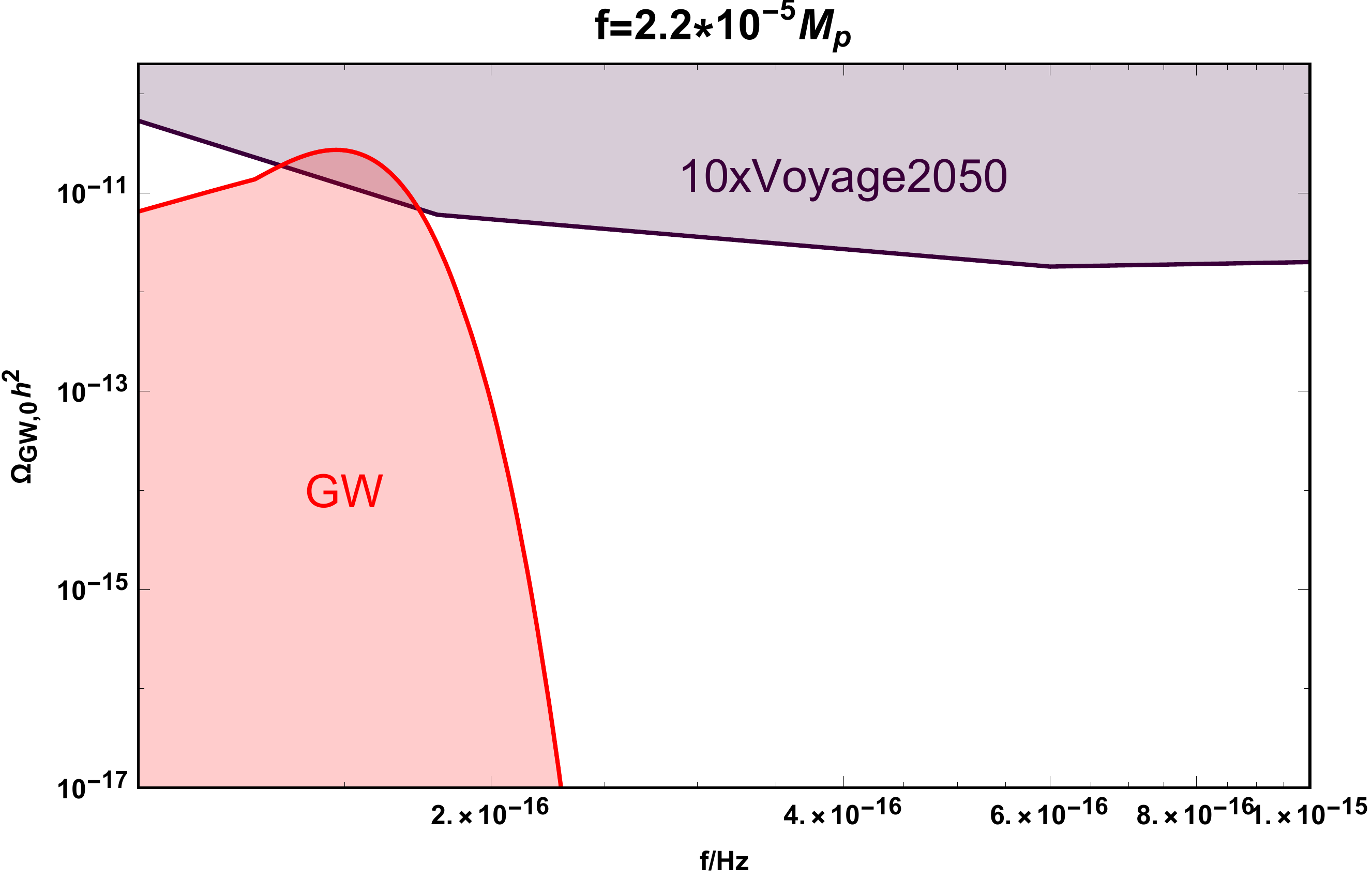}
\includegraphics[width=0.50\textwidth,angle=0]{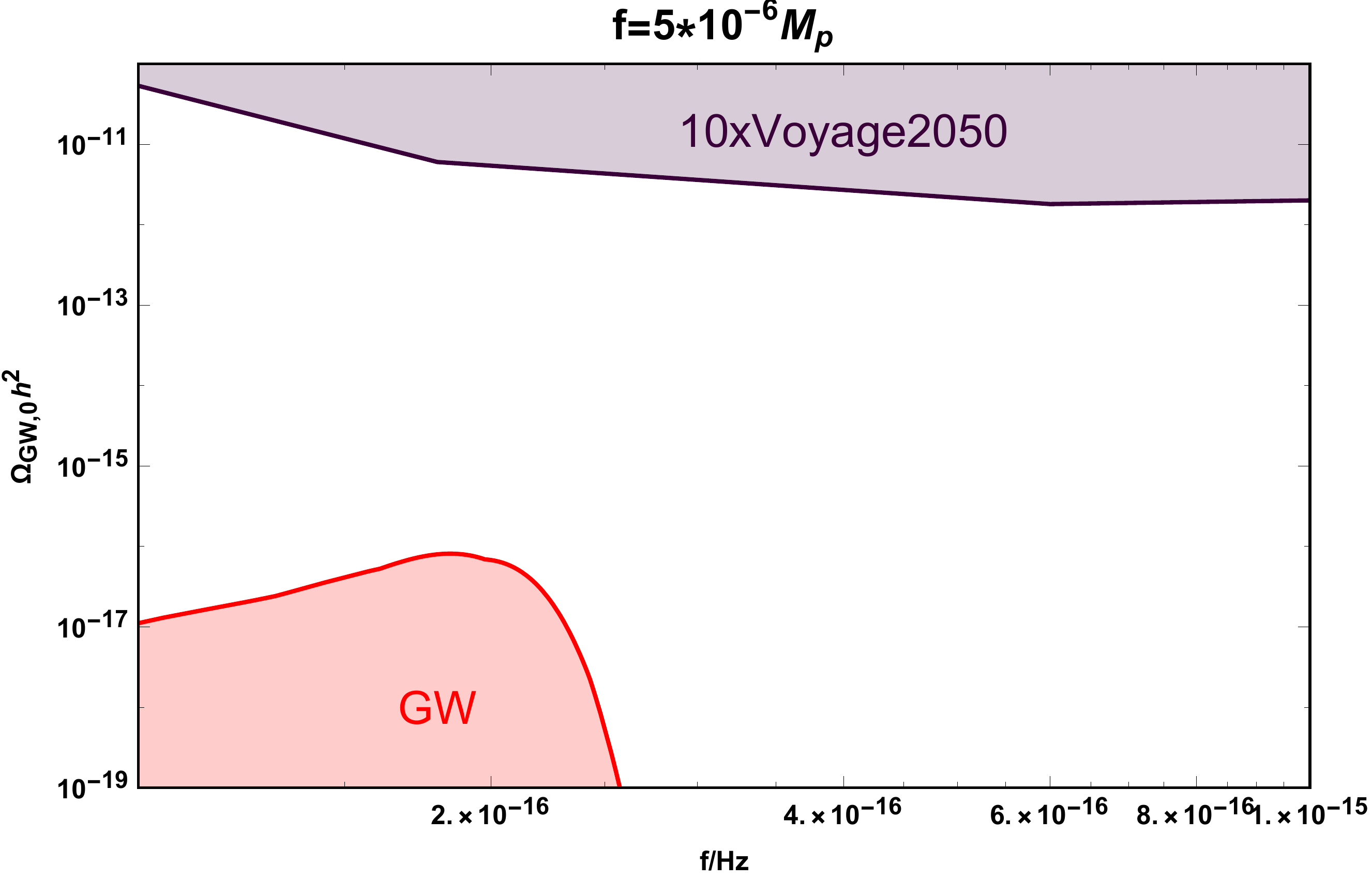}
}

\caption{Examples of the gravitational wave power spectrum for two distinct parameter choices. We display the power spectrum comparison to the sensitivity curves of the most sensitive experimental configuration that appears in figure 2 of reference \cite{Kite:2020uix}.
}
\label{fig:GWSpectra}
\end{figure}

We refer the reader to Appendix \ref{app:twopoint} for a detailed account of the derivation of the two point function using the Green function method. We skip the details here and instead display the final power spectra for two choices of parameters in figure \ref{fig:GWSpectra}. There are two examples of parameter choices in the figure. One can observe that the frequencies found are consistent with our approximate expression (\ref{eq:kmaxtrue2}) and that the coupling strength can be chosen in a way that the power spectrum overlaps with the sensitivity curves of the various $\mu$-distortion experiments. The lines for the experimental sensitivities have been taken from \cite{Kite:2020uix}.

In summary, we have performed a detailed computation of the gravitational wave power spectrum that results from the burst of gauge field particle production at the moments preceding backreaction domination in the equation of motion of the axion field and we have specified the parameter space requirements that allow for these gravitational waves to have high enough frequency to be observable by spectral distortion experiments. We would like to remind the reader that our results are valid for the choice $\lambda=1$ but one would expect similar results for values of $\lambda \simeq {\cal O}(1)$.

\section{Accelerated expansion from dissipation with an axion-SU(2) coupling} 
\label{sec:modelSU2} 

{\hskip 2em}To further our exploration of observationally compatible quintessence on a steep potential we replace the Abelian U(1) gauge field with a non-Abelian SU(2) gauge field. This should be viewed as a natural next step in gaining a better understanding of the posibility of Chern-Simons couplings to reconsile the Swampland conjectures with dark energy observations. Our analysis of the SU(2) case will follow a similar vein as in \cite{DallAgata:2019yrr}. The main mechanism is similar in the sense that slow roll will be maintained as a result of the axion field producing the gauge field at the expense of its own kinetic energy, but the two models will have some important differences as well. The most important difference is that in the SU(2) case, it can be assumed that the gauge field acquires an isotropic vacuum expectation value \cite{Maleknejad:2011jw,Maleknejad:2011sq} and as a result the friction terms in the equation of motion of the axion field will arise strictly at the level of the background in an analogous way to the model of Chromo-Natural inflation \cite{Adshead:2012kp}. In addition to studying the model at the background level, a study of the tensor perturbations will be performed and the potential of the model to produce observable gravitational waves will be explored.

This section is divided into two subsections. In subsection \ref{sec:modelSU2-M} we present the various equations of motions and solve them analytically at early times deep into the matter domination regime assuming an independent evolution for the axion and gauge field vevs. Subsequently, we display the results of a full numerical analysis and determine the parameter space for which our model is compatible with observations. In the second subsection the possibility to produce observable gravitational waves is explored.

\subsection{Axion-SU(2) Quintessence model} 
\label{sec:modelSU2-M} 

{\hskip 2em}The action of our model takes the form

\begin{equation}
S = \int d^4 x \sqrt{-g} \left[ - \frac{1}{2} \left( \partial \phi \right)^2 - V \left( \phi \right) - \frac{1}{4} F_{\mu \nu}^a F^{a\,\mu \nu} - \frac{\phi}{4 f} F_{\mu \nu}^a {\tilde F}^{a\,\mu \nu} \right] \;, 
\label{eq:Lag2}
\end{equation}

where $\phi$ is the axion-quintessence field, the SU(2) field strength tensor is defined as $F_{\mu\nu}^a=\partial_\mu A_\nu^a-\partial_\nu A_{\mu}^a -g \epsilon^{abc} A_\mu ^b A_\nu^c $ and $\tilde{F}^{a\,\mu\nu}$ is its dual. For the non-Abelian case the vector field may obtain an isotropic vev \cite{Maleknejad:2011jw,Maleknejad:2011sq}

\begin{equation}
A^a_0=0,\;\;\;\;\;A^a_i=\delta^a_i a\left(t\right) Q\left(t\right)
\end{equation}

This is the most general expression for the spatial part of the gauge field that is compatible with the isotropy of the universe. As in the previous case, we work in a FLRW geometry with line element $ds^2=a\left(\tau\right)^2\left(-d\tau^2+\delta_{ij} \,dx^i dx^j\right)$ where $\tau$
 is the conformal time. Both $\phi$ as well as $Q$ are cosmologically relevant fields, therefore they depend solely on time and their behavior is described sufficiently well by their classical equations of motion. We find the relevant dynamical equations to be
 
\begin{eqnarray}
&&\frac{\partial^2\phi}{\partial\tau^2}+\frac{2}{a}\frac{\partial a}{\partial \tau} \frac{\partial \phi}{\partial \tau} +a^2 \frac{\partial V}{\partial\phi}=-\frac{3 \,g \,Q^3}{f}\frac{\partial a}{\partial \tau}-\frac{3 \,g \,a \,Q^2 }{f}\frac{\partial Q}{\partial \tau}\label{eq:eomphi2}\\
&&\frac{\partial^2 Q}{\partial \tau^2}+2a\,H\, \frac{\partial Q}{\partial \tau}+\left(a\frac{\partial H}{\partial \tau}+2 a^2\, H^2\right)Q +2g^2\,a^2\, Q^3=\frac{g\,a\,Q^2}{f}\frac{\partial \phi}{\partial \tau}\label{eq:gaugeeom}\\
&&\left( \frac{1}{a} \, \frac{\partial a}{\partial \tau} \right)^2 = \frac{a^2}{3 M_p^2} \Bigg[ \frac{1}{2 a^2} \left( \frac{\partial \phi}{\partial \tau} \right)^2 + V + \frac{\rho_{m,{\rm in}} \, a_{\rm in}^3}{a^3}\nonumber\\
&&\quad\quad\quad\quad \quad\quad\quad\quad\quad\quad +\frac{3}{2}g^2\,Q^4 +\frac{3 Q^2}{2 a^4}\left(\frac{\partial a}{\partial \tau}\right)^2+\frac{3 Q}{a^3}\frac{\partial a}{\partial \tau}\frac{\partial Q}{\partial \tau}+\frac{3}{2 a^2}\left(\frac{\partial Q}{\partial \tau}\right)^2\Bigg]\label{eq:friedmanSU2}
\end{eqnarray}

where the first two are the equations of motion for the axion and gauge field respectively and the third one is the first Friedman equation. As with previous case, we fix the potential to be of the exponential kind

\begin{equation}
V=V_0 {\rm e}^{-\frac{\lambda \phi}{M_p}}
\end{equation}

Our plan for evaluating the potential of this model to reconcile requirements of the swampland programme with observations is the following. We initially solve the equations of motion on a fixed background deep inside the matter dominated regime. Our solutions assume that both the axion and gauge field vev have very small energy densities and initially all the terms which arise from the Chern-Simons coupling are negligibly small. In this way we can study the both the evolution of the axion as well as the gauge field vev independently. At later times when the speed of the axion becomes sufficiently high, the term on the right hand side of equation (\ref{eq:gaugeeom}) comes to dominate and the value of the gauge field vev increases exponentially rapidly and in turn, acts as a source of friction via the terms on the right hand side of (\ref{eq:eomphi2}). This latter regime will be treated purely numerically. 

\subsubsection{Early time analytic solutions} 
\label{sec:modelSU2-MAna} 

{\hskip 2em}We start at an initial time $\tau_{\rm in}$ which is assumed to be deep into the matter dominated regime. Without loss of generality we choose $a_{\rm in}=1$ and $\phi_{\rm in}=0$ since the normalization of the scale factor is arbitrary and any initial constant value of the axion field can be reabsorbed by the overall prefactor of the exponential potential. Initially both the axion and gauge fields are approximately frozen due to Hubble friction and therefore the dynamics of the background are exclusively controlled by the energy density of matter. In this regime the scale factor becomes

\begin{equation}
a=\left(\frac{\tau}{\tau_{\rm in}}\right)^2
\label{eq:scaearly}
\end{equation} 

and we define as in the previous case $\bar{\rho}_m=\frac{\rho_{m,{\rm in}}}{V_0}$ which is a quantity that is used to parametrize the initial time. We rewrite the equations of motion for the axion and gauge field vev while neglecting the terms that arise from the Chern-Simons coupling.

\begin{eqnarray}
&&\frac{\partial^2\phi}{\partial\tau^2}+\frac{2}{a}\frac{\partial a}{\partial \tau} \frac{\partial \phi}{\partial \tau} +a^2 \frac{\partial V}{\partial\phi}=0\label{eq:approxphieom}\\
&&\frac{\partial^2 Q}{\partial \tau^2}+2a\,H\, \frac{\partial Q}{\partial \tau}+\left(a\frac{\partial H}{\partial \tau}+2 a^2\, H^2\right)Q +2g^2\,a^2\, Q^3=0\label{eq:approxQeom}
\end{eqnarray}

The solution for the equation of motion of the axion field takes the same form as in the case presented in the previous section. As can be verified a-posteriori, we can neglect the variation of the potential term and then the equation of motion is solvable analytically. After fixing one initial condition to cancel a decaying term we obtain.

\begin{equation}
\phi_{\rm early}=\frac{2}{9}\frac{\lambda}{\bar{\rho}_m}M_p\left[\left(\frac{\tau}{\tau_{\rm in}}\right)^6-1\right]\label{eq:phiearly}
\end{equation}

For the case of the gauge field, the equation of motion is not solvable analytically because of the presence of the nonlinear term $2g^2\,a^2\, Q^3$. It turns out that we can ignore this term because the gauge field vev decays at early times. In this way, after waiting long enough, the nonlinear term always becomes negligibly small and then the equation without the nonlinear term is an accurate approximation of the full dynamics. When the gauge field vev is frozen due to hubble friction, we expect to have $\dot{Q}=-{\cal O}(1) H Q$. For simplicity here we select $\dot{Q}_{\rm in}=-H_{\rm in} Q_{\rm in}$ as our initial condition and obtain 

\begin{equation}
Q_{\rm early}=Q_{\rm in} \frac{a_{\rm transition}}{a(\tau)}
\label{eq:Qearly}
\end{equation}

Where we normalized the value $Q_{\rm in}$ to represent the physical value of the gauge field vev at the transition time from matter to dark energy domination. This normalization is appropriate as that is a much more meaningful time than the initial time $\tau_{\rm in}$ as the gauge field vev at the transition is very close to the gauge field vev when the friction becomes dominant.

Our early time solutions are appropriate provided that the coupling terms in the full equations of motion are negligibly small. Assuming that the initial solutions hold, we quantify this requirement by comparing the coupling terms to the terms that appear in (\ref{eq:approxphieom}) and (\ref{eq:approxQeom}). For the equation of motion of the axion this requirement takes the form

\begin{equation}
Q_{\rm in}\ll {\cal O}(1)\left(\frac{\lambda f \sqrt{V_0}}{g}\right)^{1/3}\frac{1}{\bar{\rho}_m^{1/2}}\label{eq:phicond}
\end{equation}

\begin{figure}[ht!]
\centerline{
\includegraphics[width=0.50\textwidth,angle=0]{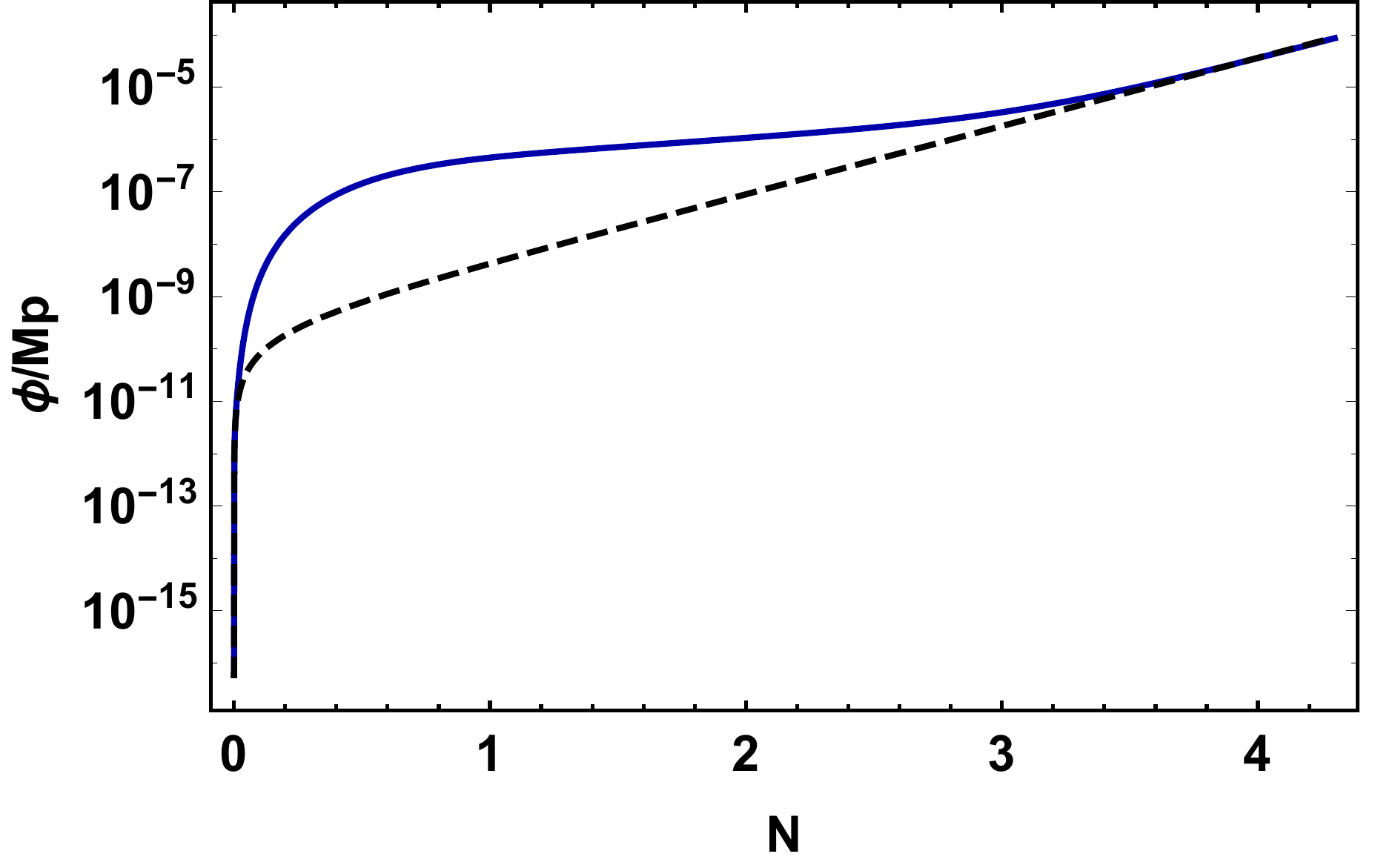}
\includegraphics[width=0.50\textwidth,angle=0]{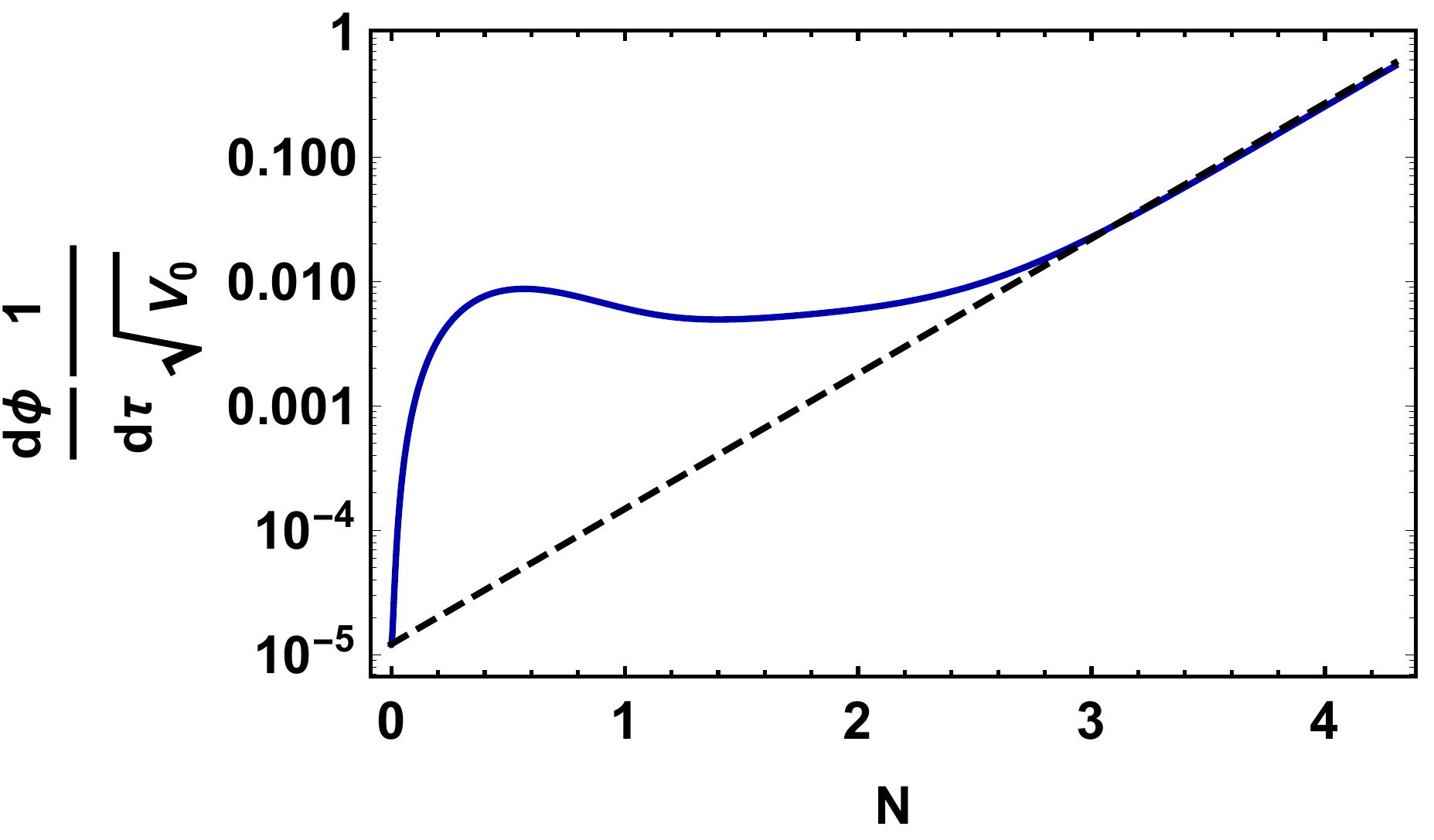}
}

\centerline{
\includegraphics[width=0.50\textwidth,angle=0]{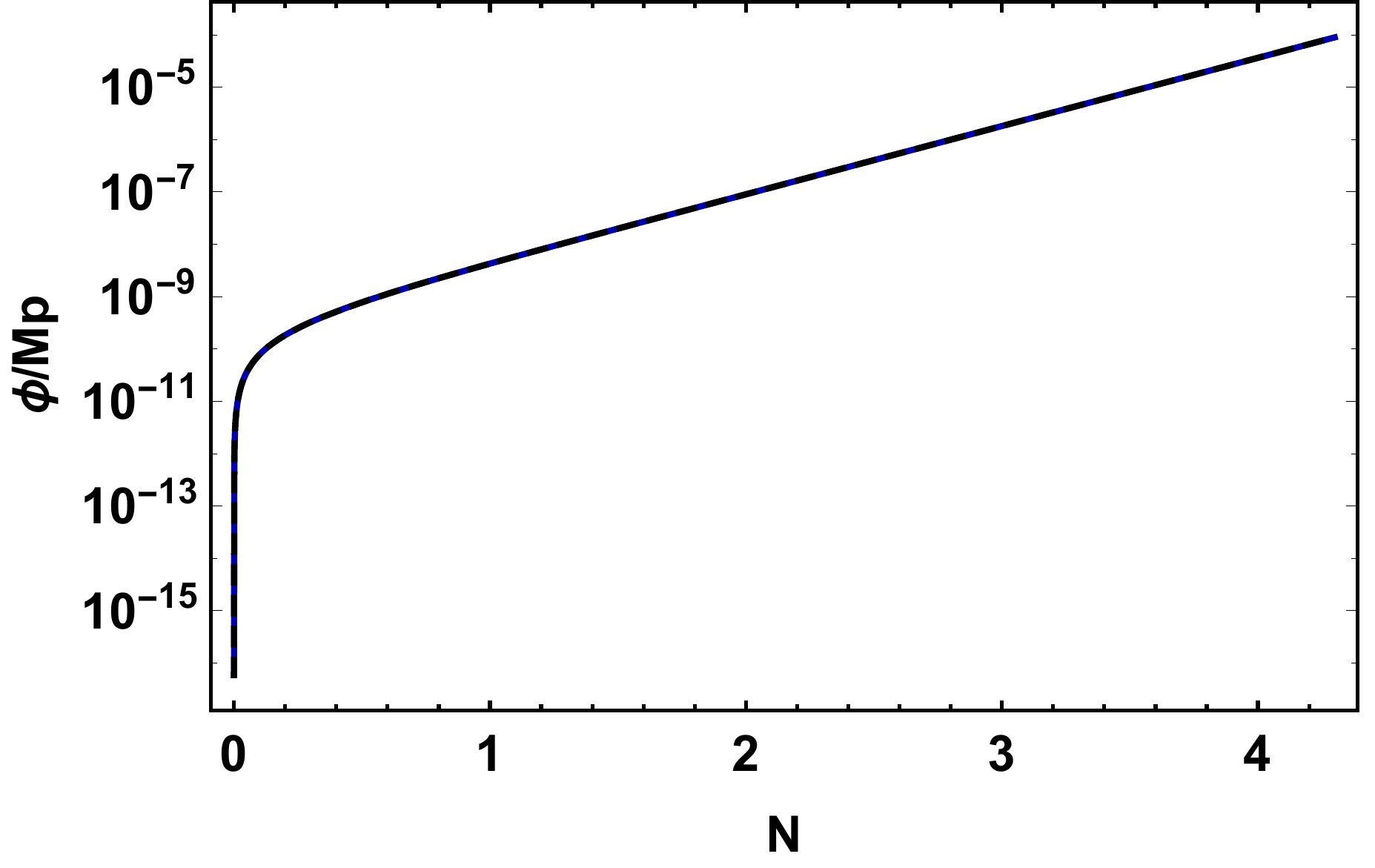}
\includegraphics[width=0.50\textwidth,angle=0]{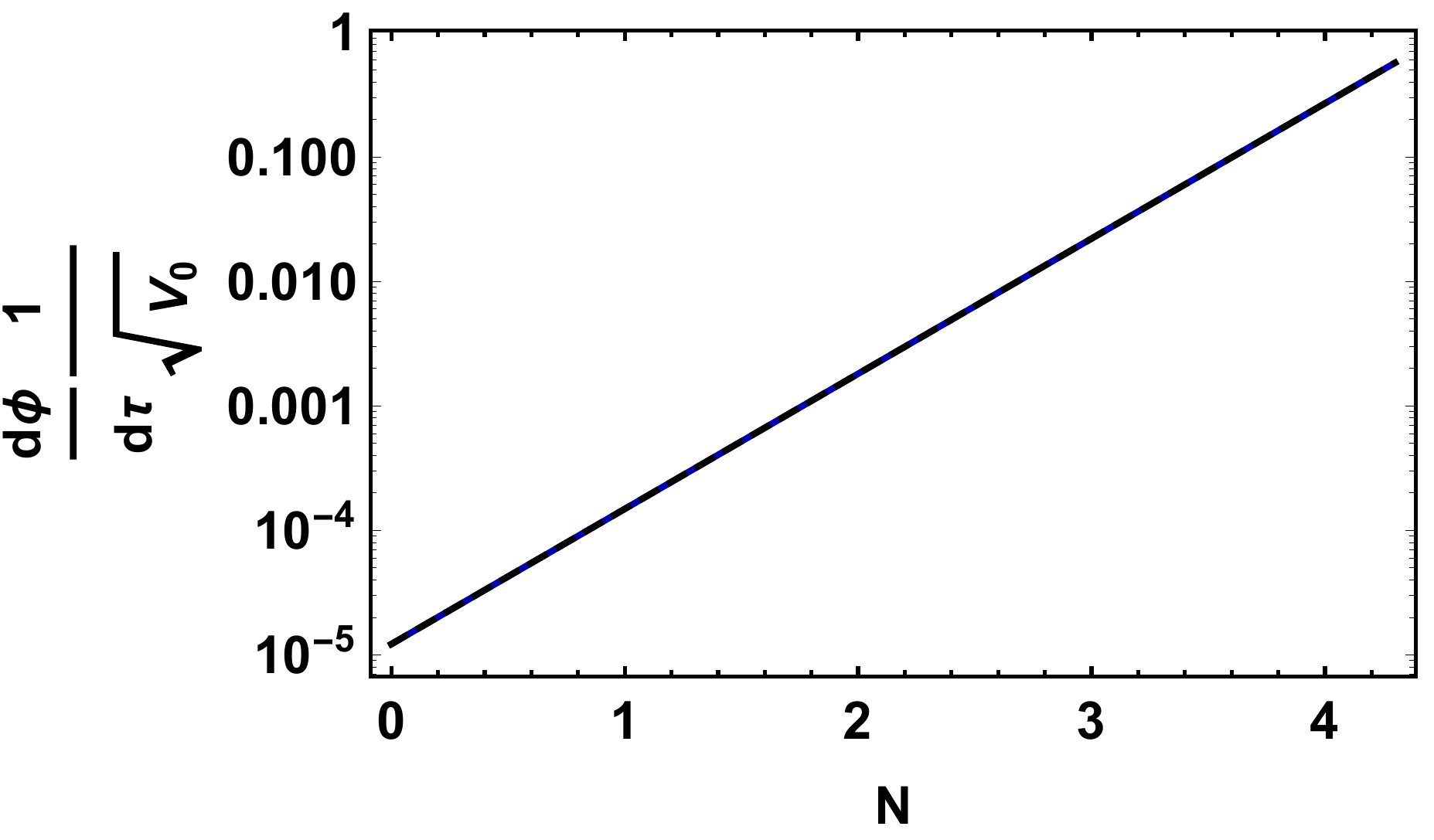}
}

\caption{Comparison between the approximate early time solution of the axion field given in (\ref{eq:phiearly}) and the full numerical solution for different values of the parameters. In the top row we choose parameters $g=9\cdot 10^{-56}$, $f/M_p= 1.26 \cdot 10^{-5}$, $\bar{\rho}_m=10^9$ and $Q_{\rm in}/M_p =3\cdot 10^{-6}$ which strongly violate condition (\ref{eq:phicond}). In the bottom panels we choose $g=9\cdot 10^{-56}$, $f/M_p= 1.26 \cdot 10^{-5}$, $\bar{\rho}_m=10^9$ and $Q_{\rm in}/M_p =3\cdot 10^{-9}$ which satisfy the requirement (\ref{eq:phicond}). As a time parameter we use the number of e-folds defined as $N\equiv\log\,a$. In both cases the parameters converge to the same late time behavior.
}
\label{fig:SU2phicomp}
\end{figure}

This requirement becomes stricter the earlier we initiate our analysis since $\bar{\rho}_m$ appears in the denominator on the right hand side. In principle we can evade this requirement for any value of $Q_{\rm in}$ as long as we choose a sufficiently small $\bar{\rho}_m$ but in practice we can violate this requirement without facing any consequences in the late time behavior of the fields. The reason for this is that the coupling terms in the equation of motion of the scalar field play the role of a friction and therefore their effect is eventually erased at later times, once the friction that arises from the gauge field becomes negligibly small. This point is illustrated in figure \ref{fig:SU2phicomp} where we compare our early time solution (\ref{eq:phiearly}) with the numerical solution of the full system for several different values of the parameters. It can be seen that whether the condition (\ref{eq:phicond}) is violated or not has no impact in the late time behavior of the system. We proceed with our analysis disregarding the requirement (\ref{eq:phicond}).

The requirement that the terms that arise from the Chern-Simons coupling are negligible in the equation of motion of the gauge field vev can be written as follows.

\begin{equation}
Q_{\rm in}\ll {\rm O}(1)\frac{f \sqrt{V_0}}{\lambda g M_p^2}\,\bar{\rho}_m^{7/6}\label{eq:Qcond}
\end{equation}

Unlike the previous case, violating this condition has a dramatic effect in the late time behavior of the system. When the coupling terms become dominant in the equation of motion of the gauge field, the gauge field value is strongly sourced by the axion field and the gauge field vev grows exponentially rapidly. We will only explore parameters that satisfy (\ref{eq:Qcond}) because we want the initial moment chosen to have no impact on the late time behavior of the system.

Finally the requirement that the non linear term in the equation of motion of the gauge field vev (\ref{eq:approxQeom}) is negligible takes the form

\begin{equation}
Q_{\rm in}\ll {\cal O}(0.1) \frac{\sqrt{V_0}}{g M_p}\bar{\rho}_m^{1/6}\label{eq:nonlincond}
\end{equation} 

We conclude this subsection by making use of our approximate analytic expressions to estimate the time that the sourcing of the axion field becomes important in the equation of motion of the gauge field vev. The relevant dynamical terms in the equation of motion of the gauge field vev are 

\begin{eqnarray}
2a\,H\, \frac{\partial Q}{\partial \tau}+\left(a\frac{\partial H}{\partial \tau}+2 a^2\, H^2\right)Q \;\;\;\; &\xrightarrow{\text{Early Time}}&\;\;\;\; \frac{72\, M_p^2\, Q_{\rm in}}{\tau^4\, V_0 \,\bar{\rho}_m}\\
\frac{g\,a\,Q^2}{f}\frac{\partial \phi}{\partial \tau} \;\;\;\; &\xrightarrow{\text{Early Time}}&\;\;\;\; \frac{\lambda\,g\, Q_{\rm in}^2 \tau^3\, V_0^2 \,\bar{\rho}_m^{5/3}}{108 \,f \,M_p^3}
\end{eqnarray}

Clearly, as time evolves, the sourcing term grows with respect to the terms that are relevant at early times. We can find the exact time the sourcing term dominates by equating the two.

\begin{equation}
\tau_{\rm src}= \left[\frac{7776\, f}{\lambda \,g \,Q_{\rm in}}\left(\frac{M_p^4}{V_0}\right)^3\right]^{1/7}\frac{1}{M_p\,\bar{\rho}_m^{1/3}}
\end{equation}

After time $\tau_{\rm src}$ the gauge field vev grows exponentially rapidly and eventually that causes the gauge friction terms to become dominant in the equation of motion of the axion field. This provides an effective friction that slows down the evolution of the axion field, which after that point maintains a slow-roll configuration.

\subsubsection{Numerical results and comparison with observations} 
\label{sec:modelSU2-MNume} 

{\hskip 2em}We now turn our attention to solving the equations (\ref{eq:eomphi2}), (\ref{eq:gaugeeom}) and (\ref{eq:friedmanSU2}) numerically. We use for our time variable the number of e-folds defined as $N\equiv\log\,a$ and we start our evolution at $N_{\rm in}=0$. The initial conditions for the various parameters are given by the early time approximate solutions we developed in subsection \ref{sec:modelSU2-MAna}. Unlike the U(1) case, developing an algorithm for the SU(2) case is relatively straightforward since the entire evolution consists of classical quantities and the slowing down of the scalar field takes place at the background level. We define the equation of state parameters

\begin{equation}
w_{\rm tot}=\frac{p_{\rm tot}}{\rho_{\rm tot}}=\frac{1}{3}-\frac{2a}{3}\frac{\partial^2 a}{\partial \tau^2}\bigg/\left(\frac{\partial a}{\partial \tau}\right)^2\;\;\;,\;\;\; w_\phi=\frac{p_\phi}{\rho_\phi}
\end{equation}

and the fractional energies

\begin{equation}
\Omega_{m}=\frac{\rho_{m}}{\rho_{\rm tot}}\;\;\;,\;\;\;\Omega_{Q}=\frac{\rho_{Q}}{\rho_{\rm tot}}\;\;\;,\;\;\;\Omega_\phi=\frac{\rho_\phi}{\rho_{\rm tot}}
\end{equation}

\begin{figure}[ht!]
\centerline{
\includegraphics[width=0.33\textwidth,angle=0]{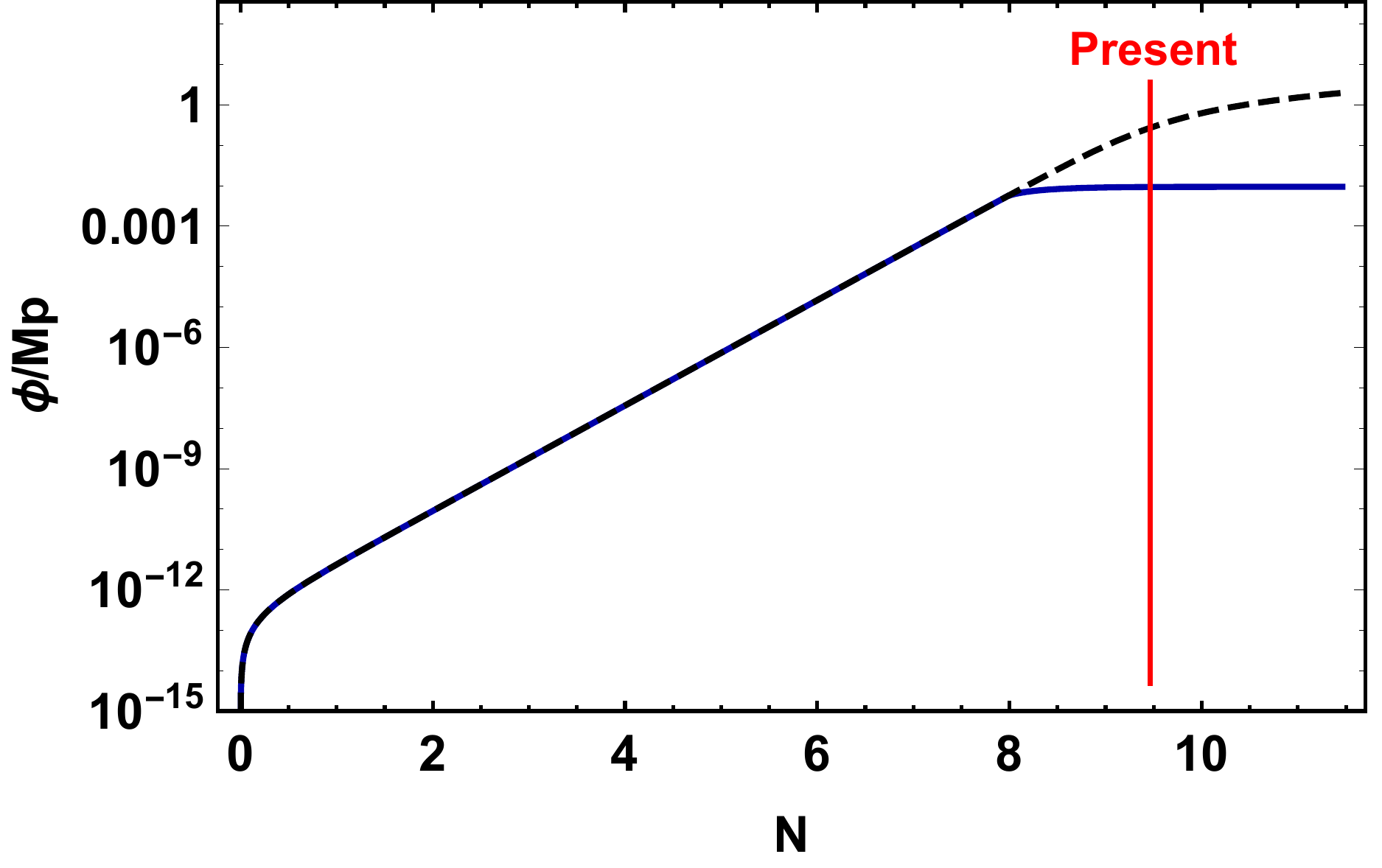}
\includegraphics[width=0.33\textwidth,angle=0]{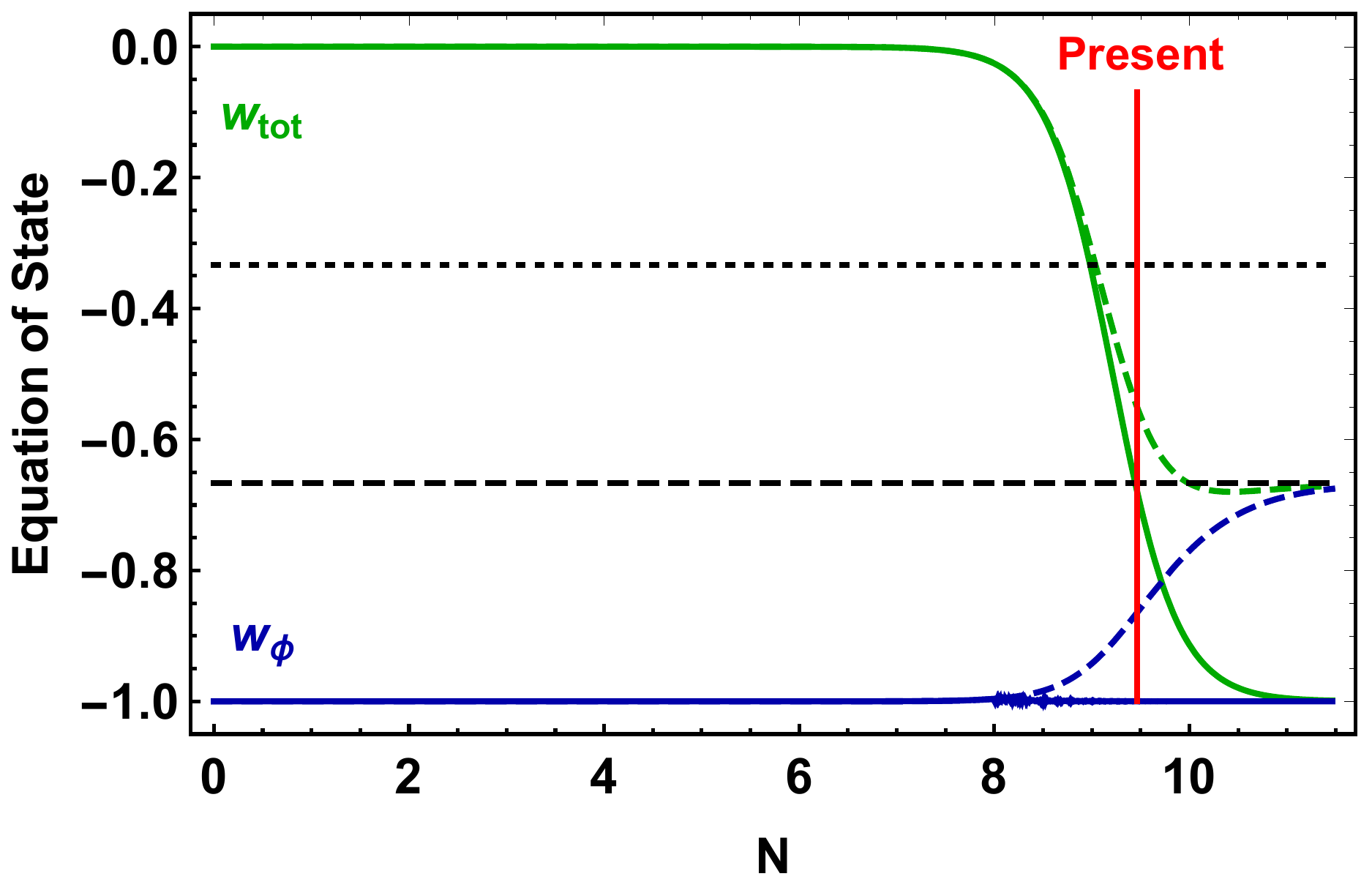}
\includegraphics[width=0.33\textwidth,angle=0]{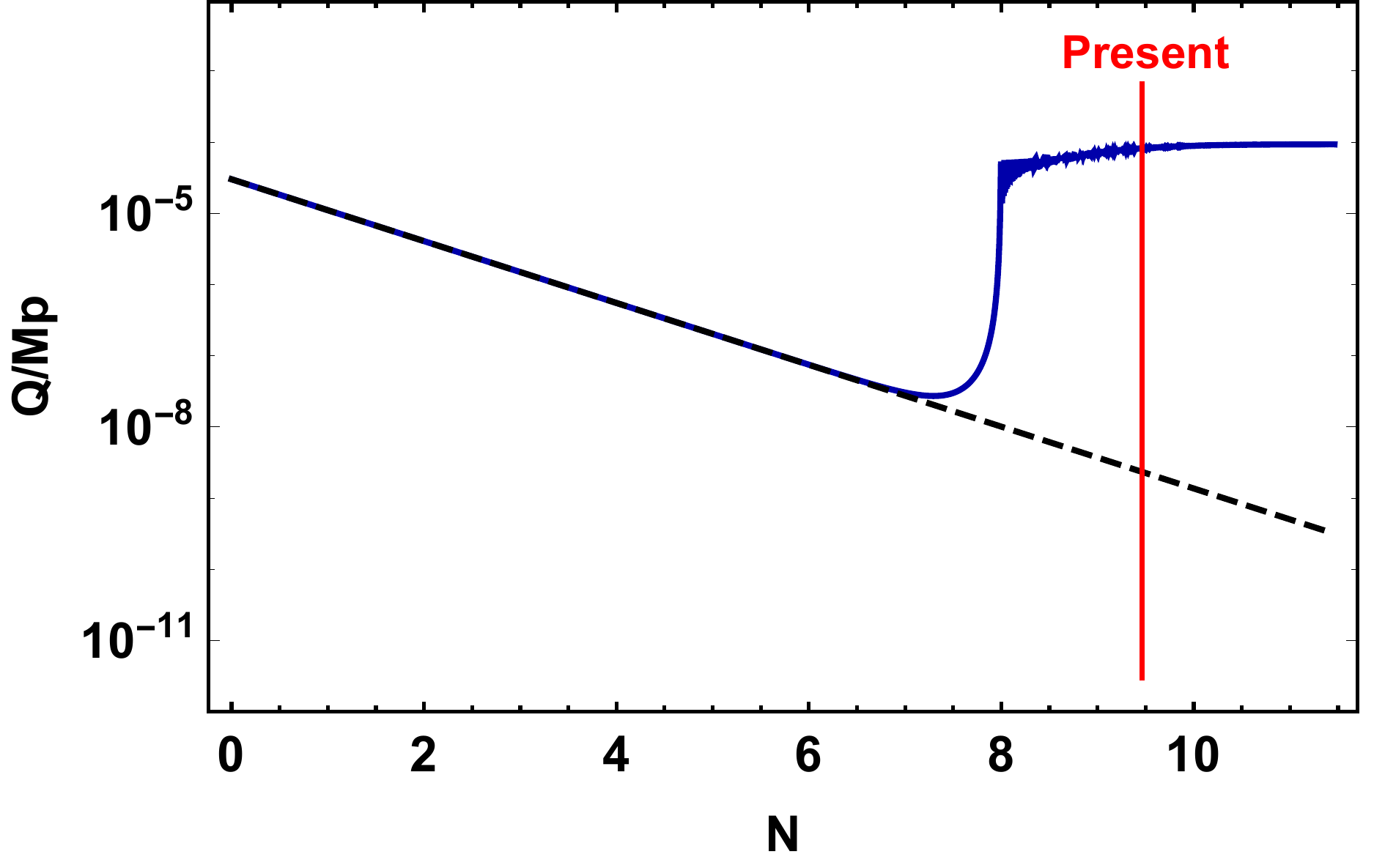}
}

\centerline{
\includegraphics[width=0.33\textwidth,angle=0]{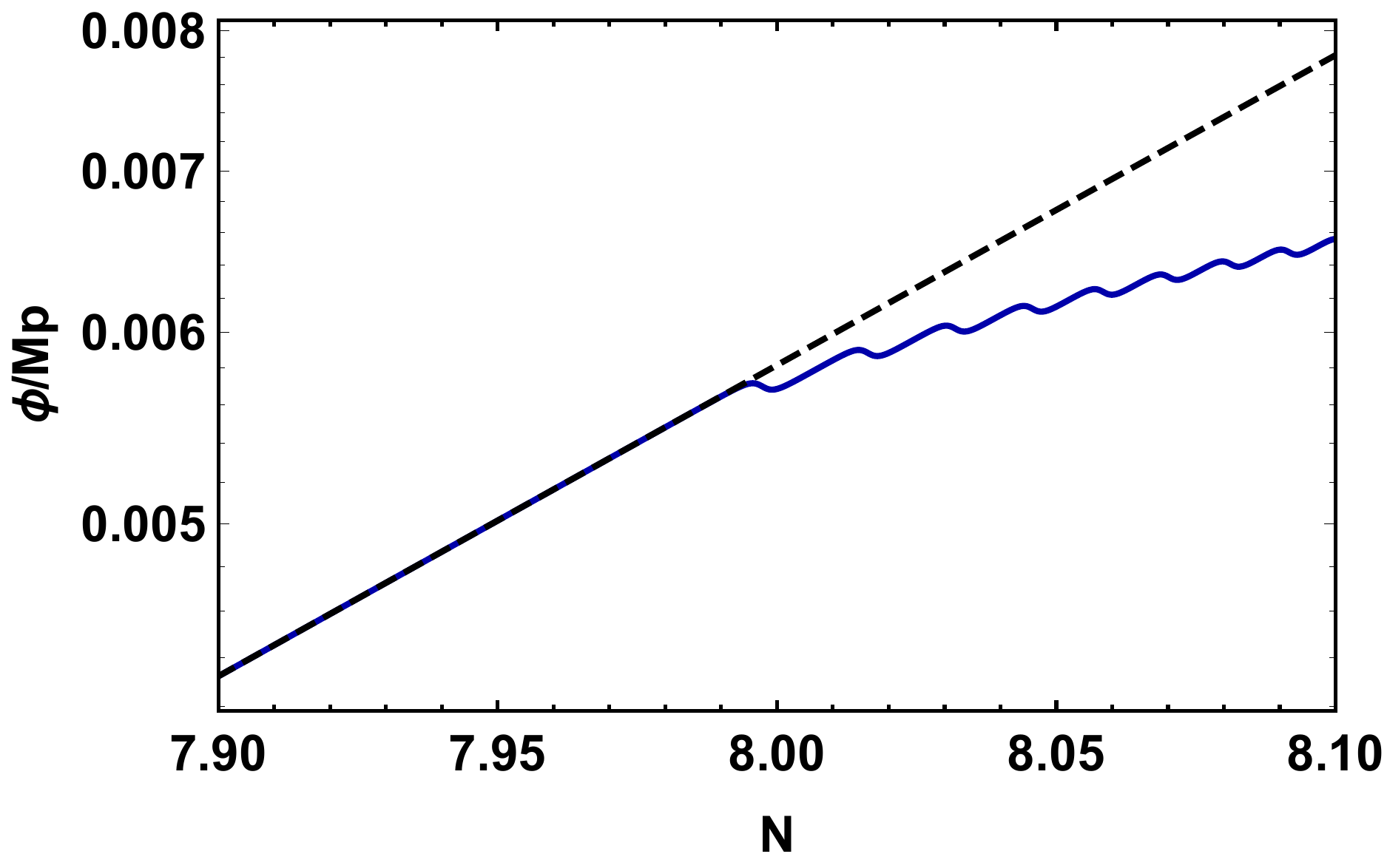}
\includegraphics[width=0.33\textwidth,angle=0]{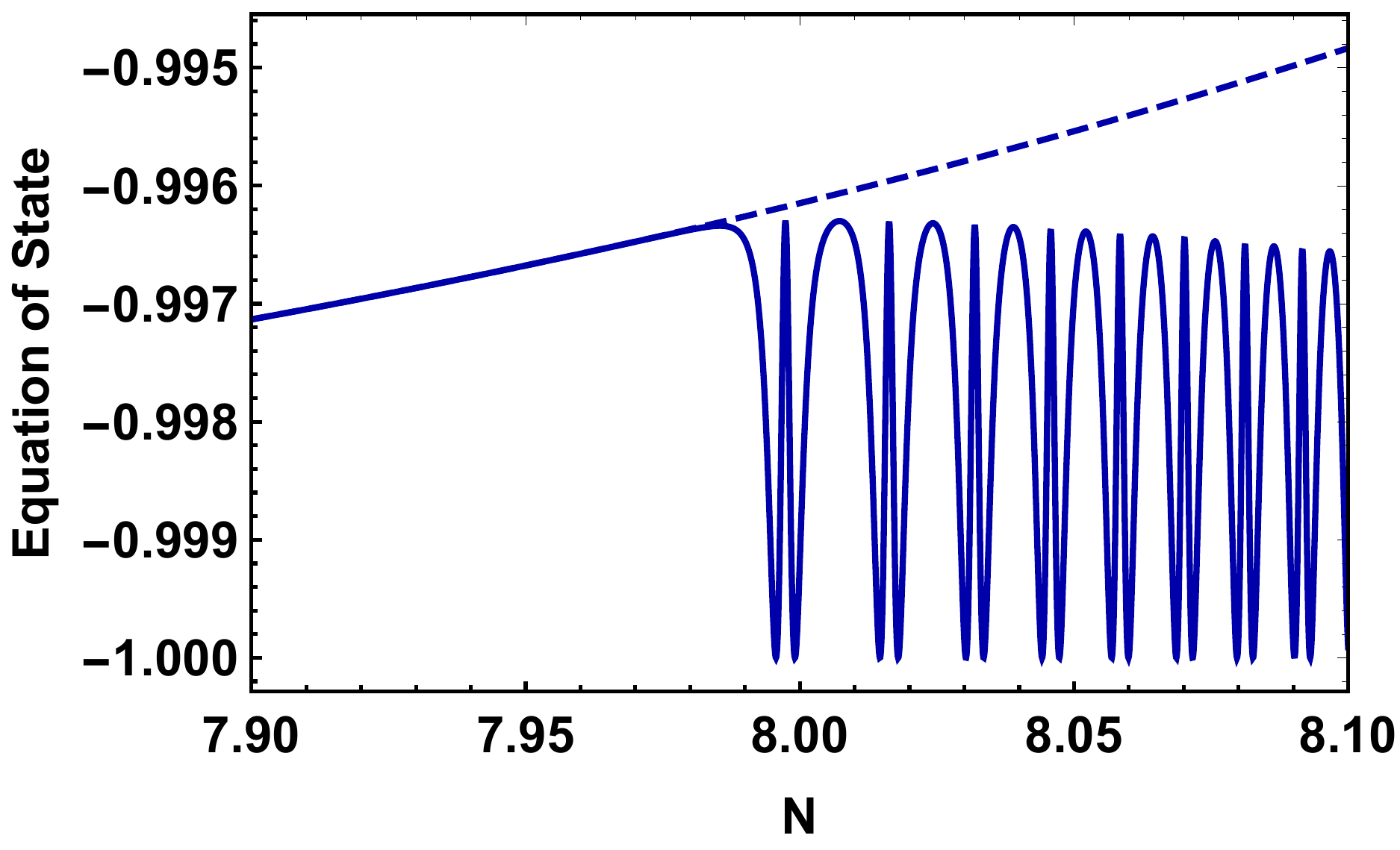}
\includegraphics[width=0.33\textwidth,angle=0]{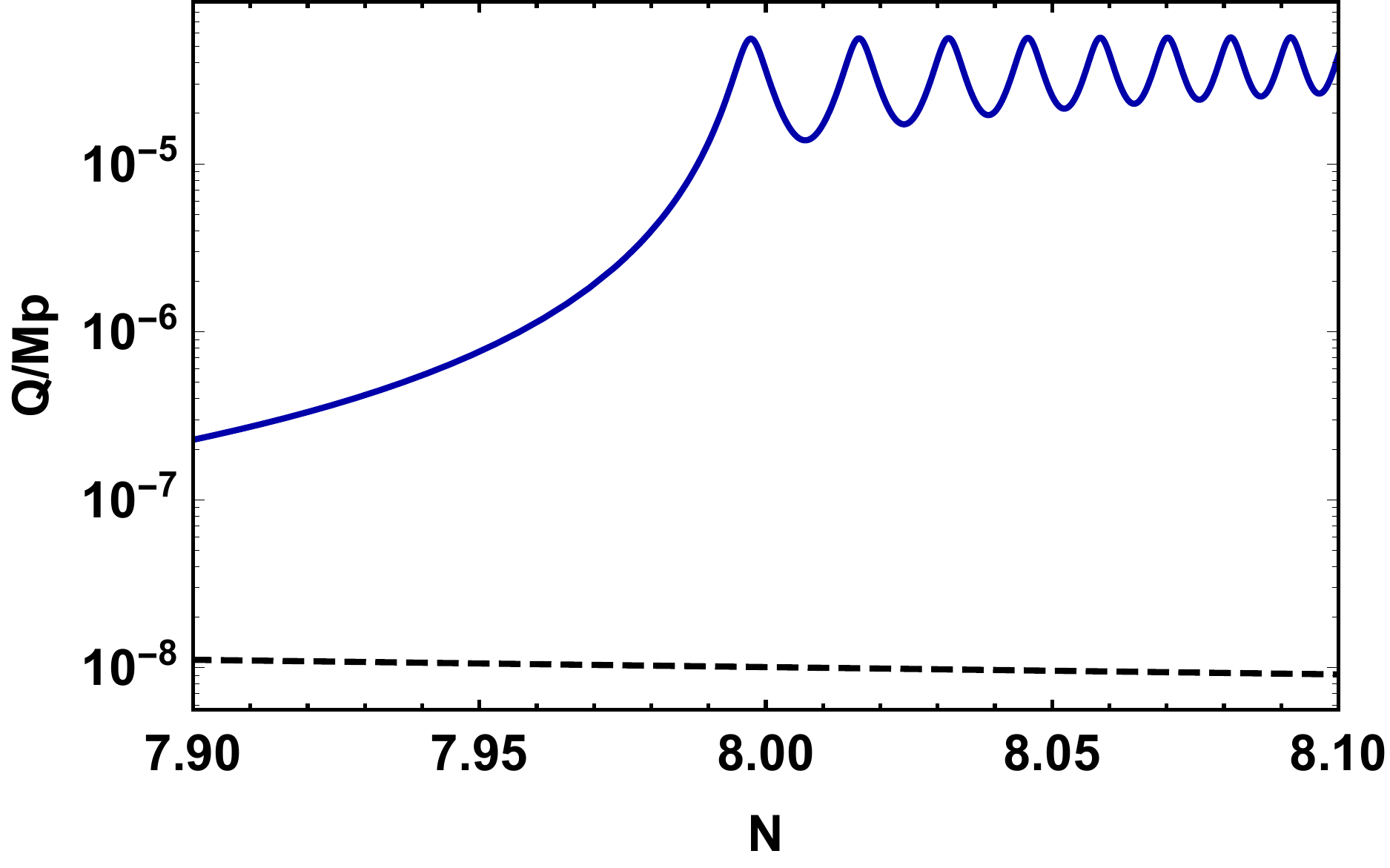}
}

\caption{The six panels display the comparison of the evolution of various quantities in the absence of the Chern-Simons coupling (dashed lines) and in the case in which the Chern Simons coupling is strong (solid lines). The top three panels show the full evolution from the start of the code until two e-folds after the present moment while the bottom three panels are focused only at the instant in which the friction terms become important in the equation of motion of the axion. The parameters used here are $\lambda=1$, $g=9\cdot 10^{-56}$, $f/M_p=1.26\cdot 10^{-7}$, $Q_{\rm in}/M_p=3\cdot 10^{-9}$ and $\bar{\rho}_m=10^{12}$.
}
\label{fig:SU2nume}
\end{figure}

\begin{figure}[ht!]
\centerline{
\includegraphics[width=0.66\textwidth,angle=0]{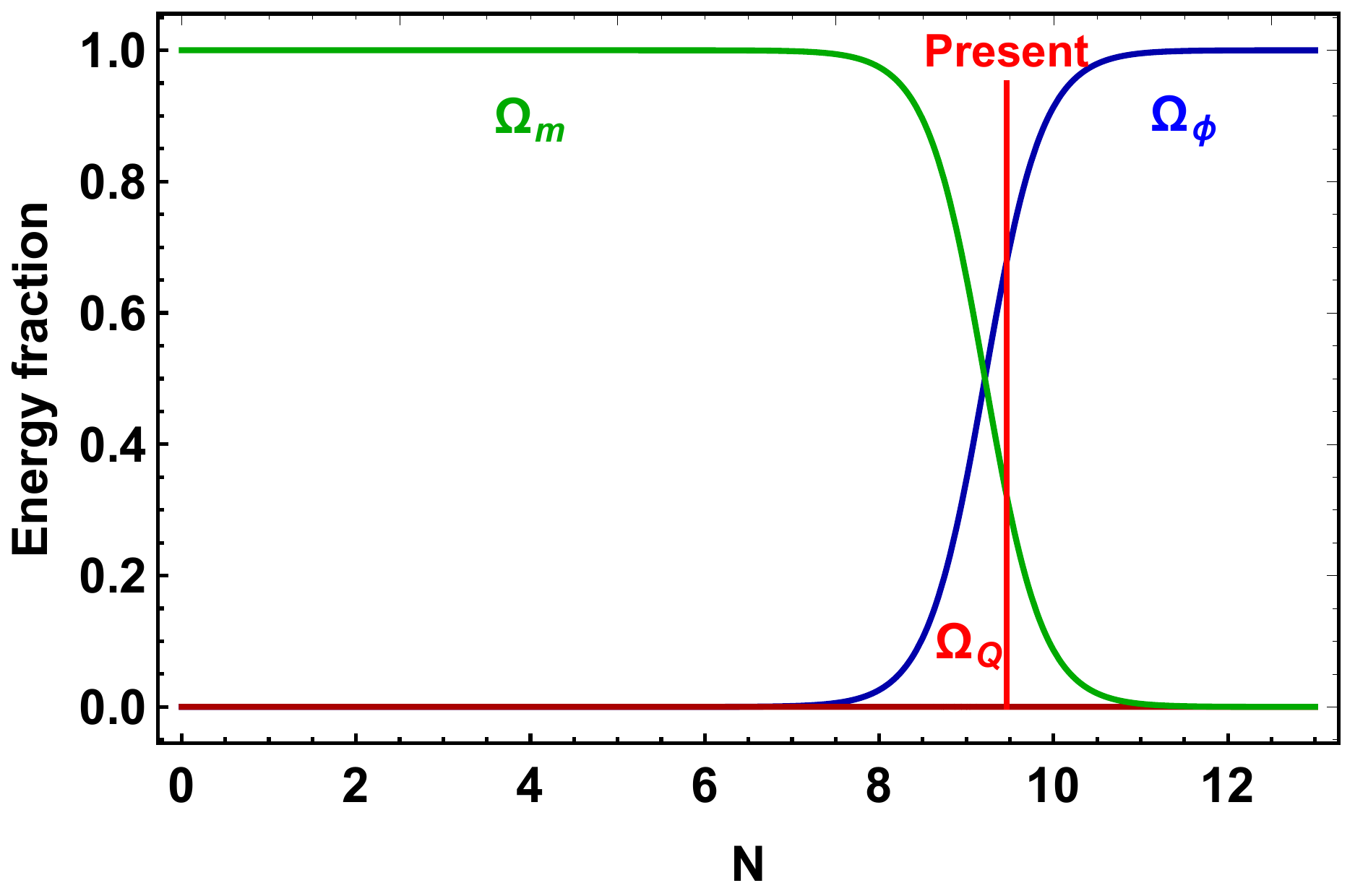}
}
\caption{The comparison of the fractional energies throught the evolution. It can be observed that the energy of the gauge field remains negligibly small throughout. The parameters used in this run are $\lambda=1$, $g=9\cdot 10^{-56}$, $f/M_p=1.26\cdot 10^{-7}$, $Q_{\rm in}/M_p=3\cdot 10^{-9}$ and $\bar{\rho}_m=10^{12}$.
}
\label{fig:SU2energy}
\end{figure} 

Our numerical results for a particular choice of parameters is displayed in figures \ref{fig:SU2nume} and \ref{fig:SU2energy}. From the top right panel of \ref{fig:SU2nume} it is possible to observe that at time $\tau_{\rm src}$, found in the previous subsection, the terms that arise from the Chern-Simons coupling start to become important and the gauge field vev increases exponentially. Eventually this exponential increase stops because the gauge field vev becomes so large that it activates the friction terms in the equation of motion of the axion field. After that moment both fields become approximately frozen. The net effect is the slow down of the axion field and its subsequent slow-roll configuration. In the bottom row of \ref{fig:SU2nume} we have "zoomed-in" on the instant when the friction terms become relevant in the equation of state of the axion field. This row is completely analogous to the bottom row of figure \ref{fig:comparison} which was plotted for the U(1) case. We see that in both cases, high frequency oscillations occur but the pattern of oscillation is considerably different in the two cases. This difference is expected as the mechanism that causes the slow down is fundamentally different. 

\vspace*{1 cm}

The next step in our analysis is to compare with observational data. In lieu of a full data analysis our comparison will use the same conservative approach as in the analysis of the U(1) case presented in the previous section. In order to simplify our analysis we will fix the value of the slope of the potential to $\lambda=1$ which, as explained in the previous section, lies at the interface between compatibility with observational data and the requirements of the swampland conjectures. Observing all the terms that appear in the set of equations (\ref{eq:eomphi2}), (\ref{eq:gaugeeom}) and (\ref{eq:friedmanSU2}), it becomes clear that the predictions of the model depend on the following set of parameters. 

\begin{equation}
\left(g,\frac{g}{\tilde{f}\sqrt{\tilde{V}_0}},\tilde{Q}_{\rm in}\right)
\label{eq:parameters}
\end{equation}

where parameters with a tilde are the dimensionless equivalents of the respective parameters without the tilde in units of reduced planck mass. The first parameter in the set (\ref{eq:parameters}) captures the strength of the nonlinear term $2g^2\,a^2\, Q^3$ in the equation of motion of the gauge field vev, the second captures the strength of the Chern-Simons coupling and the last one captures the overall initial energy scale of the gauge field, which is a free parameter in our model. In order to simplify our analysis, we will work in a regime in which the nonlinear term $2g^2\,a^2\, Q^3$ never becomes relevant in the equation of motion of the gauge field vev. This will be true as long as the gauge coupling is less than a threshold value that is dependent on $\tilde{Q}_{\rm in}$ which we call $g_{\rm th}\left(\tilde{Q}_{\rm in}\right)$. As long as that regime holds, the comparison with the data depends only on the parameters 

\begin{equation}
\left(\frac{g}{\tilde{f}\sqrt{\tilde{V}_0}},\tilde{Q}_{\rm in}\right)
\label{eq:parameters2}
\end{equation}

and the numerical solution can be represented in a very elegant way. We again make use of the CPL parametrization as done in the previous section

\begin{equation}
w_\phi(a)=w_0+\left(1-\frac{a}{a_0}\right)w_a
\end{equation}

and we use the same criteria to facilitate our comparisonn. Namely we define:

\begin{itemize}
\item Compatibility with the data: Parameters $(w_\phi,\frac{d w_\phi}{da})$ are within the 1-$\sigma$ contour of figure 30 of \cite{Akrami:2018odb}.

\item Incompatibility with the data: Any effects in the motion of the axion that arise from the Chern-Simons coupling take place after the present moment.
\end{itemize}

\begin{figure}[ht!]
\centerline{
\includegraphics[width=0.99\textwidth,angle=0]{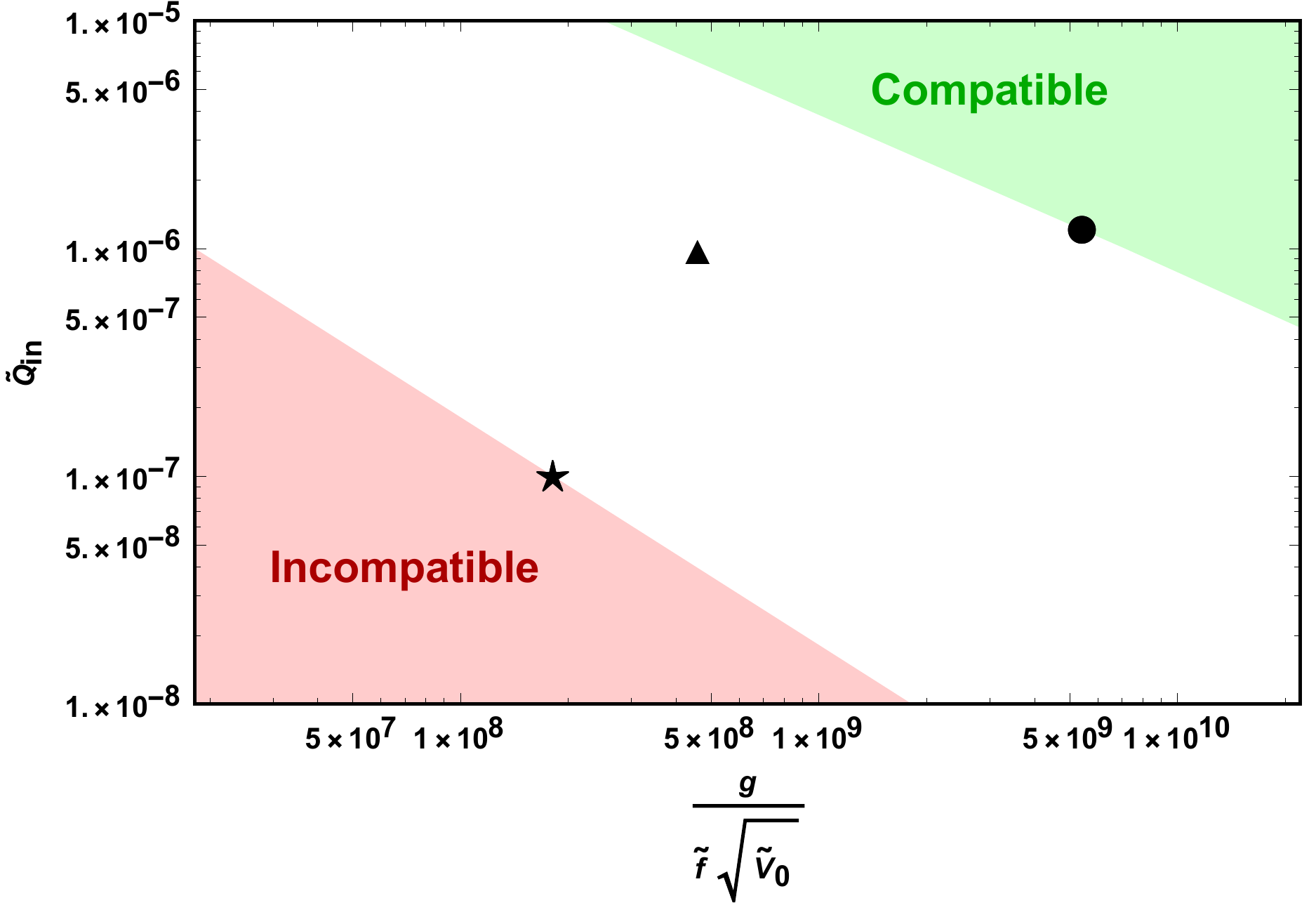}
}
\caption{The green (red) region represents the parameter space that is compatible (incompatible) with observations as defined in the criteria listed in the main text. The three black markers are three sample choices of parameters for which we show the numerical evolution in detail in figure \ref{fig:SU2examples}.
}
\label{fig:parameterspace}
\end{figure}

Our criterion for compatibility is conservative in that it ensures that the fast oscillations that occur after the Chern-Simons terms become relevant have decayed sufficiently enough so that the evolution of the equation of state parameter can be considered linear on average. This criterion justifies the use of the CPL parametrization. The true threshold between compatibility and incompatibility lies somewhere between the two criteria defined above, but probing it would require a dedicated data analysis which we do not attempt to undertake in this work. The results of our analysis are displayed in figure \ref{fig:parameterspace}.

We have performed a fitting for the lines that appear in figure \ref{fig:parameterspace} and we have found that the lines are accurately captured by

\begin{eqnarray}
\tilde{Q}_{\rm in}&>&12.24 \left(\frac{\tilde{f}\sqrt{\tilde{V}_0}}{g}\right)^{0.721}\;\;\;\;\;\;\;{\rm (Compatible\;with\;data)}\label{eq:SU2compatible}\\
\tilde{Q}_{\rm in}&<&18.18 \left(\frac{\tilde{f}\sqrt{\tilde{V}_0}}{g}\right)\;\;\;\;\;\;\;\;\;\;\;\;\;\;{\rm (Incompatible\;with\;data)}\label{eq:SU2incompatible}
\end{eqnarray}

\begin{figure}[ht!]
\centerline{
\includegraphics[width=0.5\textwidth,angle=0]{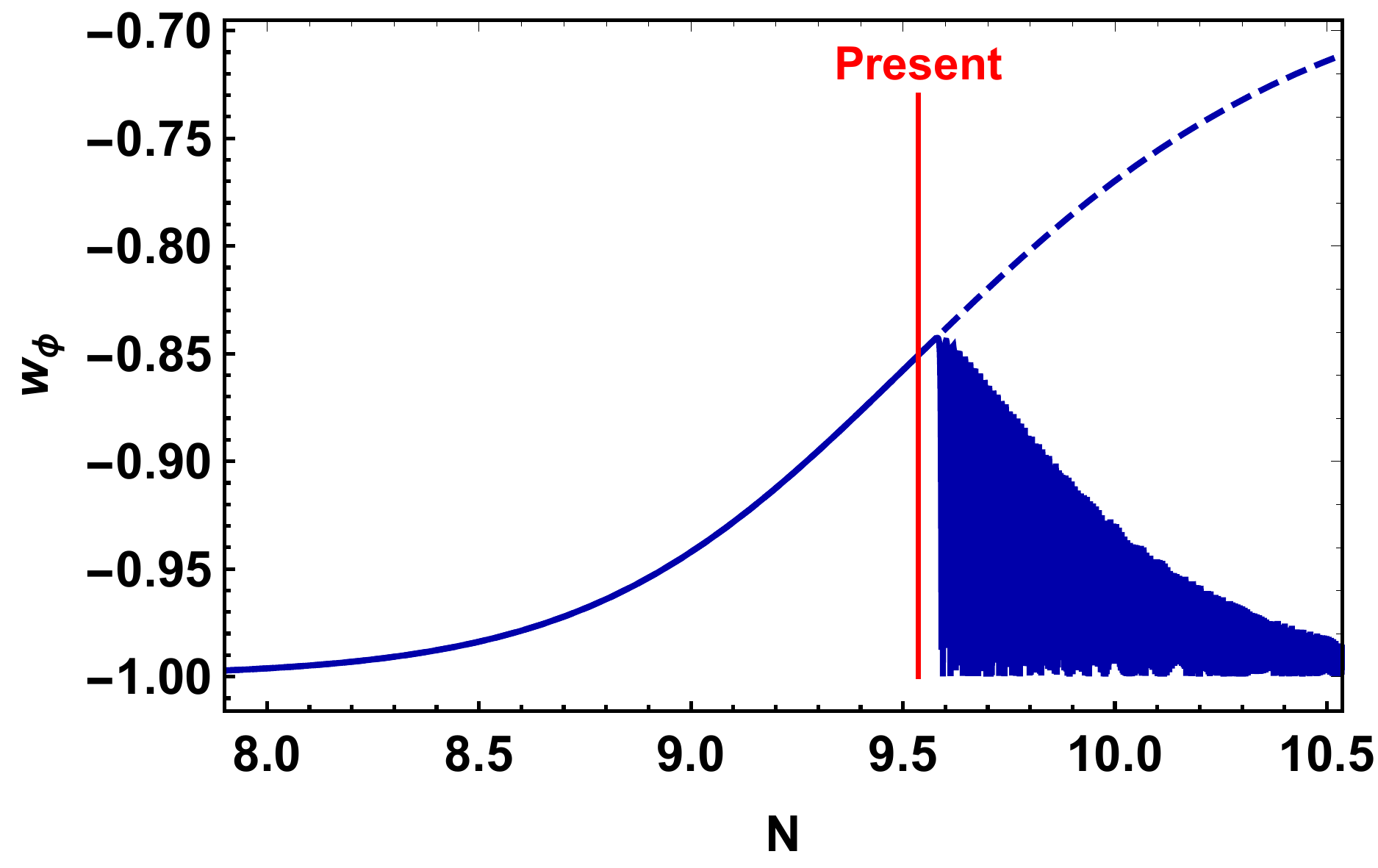}
\includegraphics[width=0.5\textwidth,angle=0]{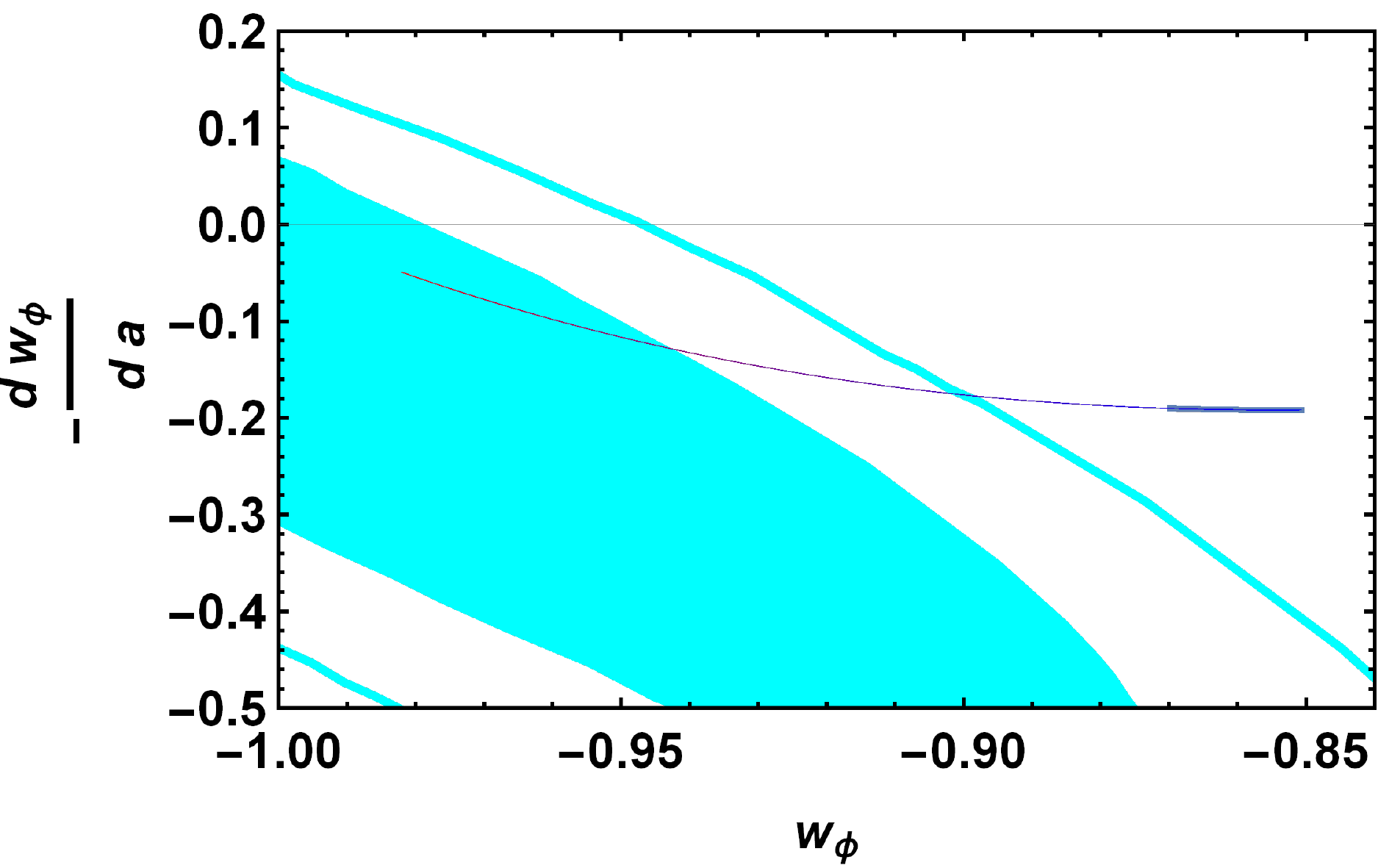}
}
\centerline{
\includegraphics[width=0.5\textwidth,angle=0]{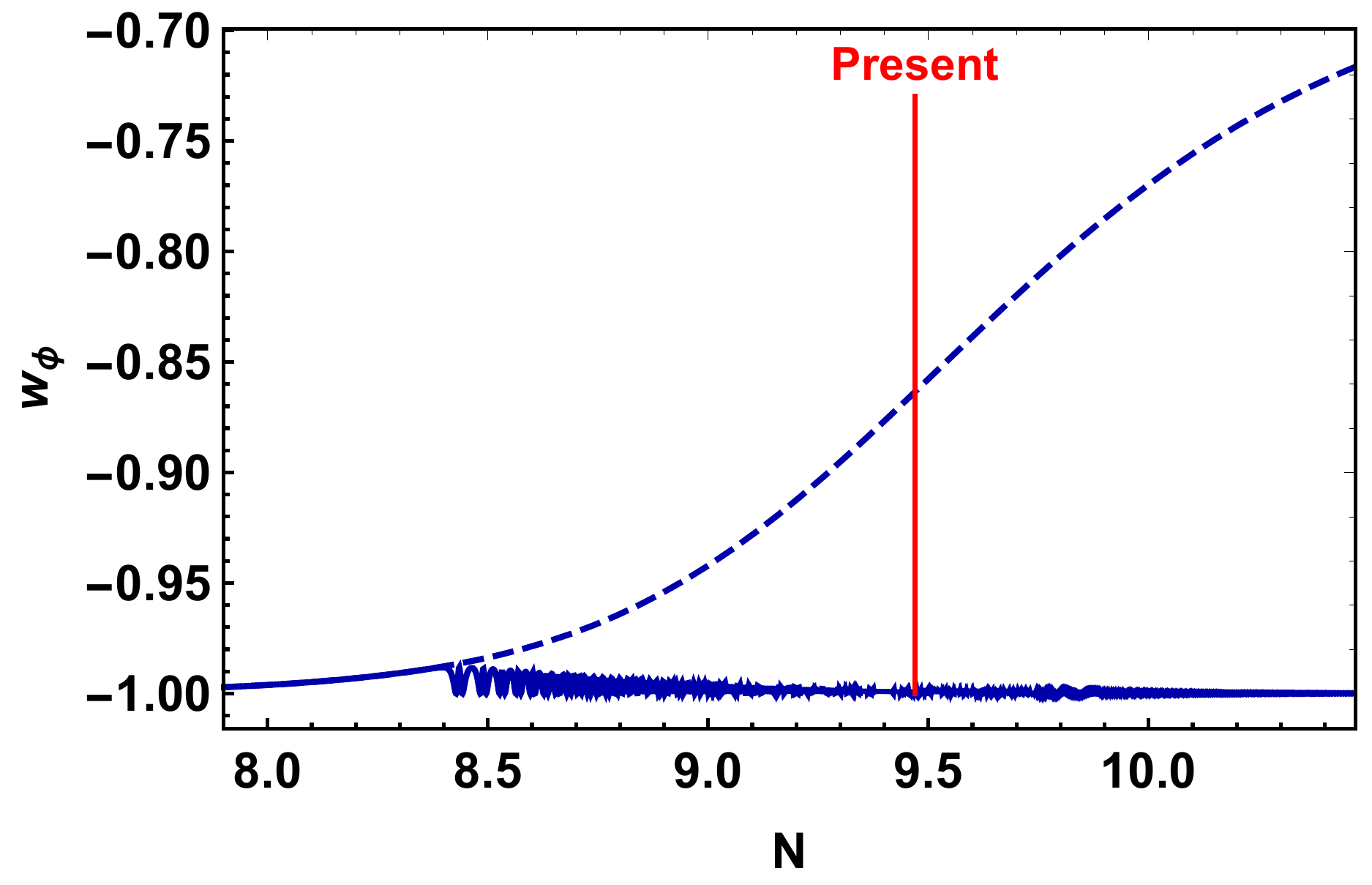}
\includegraphics[width=0.5\textwidth,angle=0]{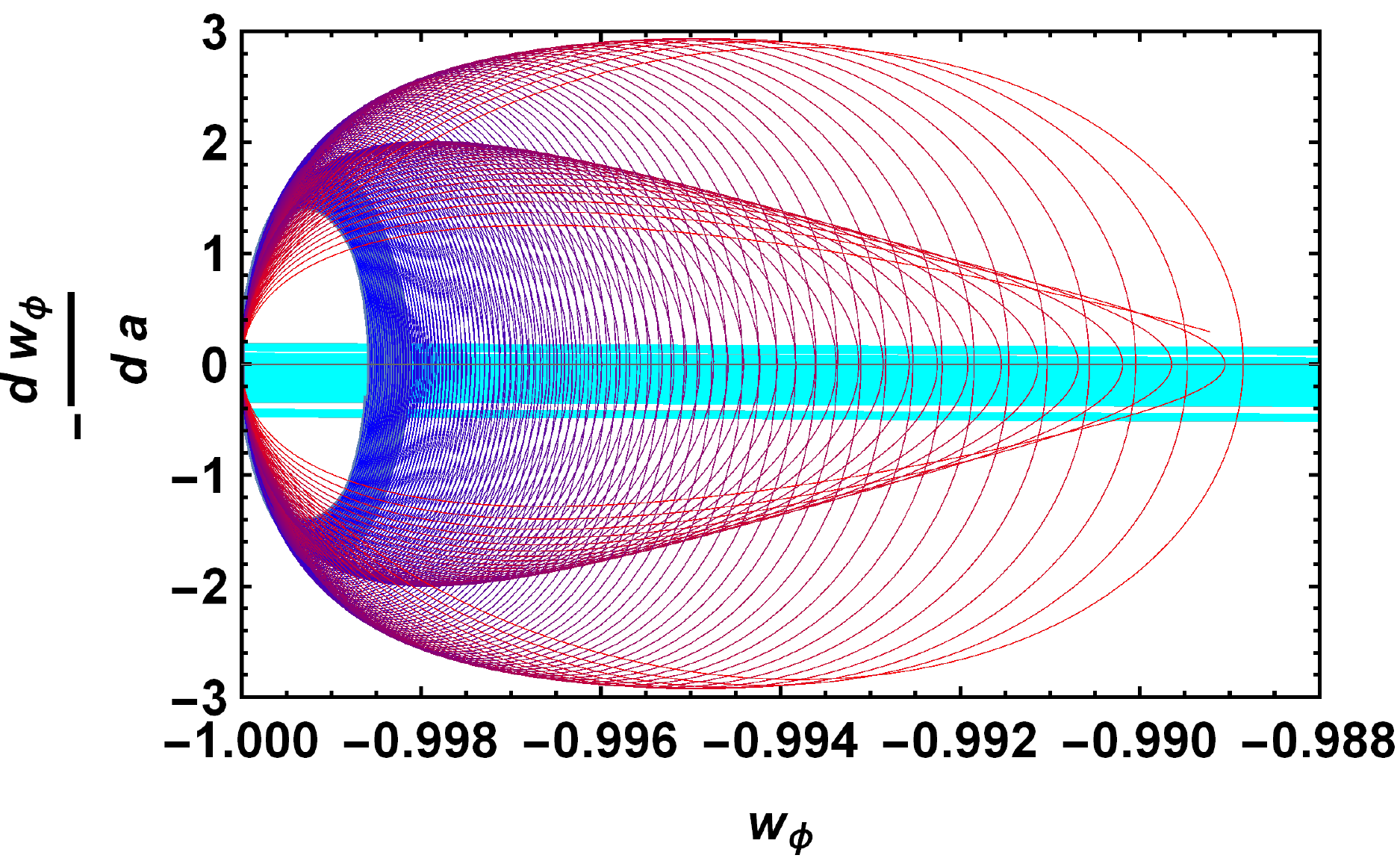}
}
\centerline{
\includegraphics[width=0.5\textwidth,angle=0]{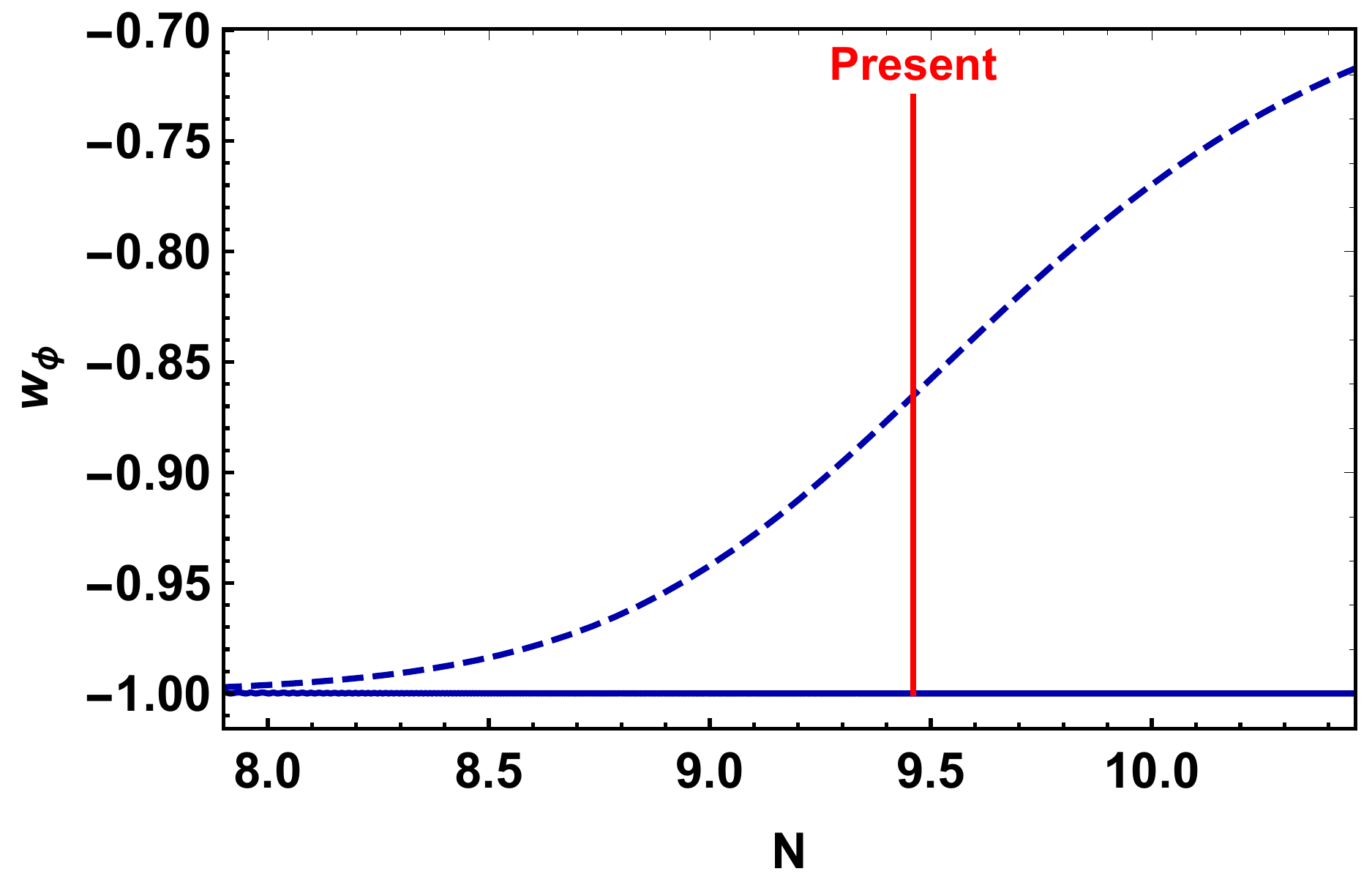}
\includegraphics[width=0.5\textwidth,angle=0]{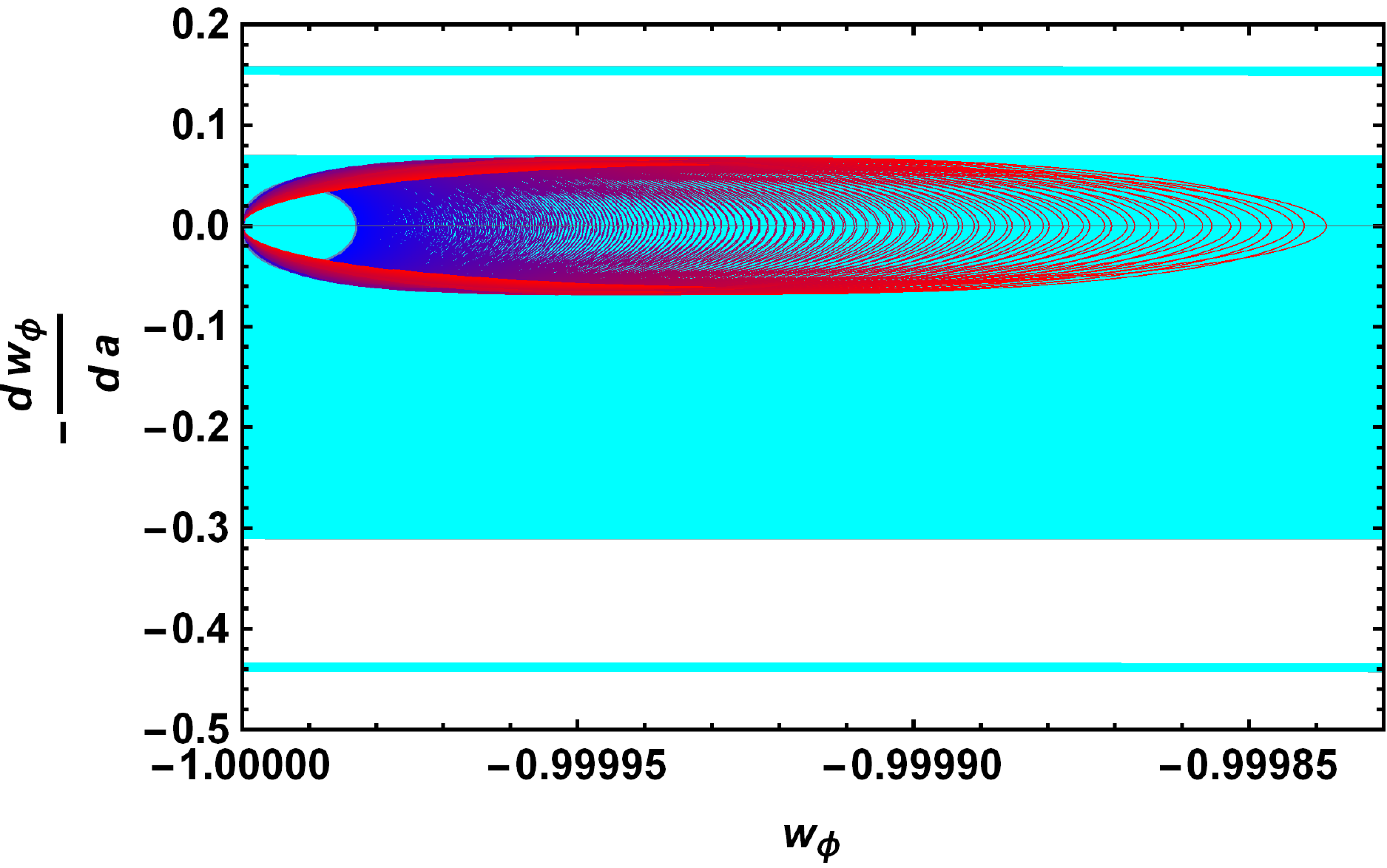}
}
\caption{We show the evolution of the equation of state parameter of the axion for the three choices of parameters displayed by the three markers in figure \ref{fig:parameterspace}. The top panels are labeled with a star with parameter choices $\frac{g}{\tilde{f}\sqrt{\tilde{V}_0}}=1.8\cdot 10^8$ and $\tilde{Q}_{\rm in}=10^{-7}$. The middle panels are labeled with a triangle with parameter choices $\frac{g}{\tilde{f}\sqrt{\tilde{V}_0}}=4.6\cdot 10^8$ and $\tilde{Q}_{\rm in}=10^{-6}$ and the bottom panels are labeled with a circle with parameter choices $\frac{g}{\tilde{f}\sqrt{\tilde{V}_0}}=5.4\cdot 10^{9}$ and $\tilde{Q}_{\rm in}=1.25\cdot 10^{-6}$. The right panels display the last e-fold of the evolution of the equation of state parameter until the present moment. The color transitions from red (past) to blue (present).
}
\label{fig:SU2examples}
\end{figure}

In order to illustrate how the various parameter choices affect the numerical evolution we choose three values in the $\left(\frac{g}{\tilde{f}\sqrt{\tilde{V}_0}},\tilde{Q}_{\rm in}\right)$ plane which are labeled in figure \ref{fig:parameterspace} by a star, a triangle and a circle. For these three choices we display the evolution of the equation of state as well as of the set $(w_\phi,\frac{d w_\phi}{da})$ superimposed with the cyan contours in figure 30 of \cite{Akrami:2018odb}. The results are shown in figure \ref{fig:SU2examples}.

Before concluding this subsection we revisit the assumption that the nonlinear term $2g^2\,a^2\, Q^3$ is always negligible throughout the evolution. We want to demonstrate the effect of this term and come up with the exact threshold $g_{\rm th}(\tilde{Q}_{\rm in})$ above which the nonlinear term becomes relevant. We initiate this analysis by defining the quantity 

\begin{equation}
{\cal R}\equiv\frac{\frac{g\,a\,Q^2}{f}\frac{\partial \phi}{\partial \tau}}{2g^2\,a^2\, Q^3}
\label{eq:SU2nonlin}
\end{equation}

\begin{figure}[ht!]
\centerline{
\includegraphics[width=0.5\textwidth,angle=0]{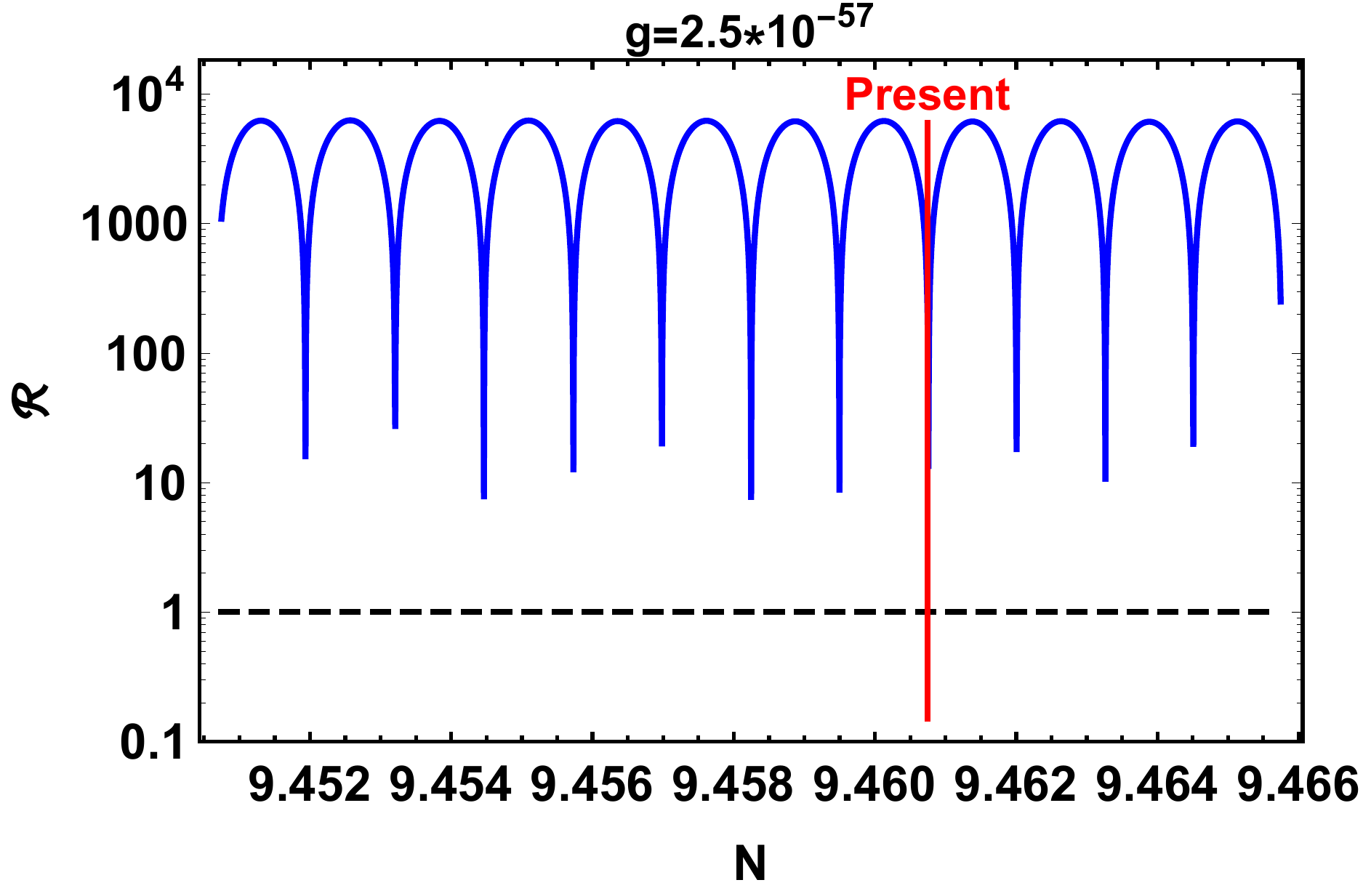}
\includegraphics[width=0.5\textwidth,angle=0]{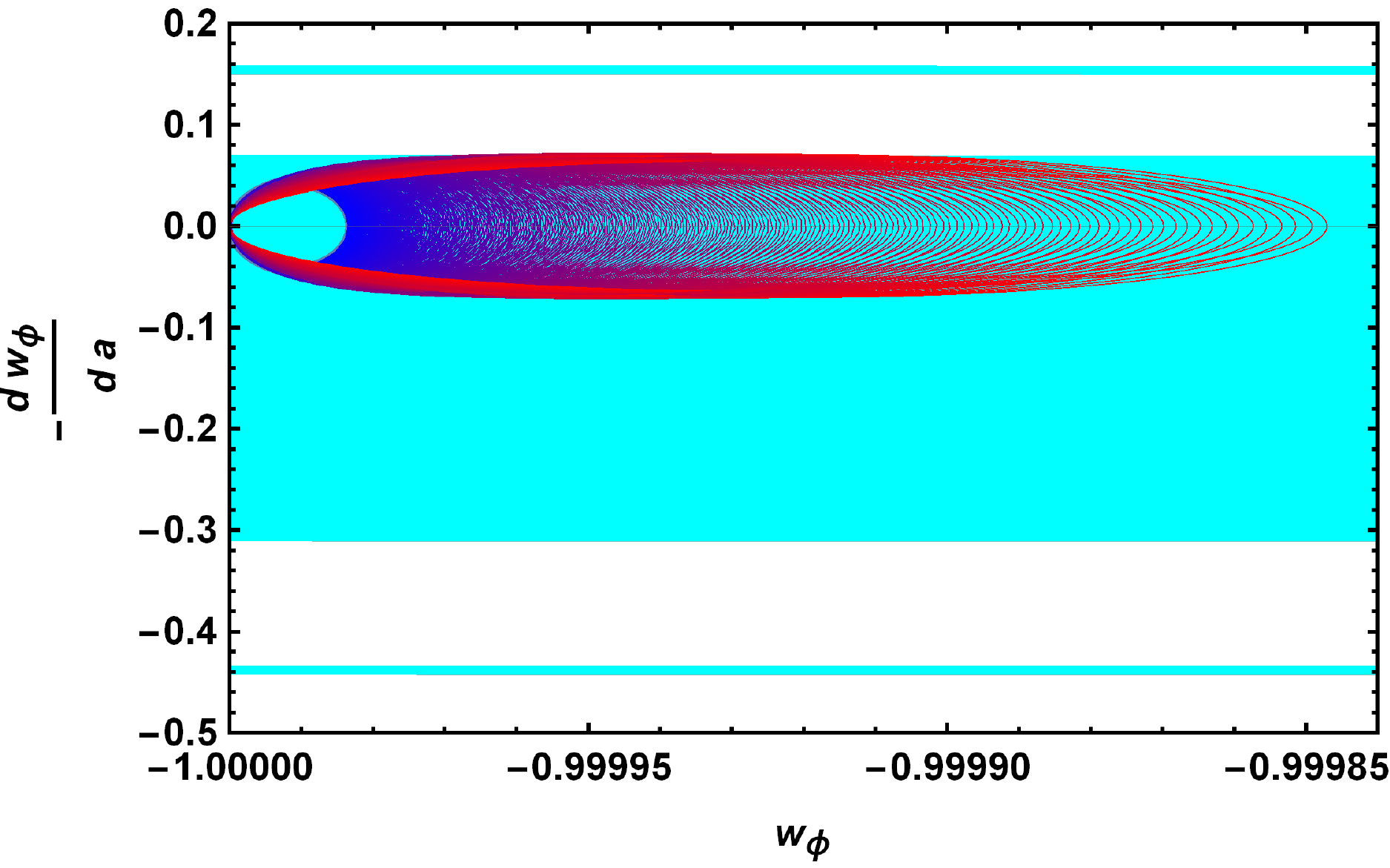}
}
\centerline{
\includegraphics[width=0.5\textwidth,angle=0]{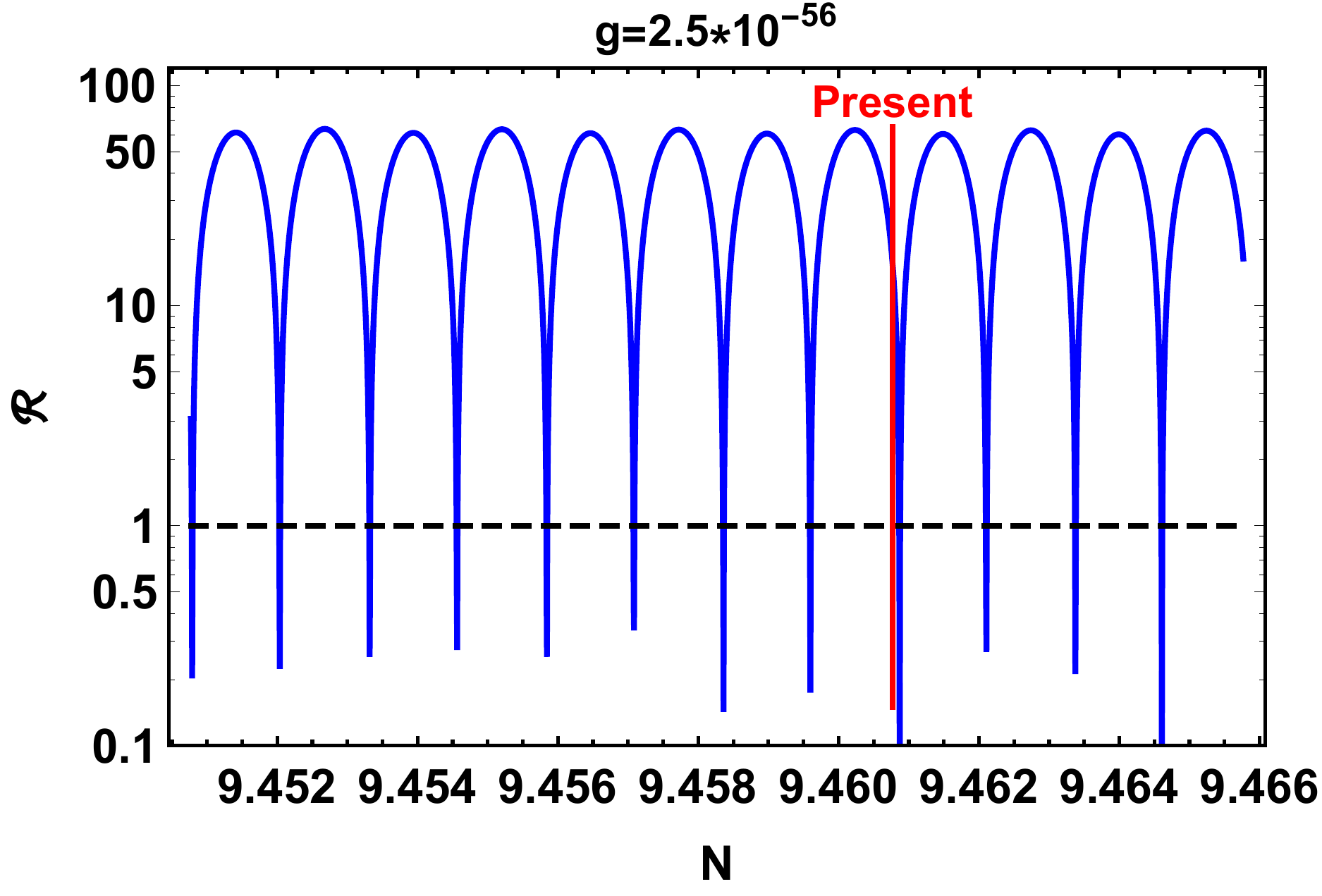}
\includegraphics[width=0.5\textwidth,angle=0]{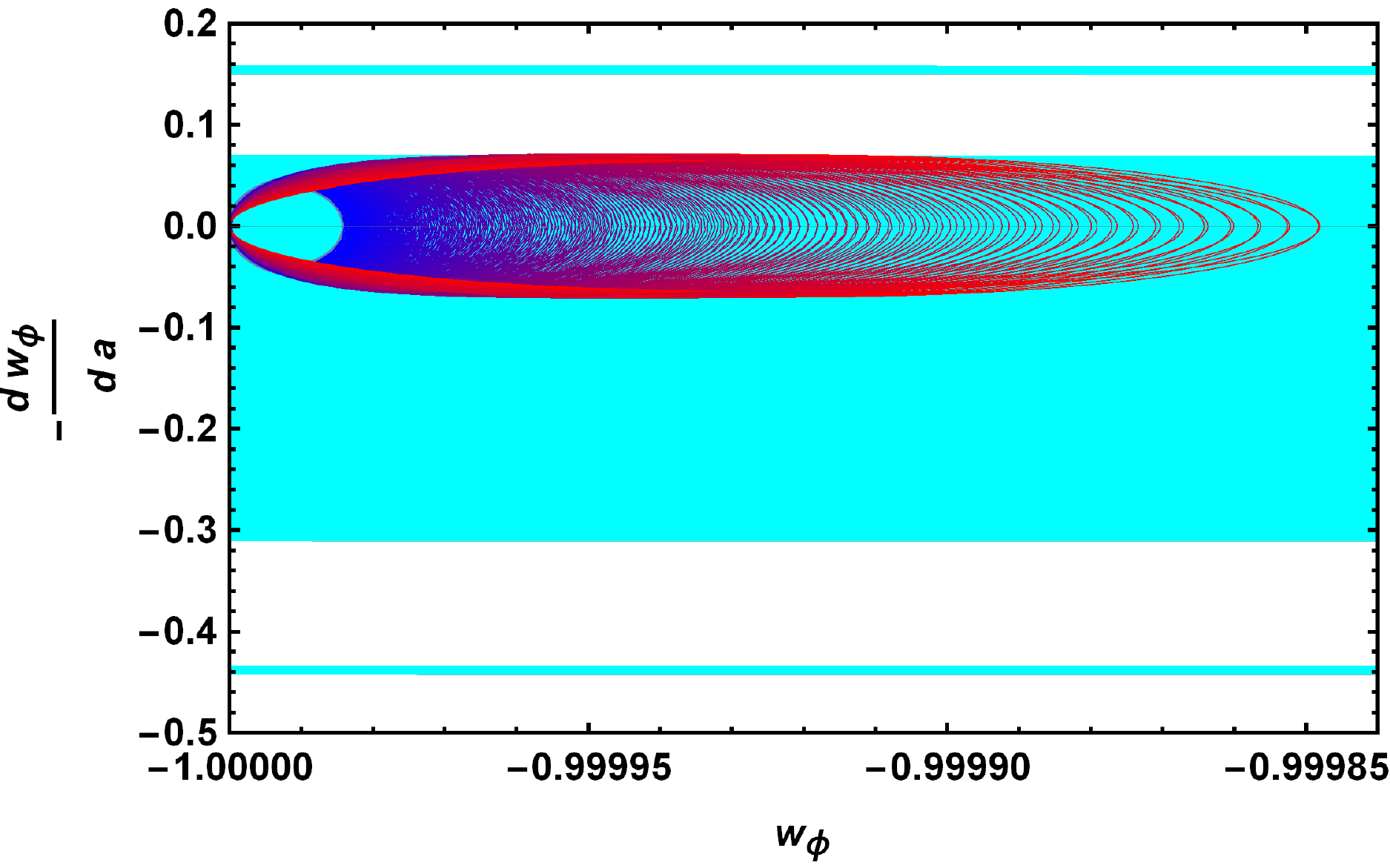}
}
\centerline{
\includegraphics[width=0.5\textwidth,angle=0]{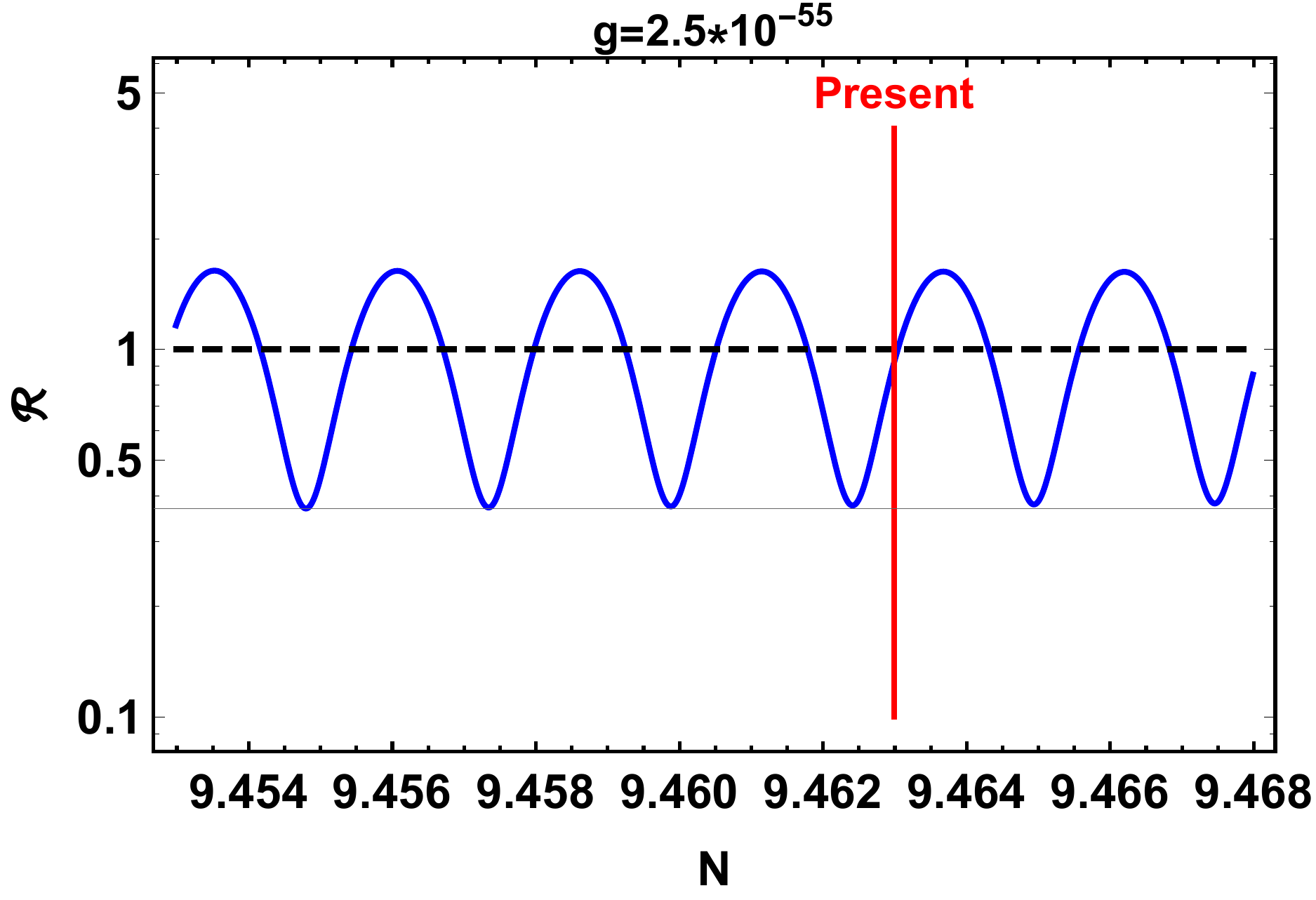}
\includegraphics[width=0.5\textwidth,angle=0]{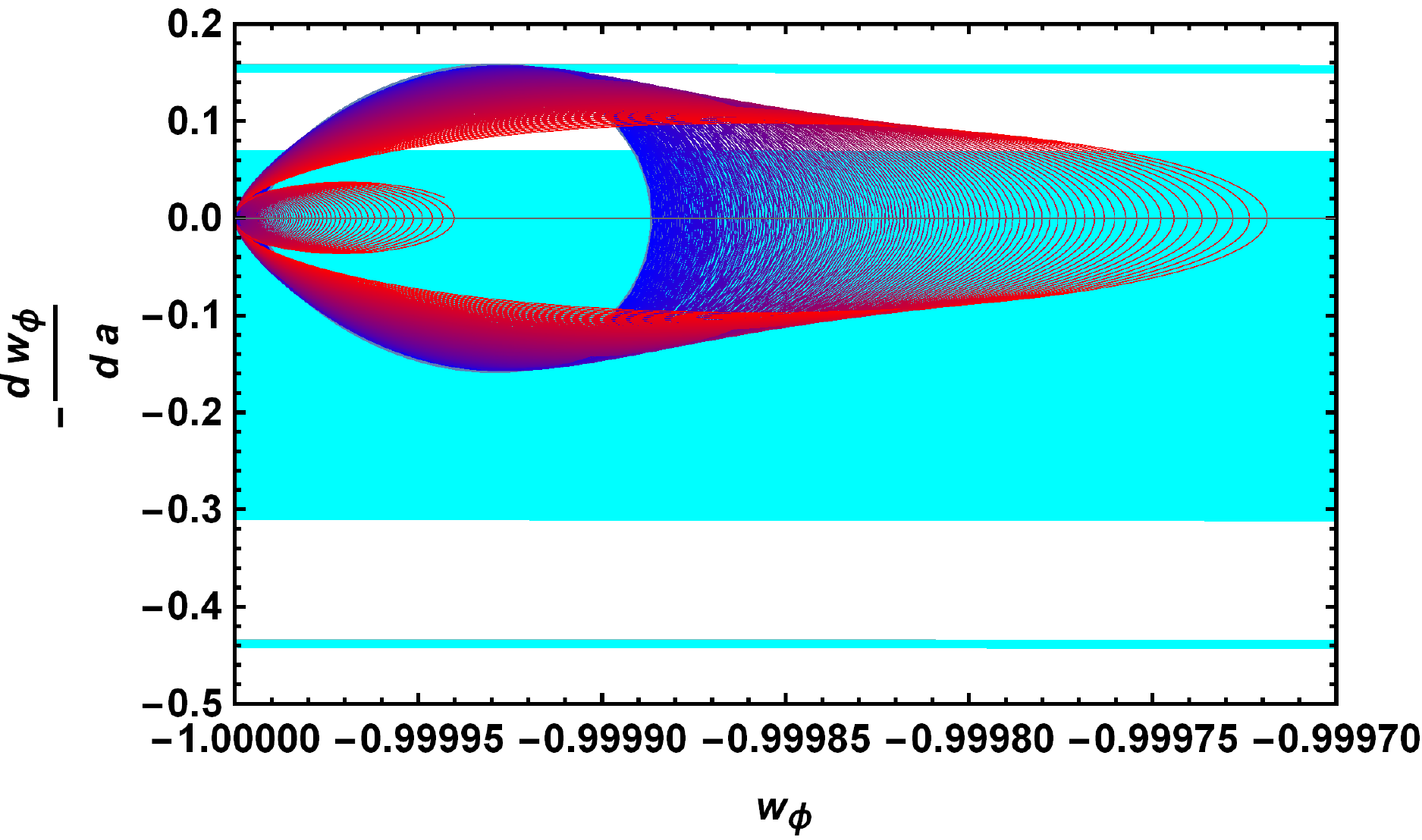}
}
\caption{The left panels display the value of ratio ${\cal R}$ defined in (\ref{eq:SU2nonlin}) for multiple values of $g$ but for fixed $\frac{g}{\tilde{f}\sqrt{\tilde{V}_0}}=7.14\cdot 10^9$ and $\tilde{Q}_{\rm in}=10^{-6}$. The panels on the right display the corresponding evolution of the equation of state parameter for the last e-fold before the present moment from red (past) to blue (present). We see that in the first two rows, the equation of state parameter is insensitive to the value of $g$ because the nonlinear term is always negligible (on average) compared to the Chern-Simons term. On the other hand for the bottom row, the nonlinear term is of the same order of magnitude as the Chern-Simons coupling and that induces a more violent oscillation in the equation of state parameter. 
}
\label{fig:SU2nonlin}
\end{figure}

which quantifies the relative size of the nonlinear term $2g^2\,a^2\, Q^3$ and the term that arises from the Chern-Simons coupling $\frac{g\,a\,Q^2}{f}\frac{\partial \phi}{\partial \tau}$. As long as ${\cal R}\gg 1$ the nonlinear term is negligibly small and the only relevant parameters are $\left(\frac{g}{\tilde{f}\sqrt{\tilde{V}_0}},\tilde{Q}_{\rm in}\right)$ as we have demonstrated above. If instead parameter $g$ is high enough so that ${\cal R}\simeq 1$ in the last e-fold, then the oscillation of the equation of state parameter of the axion becomes very violent and as a result, compatibility with observations becomes less likely. figure \ref{fig:SU2nonlin} demonstrates how the equation of state parameter changes with respect to a variation of $g$. All three panels are plotted for a fixed $\frac{g}{\tilde{f}\sqrt{\tilde{V}_0}}=7.14\cdot 10^9$ and $\tilde{Q}_{\rm in}=10^{-6}$. The two top rows are both plotted for the case $g<g_{\rm th}$ and therefore on average ${\cal R}\gg 1$. This implies that the nonlinear term never affects the evolution and in that case the predictions of the model depend only on the set $\left(\frac{g}{\tilde{f}\sqrt{\tilde{V}_0}},\tilde{Q}_{\rm in}\right)$. This is demonstrated by the fact that despite the different value of $g$, the equation of state parameter in the top two rows of figure \ref{fig:SU2nonlin} is identical. On the other hand, in the bottom row, we choose a value of $g > g_{\rm th}$ which makes ${\cal R}\simeq 1$ close to the present moment. In that case, as is seen in the bottom right plot, the degeneracy is broken and the outcome depends on the set $\left(g,\frac{g}{\tilde{f}\sqrt{\tilde{V}_0}},\tilde{Q}_{\rm in}\right)$. Since violating the requirement $g<g_{\rm th}$ results in an equation of state that oscillates more violently, we disregard this possibility and only limit our analysis to the case $g<g_{\rm th}$.

We conclude this subsection by providing the reader with an approximate formula which gives the value $g_{\rm th}(\tilde{Q}_{\rm in})$ below which our analysis is valid. We find this formula by performing multiple runs of varying parameters $g$ and $Q_{\rm in}$ and then find the value $g_{\rm th}$ for each $Q_{\rm in}$. We then fit the data and find 

\begin{equation}
g_{\rm th}\left(\tilde{Q}_{\rm in}\right)=\frac{2.75\sqrt{\tilde{V}_0}}{\tilde{Q}_{\rm in}^{0.7657}}
\label{eq:gth}
\end{equation}

As long as $g<g_{\rm th}$ the acceptable parameter space is given by figure \ref{fig:parameterspace}. On a last note, we want to stress that this analysis does not take into account possible backreaction of the various perturbations in the equations of motion. The issue of backreaction will be addressed in the next subsection where we compute the backreaction terms and we find the parameter space that guarantees that the backreaction is negligible until the present moment.

\subsection{Gravitational wave production in the axion-SU(2) case and backreaction} 
\label{sec:modelSU2-gw} 

{\hskip 2em}We tackle here the issue of gravitational wave production for the axion-SU(2) model. We write the perturbed metric and gauge field dropping all perturbations aside from the tensor ones

\begin{eqnarray}
A^a_i&=&a \left(\,Q \,\delta_{ia} +t_{i a}\right)\nonumber\\
g_{ij}&=& a^2 \left( \delta_{ij}+h_{ij}\right)
\end{eqnarray}

In this expression, $a=1,2,3$ is the SU(2) index whereas $i=1,2,3$ denotes the spatial coordinates. We name the $t_{ia}$ perturbations as "tensor modes" because they couple to the true tensor modes of the metric at the linearized level under the presence of the isotropic vev. We require that both perturbations be transverse and traceless 

\begin{equation}
\partial_i h_{ij}=\partial_i t_{ia}=\partial_a t_{ia}=0 \;\;\;,\;\;\; h_{ii}=t_{ii}=0
\end{equation}

The decomposition of the perturbations into fourier modes and left/right handed polarizations follows the convention

\begin{eqnarray}
h_{ij}\left(\tau,\vec{x}\right)&=&\int\frac{d^3 k}{\left(2\pi\right)^{3/2}}{\rm e}^{i \vec{k}\cdot\vec{x}}\sum_{\lambda=+,-}\Pi^*_{ij,\lambda}\left(\hat{k}\right)h_\lambda\left(\tau,\vec{k}\right)\\
t_{ia}\left(\tau,\vec{x}\right)&=&\int\frac{d^3 k}{\left(2\pi\right)^{3/2}}{\rm e}^{i \vec{k}\cdot\vec{x}}\sum_{\lambda=+,-}\Pi^*_{ia,\lambda}\left(\hat{k}\right)t_\lambda\left(\tau,\vec{k}\right)
\end{eqnarray}

\begin{figure}[ht!]
\centerline{
\includegraphics[width=0.3\textwidth,angle=0]{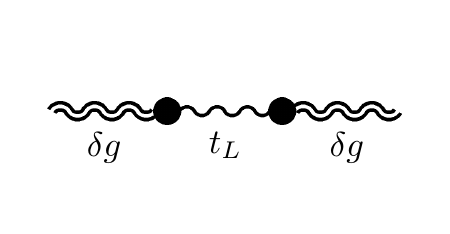}
}

\caption{Diagramatic representation of the linear sourcing of gravitational waves by the unstable polarization of the gauge field $\tilde{t}_+$
}
\label{fig:FeynmanSU(2)}
\end{figure}

The canonical normalization of the quadratic action of the tensor perturbations requires the rescaling

\begin{equation}
\begin{pmatrix}
\tilde{h}_{\pm} \\
\tilde{t}_{\pm} 
\end{pmatrix}=
\begin{pmatrix}
\frac{a M_p}{\sqrt{2}} h_{\pm} \\
\sqrt{2}a t_{\pm}
\end{pmatrix}
\end{equation}

under this rescaling we write the equations of motion for the two canonical perturbations

\begin{eqnarray}
&&\tilde{t}_{\pm}''+\left(k^2+\frac{g\,a\,Q\,\partial_{\tau}\phi}{f}\mp 2\,g\,k\,a\,Q\mp\frac{k\,\partial_{\tau}\phi}{f}\right)\tilde{t}_{\pm}=\left(\frac{2\,Q\,\partial_\tau a}{a\, M_p}+\frac{2\,\partial_\tau Q}{M_p}\right)\tilde{h}_{\pm}'\nonumber\\
&&\quad\quad\quad\quad\quad +\left(\frac{2\,g\,a\,Q^2\,\partial_\tau\phi}{f\,M_p}-\frac{2\, \partial_\tau a\, \partial_\tau Q}{a\, M_p}-\frac{2\,Q\,\left(\partial_\tau a\right)^2}{a^2\, M_p}-\frac{2\,g^2\,a^2\,Q^3}{M_p}\mp\frac{2\,g\,k\,a\,Q^2}{M_p}\right)\tilde{h}_\pm\label{eq:eomh}\\
&&\tilde{h}_\pm''+\left(k^2+\frac{a_{\rm in}^3 \rho_{\rm in}}{a^3}+\frac{3\,g^2\,a^2\,Q^4}{M_p^2}+\frac{\left(\partial_\tau \phi\right)^2}{2\,M_p^2}-\frac{2 \left(\partial_\tau a\right)}{a^2}-\frac{Q^2 \left(\partial_\tau a\right)^2}{a^2\, M_p^2}-\frac{\left(\partial_\tau Q\right)^2}{M_p^2}-\frac{2\,Q\,\partial_\tau a\,\partial_\tau Q}{a\,M_p^2}\right)\tilde{h}_\pm\nonumber\\
&&\quad\quad\quad\quad\quad\quad\quad\quad\quad\quad\quad\quad=\left(\frac{2\,g^2\,a^2\,Q^3}{M_p}\mp\frac{2\,g\,k\,a\,Q^2}{M_p}\right)\tilde{t}_\pm-\left(\frac{2\,Q\,\partial_\tau a}{a\, M_p}+\frac{2\,\partial_\tau Q}{M_p}\right)\tilde{t}_{\pm}'
\label{eq:eomt}
\end{eqnarray}

where we have separated the homogeneous and inhomogeneous terms to be respectively on the left and right hand sides. It is evident from the homogeneous part of (\ref{eq:eomt}) that the $\tilde{t}_{+}$ polarization of the gauge field tensor mode can become unstable for a range of wavenumbers under the right circumstances. This instability causes the gauge field mode functions to increase exponentially rapidly and in turn, these enhanced modes source the $\tilde{h}_+$ polarization of gravitational waves through the inhomogeneous terms on the right hand side of (\ref{eq:eomh}). The overall process of linear sourcing of gravitational waves can be diagrammatically represented as shown in figure \ref{fig:FeynmanSU(2)}. 

\subsubsection{Frequency diagnostics} 
\label{sec:modelSU2-gwApprox} 

{\hskip 2em}In order to gain a better understanding of the gravitational wave frequencies that are produced in the case of the axion-SU(2) model, we will produce a number of diagnostic formulas. The starting point is the equation of motion of the $\tilde{t}_{+}$ mode. We set the coefficient of $\tilde{t}_+$ on the left hand side equal to zero and we solve for the wavenumber $k$ to find the threshold values. We find

\begin{equation}
k_{\rm thr,\pm}\left(\tau\right)=\frac{1}{2}\left(2\,g\,a\,Q+\frac{\partial_\tau \phi}{f}\pm\frac{\sqrt{4\,f^2\,g^2\,a^2\,Q^2+\left(\partial_\tau \phi\right)^2}}{f}\right)
\label{eq:kthr}
\end{equation}

These threshold values designate the range of unstable modes at each instant in time. Since we want to maximize the wavenumbers of the produced modes we focus here on approximating only the upper limit $k_{\rm thr,+}$. we replace in the expression above the early time solutions (\ref{eq:scaearly}), (\ref{eq:phiearly}) and (\ref{eq:Qearly}).

\begin{equation}
k_{\rm thr,+,early}\left(\tau\right)\simeq\sqrt{\frac{V_0\,\lambda^2}{27\,f^2\,\bar{\rho}_m}\left(\frac{\tau}{\tau_{\rm in}}\right)^{10}+g^2\,Q_{\rm in}^2\,\bar{\rho}_m^{2/3}}+\frac{\sqrt{V_0}\,\lambda}{3\sqrt{3}\,f\,\bar{\rho}_m^{1/2}}\left(\frac{\tau}{\tau_{\rm in}}\right)^{5}+g\,Q_{\rm in}\,\bar{\rho}_m^{1/3}
\end{equation}

the growing terms always quickly overtake the constant terms therefore our final expression is

\begin{equation}
k_{\rm thr,+,early}\left(\tau\right)\simeq\frac{2 \sqrt{V_0}\,\lambda}{3\sqrt{3}\,f\,\bar{\rho}_m^{1/2}}\left(\frac{\tau}{\tau_{\rm in}}\right)^{5}
\end{equation}

which is exactly the same expression as what we found in the Abelian case. The maximum wavenumber will be achieved when the gauge friction term dominates the equation of motion of the axion field. After that moment $k_{\rm thr,+}$ seizes to increase and our approximations break down. We denote that moment as $\tau_{\rm fr}$. The maximum possible wavenumber is then 

\begin{equation}
k_{\rm max}\left(\tau_{\rm fr}\right)\simeq\frac{2 \sqrt{V_0}\,\lambda}{3\sqrt{3}\,f\,\bar{\rho}_m^{1/2}}\left(\frac{\tau_{\rm fr}}{\tau_{\rm in}}\right)^{5}
\label{eq:SU2kmax1}
\end{equation}

Following along similar lines as the axion-U(1) case, we compute the moment $\tau_{\rm fr}$ for various parameters in $\left(\frac{g}{\tilde{f}\sqrt{\tilde{V}_0}},\tilde{Q}_{\rm in}\right)$ space and we perform a two dimensional fitting. For $\lambda=1$ we find an expression that gives $\tau_{\rm fr}$ with about $5\%$ accuracy which is sufficient for our estimation

\begin{equation}
\tau_{\rm fr}\left(\frac{g}{\tilde{f}\sqrt{\tilde{V}_0}},\tilde{Q}_{\rm in}\right)\simeq D_1 \left(\frac{g}{\tilde{f}\sqrt{\tilde{V}_0}}\right)^{D_2}\, \tilde{Q}_{\rm in}^{D_3}\;\tau_{\rm in} \, \bar{\rho}_m^{1/6}
\label{eq:timefric}
\end{equation}

with the constants given by 

\begin{equation}
D_1=1.383\;\;\;\;,\;\;\;\;D_2=-0.1612\;\;\;\;,\;\;\;\;D_3=-0.1774
\end{equation}

Formula (\ref{eq:timefric}) should be understood to be the equivalent of (\ref{eq:timeback}) for the U(1) case. Replacing (\ref{eq:timefric}) in (\ref{eq:SU2kmax1}) and collecting the various parameters we have

\begin{equation}
k_{\rm max}\left(\tau_{\rm fr}\right)\simeq \frac{2 \, D_1^5\, \tilde{Q}^{5\,D_3}_{\rm in}\,\bar{\rho}_m^{1/3}\sqrt{V_0}}{3\sqrt{3}\,\tilde{f}\,M_p}\left(\frac{g}{\tilde{f}\sqrt{V_0}}\right)^{{D_2}}
\end{equation}

and converting to the physical frequency evaluated at the present moment we get

\begin{equation}
f_{\rm max}\left(\tau_{\rm fr}\right)\simeq \frac{1.517}{2\pi\;\tilde{f}^{0.194}\;\tilde{Q}_{\rm in}^{0.887}}\left(\frac{\sqrt{\tilde{V}_0}}{g}\right)^{0.806}\frac{\sqrt{V_0}}{M_p}
\label{eq:SU2kmax2}
\end{equation}

\begin{figure}[ht!]
\centerline{
\includegraphics[width=1\textwidth,angle=0]{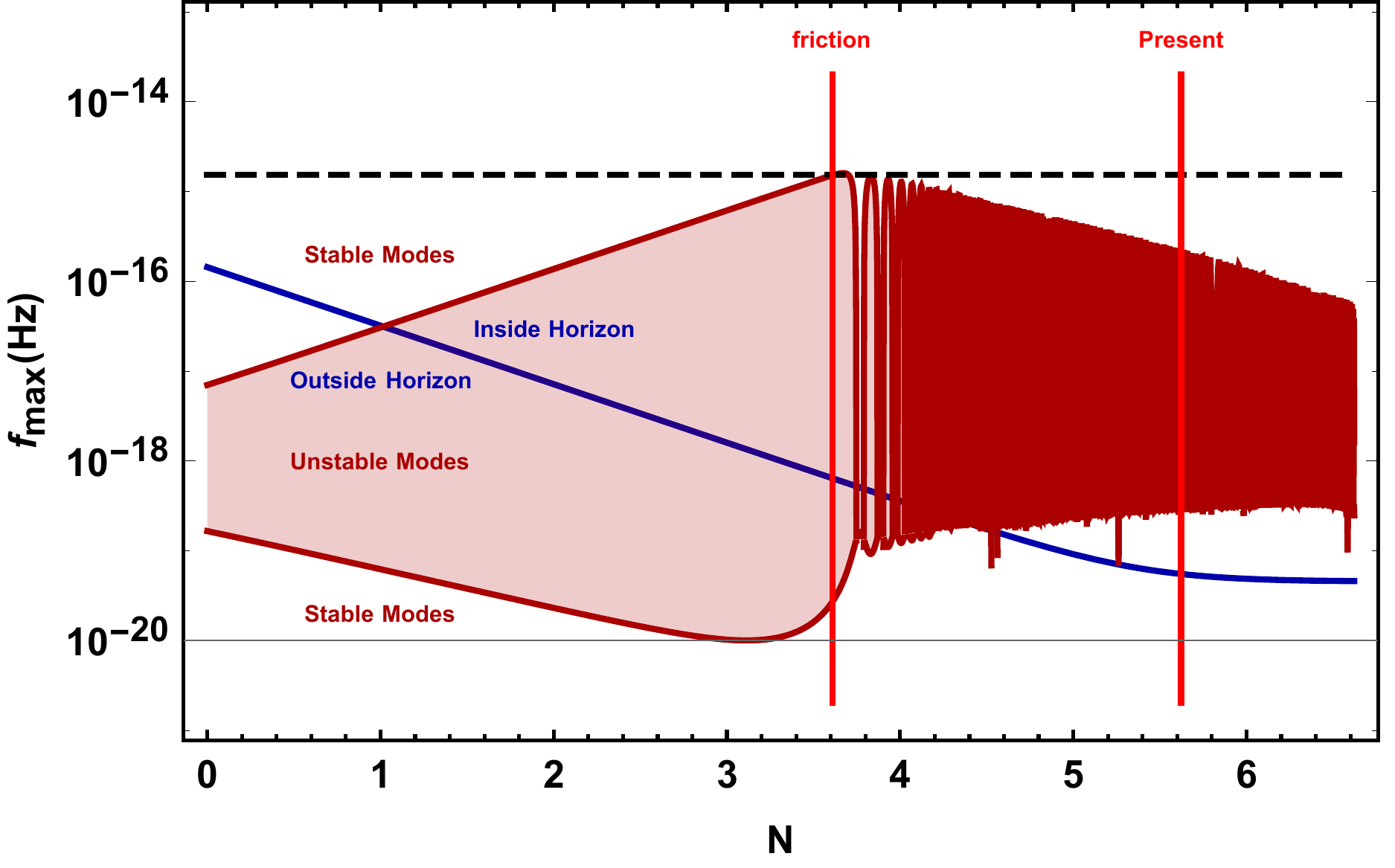}
}

\caption{Comparison between the physical unstable frequency thresholds (maroon lines), Hubble rate (blue line) and our analytic prediction (\ref{eq:SU2kmax2}) (black dashed line). The first two quantities have been converted into physical frequencies and have been normalized by dividing them by the ratio of the value of the scale factor today over the value of the scale factor at $\tau_{\rm fr}$. The two red vertical lines point out the instants when the friction becomes important, and the present moment.
}
\label{fig:SU2Comparisonfreq}
\end{figure}

This formula can be used as a diagnostic tool to compute the physical wavenumbers or frequencies of gravitational waves that are excited in our mechanism. In figure \ref{fig:SU2Comparisonfreq}, we plot the comparison between our simple analytic prediction (\ref{eq:SU2kmax2}) (black dashed line) and the numerically evaluated (\ref{eq:kthr}) our prediction is a very accurate estimate of the maximum unstable frequencies.

A few important comments are in order. Firstly, all the formulas used in this subsection referred to the unstable modes of the $\tilde{t}_{+}$ mode. Since the process that sources the gravitational waves is linear, the wavenumbers of gravitational waves that are enhanced will be the same. Secondly the formulas assume that the gauge friction becomes important in the equation of motion of the axion deep into matter domination. This will always be true as long as one considers parameters that satisfy the compatibility condition (\ref{eq:SU2compatible}) derived in the previous section. Lastly it may seem at first sight that one can produce gravitational waves of very high frequencies by choosing the various parameters appropriately. However, if one chooses really high values for the couplings or for the initial value of the gauge field it is possible that the backreaction of the enhanced tensor modes in the various dynamical equations might become important and then neither our frequency diagnostics or our conclusions in regards to compatibility with observations should be regarded as valid. We comment on the issue of backreaction in the next subsection.

\subsubsection{Gravitational wave power spectrum and backreaction} 
\label{sec:modelSU2-gwPS} 

{\hskip 2em}Unlike the U(1) case, the gravitational wave power spectrum for the SU(2) case will be computed using only numerical means. The reason for this is that the onset of friction takes place at considerably earlier times compared to the U(1) case. This implies that, if we only compute the gravitational waves produced until the onset of friction, we will be severely underestimating the total gravitational wave power spectrum as those gravitational waves will have decayed substantially until the present moment.

Additionally, computing the gravitational wave power spectrum cannot be done in a consistent way unless the issue of backreaction is addressed concurrently. For other work on the issue of backreaction in axion-SU(2) couplings in the context of inflation see \cite{Dimastrogiovanni:2016fuu,Mirzagholi:2019jeb,Fujita:2017jwq}. The backreaction of the unstable $\tilde{t}_{+}$ modes on the background equations of motion is given by

\begin{eqnarray}
\rho_{\tilde{t}}&=&\frac{1}{2\,a^2}\int\,\frac{d^3k}{(2\pi)^3}\left[\left|\frac{\partial_{\tau}\tilde{t}_+}{a}\right|^2+\left(\frac{k^2}{a^2}-2\,g\,Q\,\frac{k}{a}\right)\left|\tilde{t}_{+}\right|^2\right]\\
{\cal T}^\phi_{\rm BR}&=& -\frac{1}{2\,a^4\,f}\partial_\tau\int\,\frac{d^3k}{(2\pi)^3}\,\left(a\,g\,Q-k\right)\left|\tilde{t}_{+}\right|^2\\
{\cal T}^Q_{\rm BR}&=& \frac{g}{3a^2}\int\frac{d^3k}{(2\pi)^3}\left(\frac{\partial_\tau \phi}{2\,f\,a}-\frac{k}{a}\right)\left|\tilde{t}_{+}\right|^2
\end{eqnarray}

We define the ratios

\begin{equation}
\tilde{\rho}_{\tilde{t}}=\frac{\rho_{\tilde{t}}}{\rho_{\rm tot}}\;\;\;,\;\;\;
\tilde{{\cal T}}^Q_{\rm BR}=\frac{{\cal T}^Q_{\rm BR}}{\partial^2_\tau Q}\;\;\;,\;\;\;
\tilde{{\cal T}}^\phi_{\rm BR}=\frac{{\cal T}^\phi_{\rm BR}}{\partial^2_\tau \phi}
\label{eq:SU2backcheck}
\end{equation}

Which can be used to gauge the relative contribution of the backreaction to the equations of motion.

\begin{figure}[ht!]
\centerline{
\includegraphics[width=0.5\textwidth,angle=0]{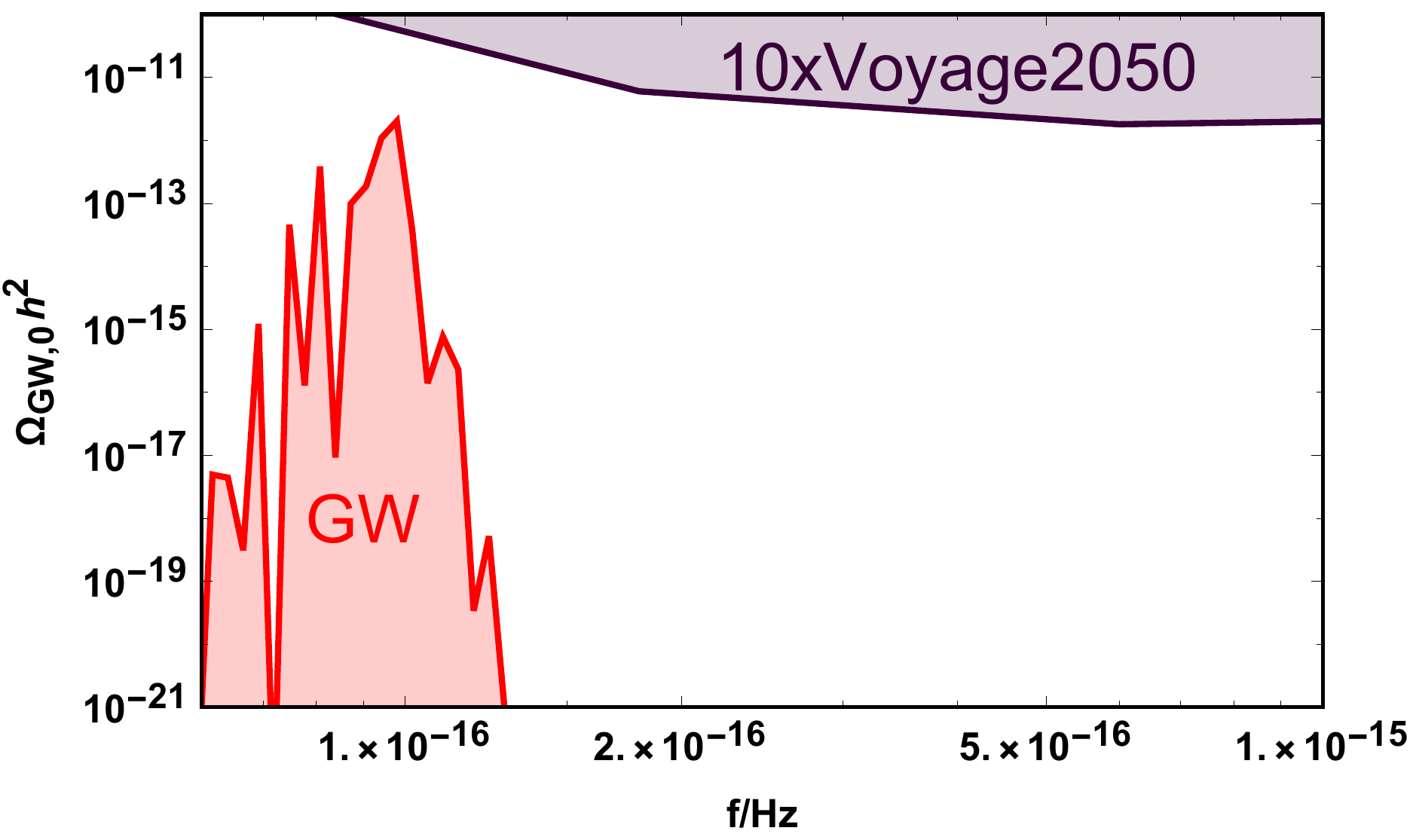}
\includegraphics[width=0.5\textwidth,angle=0]{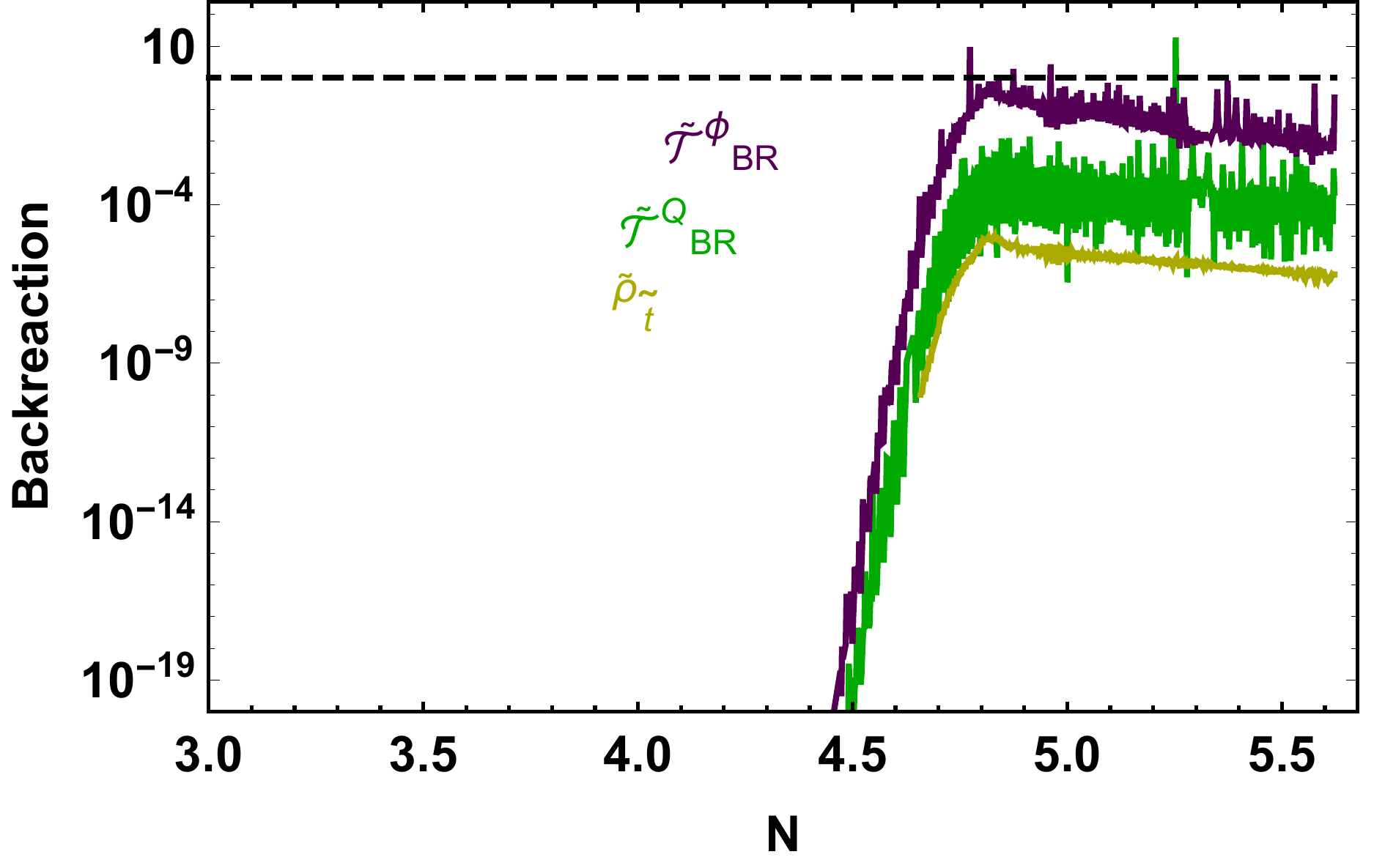}
}

\caption{\textit{Left Panel}: We plot the gravitational wave power spectrum and compare with the most sensitive experimental projections shown in figure 2 of \cite{Kite:2020uix}. \textit{Right Panel}: The backreaction criteria (\ref{eq:SU2backcheck}) are plotted. As long as they are less than one, the backreaction has no impact on the background equations of motion and the solution is consistent. We choose parameters that maximize the gravitational wave power spectrum while saturating the backreaction limits. The parameters used in this run are $g=9.06\cdot 10^{-56}$, $f/M_p=1.27\cdot 10^{-5}$, $Q_{\rm in}/M_p=10^{-6}$ and $\lambda=1$.
}
\label{fig:SU2GW}
\end{figure}

We solve a system of differential equations that consists of the background equations (\ref{eq:eomphi2}), (\ref{eq:gaugeeom}) and (\ref{eq:friedmanSU2}) combined with the equations of motion for the perturbations (\ref{eq:eomh}) and (\ref{eq:eomt}). The backreaction terms are not included in the evolution. Instead they are reconstructed at the end and we check if the quantities (\ref{eq:SU2backcheck}) are less than one. The perturbations are initialized at early times when the solutions are positive frequency plane waves with amplitude $\frac{1}{\sqrt{2k}}$. We include in our system ${\cal O}(10)$ number of modes spanning wavelengths up to a maximum frequency given by (\ref{eq:SU2kmax2}) which are equally separated in $\log k$ space. After scanning the parameter space we find that it is very difficult to find parameters for which the backreaction conditions (\ref{eq:SU2backcheck}) are less than one while simultaneously resulting in a power spectrum that overlaps with CMB spectral distortion curves. The best case scenario is shown in figure (\ref{fig:SU2GW}). One can observe that the gravitational wave spectrum is highly oscillating in $k$-space. This is a result of the fast oscillations of $Q(\tau)$ and $\partial_\tau \phi$ which take place when friction becomes strong. 

Overall it appears that it is highly unlikely that gravitational waves produced in this scenario would be detectable in spectral distortions of the CMB, at least in the low backreaction limit. Exploring the strong backreaction limit for the non-Abelian case is beyond the scope of this work but we are hoping to revisit this issue in the future.

\section{Conclusions} 
\label{sec:conclusions} 

{\hskip 2em}The universe is currently undergoing a period of accelerated expansion. This is the second period of accelerated expansion aside from inflation which took place in the primordial universe and set the initial conditions for the standard "Big-Bang" Cosmology. Drawing inspiration from inflation, Quintessence models aspire to explain the late period of accelerated expansion by assuming that there is a scalar field in a slow-roll configuration whose energy density dominates at late times. In particular, models of tracking quintessence are particularly interesting since the evolution of the scalar field tracks that of the dominant background source (matter and radiation) for most of the cosmic history, and emerges only at late times \cite{Steinhardt:1999nw}. The tracking mechanism has been proposed as a possible explanation of the coincidence problem, namely why the current value of the dark energy is comparable to that of dark matter. 

Observations constrain the slope of the potential to be very flat. This requirement is in tension with conjectures that arise from String Theory and Quantum Gravity which suggest that the scalar potential should be steeper. Looking to reconcile these two requirements, a model titled "warm dark energy" \cite{DallAgata:2019yrr} was recently proposed in which an axion field is coupled to a U(1) gauge field by means of a topological term. In this instance slow-roll is achieved owning to the fact that the axion field produces the gauge field at the expense of its own kinetic energy. This model is a dark energy realization of models of inflation in which slow-roll is achieved as a result of a process that provides additional effective friction aside from the standard Hubble one, such as the original model of warm inflation \cite{Berera:1995ie}, trapped inflation \cite{Green:2009ds,Pearce:2016qtn}, the Anber-Sorbo mechanism \cite{Anber:2009ua} and Chromo-Natural Inflation \cite{Adshead:2012kp,Dimastrogiovanni:2012ew,Adshead:2013nka}. 

In this work we undertake the natural next step after the study of quintessence with an axion-U(1) coupling. We couple the axion with a non-Abelian SU(2) gauge field and we perform a full analysis of the background and specify the parameter space that allows the swampland programme to be reconciled with observational data. Our approach assumes that early, deep in the matter domination the two fields are not interacting and that the gauge field vev has an non-zero but small, isotropic vev. Our results suggest that at late times the two fields interact in such a way that additional friction is produced in the equation of motion of the axion field. Our main results at the background level are shown in figure \ref{fig:parameterspace} and are complemented by formulas (\ref{eq:SU2compatible}), (\ref{eq:SU2incompatible}) and (\ref{eq:gth}). These formulas define the allowed parameter space. We find that this scenario can successfully reconcile requirements of the swampland conjectures and observational data. We stress that the mechanism employed here is fundamentally different than the axion-U(1) case. In the axion-U(1) case the friction is provided by tachyonic production and backreaction of U(1) gauge quanta whereas in the axion-SU(2) case the friction arises at the classical level using only the background equations of motion of the axion and of a classical gauge field vev. 

Additionally, the recent proposal that gravitational waves of small frequencies, in the range $f\sim \left(10^{-15} - 10^{-9}\right)\,{\rm Hz}$ can be probed by CMB spectral distortions \cite{Kite:2020uix} provides motivation for an analysis of the gravitational waves produced by both the axion-U(1) and axion-SU(2) couplings in the context of quintessence. We have derived analytic formulas that allow for the estimation of the gravitational waves produced and we found that frequencies of order $f\sim {\cal O}(10^{-16}) \;{\rm Hz}$ can be produced in both models for a reasonable range of parameters. We subsequently computed the power spectrum for both models. 

In the axion-U(1) case,  we find that in the most favorable scenario there can be a small overlap between the gravitational wave power spectrum produced and the most generous sensitivity curves shown in \cite{Kite:2020uix}. Our analysis relies on analytic formulas and only computes gravitational waves produced up to the onset of backreaction. As such it should be considered as a lower limit. Despite that we believe it is unlikely that such gravitational waves would leave a significant imprint on CMB spectral distortions and we do not report any constraints on the parameter space as a result of overproduction of gravitational waves.

In the axion-SU(2) case, we instead perform a full numerical analysis that computes the gravitational wave production until the present moment. In that case we find that even in the best case scenario there is no overlap between experimental sensitivities and our computed power spectrum because the production of gravitational waves is typically accompanied by the production of high amplitude perturbations which backreact strongly on the background equations of motion. As a result these gravitational waves would be inaccessible to measurements of CMB spectral distortions.

Additionally we would like to comment on similarities and differences between the present work and closely related models that appear in \cite{Alexander:2016nrg,Alexander:2016mrw}. In both of those works, as well as the present one, an axion field coupled to an SU(2) gauge field with a classical isotropic vev plays the role of the quintessence field responsible for the late time acceleration of the universe. The Chern Simons coupling leads to a growing contribution to the gauge field vev which then provides additional friction in the equation of motion of the axion. In the present work we made use of a massless gauge field which, immediatelly before the interactions become important, scales as $\rho_{\rm gauge}\sim a^{-4}$. On the other hand in both \cite{Alexander:2016nrg,Alexander:2016mrw} a massive gauge field was used instead which scales like $\rho_{\rm gauge}\sim a^{-3}$ in the moments before the growing term emerges. The presence of the mass changes both the form of the growing term as well as the late time behavior of the solutions. Another important difference is that, in addition to studying the background dynamics, we also perform a study of the gravitational waves produced in the axion-gauge quintessence scenario.

Finally, we find that both an axion-U(1) coupling as well as an axion-SU(2) coupling can lead to slow-roll of a quintessence field in a way that satisfies string theory and quantum gravity conjectures while remaining consistent with observations. One regime that we have not yet probed is the high backreaction regime of the axion-SU(2) case. We believe that this is possible to do using the techniques developed in \cite{DallAgata:2019yrr}. Exploring the strong backreaction regime in this model may have important implications for models of inflation such as chromo-natural inflation or spectator chromo-natural inflation. We hope to revisit this subject in a future work.

\vskip.25cm
\section*{Acknowledgements} 

{\hskip 2em}The author would like to thank Marco Peloso for valuable discussions and feedback at various stages of this work.

\vskip.25cm

\appendix

\section{Details of two point function}
\label{app:twopoint}

{\hskip 2em}We devote this Appendix to presenting the details of the computation of the gravitational wave power spectrum since their presence in the main body might be distracting to the reader. We start with the equation of motion of the tensor modes

\begin{equation}
G_{ij}=\frac{1}{2a^2}\left(\partial_\tau^2-\Delta+\frac{2a'}{a}\partial_\tau\right)h_{ij}=\frac{1}{M_p^2}T_{ij}
\end{equation}

We are working in the transverse-traceless gauge and therefore we are only interested in the TT contribution of the energy-momentum tensor for the gauge field

\begin{equation}
T_{ij}^{TT}=F_{i}^\mu F_{j\nu}=\frac{1}{a^4}\left(-F_{i0}F_{j0}+F_{ik}F_{jk}\right)^{TT}=(-E_i E_j - B_i B_j)^{TT}
\end{equation}

The equation of motion becomes

\begin{equation}
\left(\partial_\tau^2-\Delta+\frac{2a'}{a}\partial_\tau\right)h_{ij}=-\frac{2 a^2}{M_p^2}\left(E_i E_j+B_i B_j\right)^{TT}\equiv{\cal S}_{ij}^{TT}
\end{equation}

We expand both sides into a fourier basis. 

\begin{equation}
h_{ij}\left(\tau,\vec{x}\right)=\int \frac{d^3 k}{\left(2 \pi\right)^{3/2}} {\rm e}^{i \vec{k}\cdot \vec{x}}h_{ij}\left(\tau,\vec{k}\right)
\end{equation}

\begin{equation}
{\cal S}_{ij}\left(\tau,\vec{x}\right)=\int \frac{d^3 k}{\left(2 \pi\right)^{3/2}} {\rm e}^{i \vec{k}\cdot \vec{x}}{\cal S}_{ij}\left(\tau,\vec{k}\right)
\end{equation}

The equation of motion is then transformed

\begin{equation}
\left(\partial_\tau^2+k^2+\frac{2a'}{a}\partial_\tau\right)h_{ij}\left(\tau,\vec{k}\right)={\cal S}_{ij}^{TT}\left(\tau,\vec{k}\right)
\end{equation}

Next we write the transverse traceless part of the source in terms of the original source using the projection operator 

\begin{equation}
{\cal S}_{ij}^{TT}\left(\tau,\vec{k}\right)=\Lambda_{ij,lm}\left(\hat{k}\right)\,{\cal S}_{lm}\left(\tau,\vec{k}\right)
\end{equation}

Where the projection operator is defined as

\begin{equation}
\Lambda_{ij,lm}\left(\hat{k}\right)\equiv P_{il}\left(\hat{k}\right)P_{jm}\left(\hat{k}\right)-\frac{1}{2}P_{ij}\left(\hat{k}\right)P_{lm}\left(\hat{k}\right)\;\;\;\;,\;\;\;\; P_{ij}\left(\hat{k}\right)=\delta_{ij}-\hat{k}_1\hat{k}_2
\end{equation}

We rewrite the equation of motion

\begin{equation}
\left(\partial_\tau^2+k^2+\frac{2a'}{a}\partial_\tau\right)h_{ij}\left(\tau,\vec{k}\right)=\Lambda_{ij,lm}\left(\hat{k}\right){\cal S}_{lm}\left(\tau,\vec{k}\right)
\end{equation}

and then we operate on both sides with $\Pi^*_{ij,\lambda}\left(\hat{k}\right)$ in order to decompose the fields into a chiral basis. We can conveniently use the property

\begin{equation}
\Pi_{ij,\lambda}^*\left(\hat{k}\right)\Lambda_{ij,lm}\left(\hat{k}\right)=\Pi^*_{lm,\lambda}\left(\hat{k}\right)
\end{equation}
 
so that

\begin{equation}
\left(\partial_\tau^2+k^2+\frac{2a'}{a}\partial_\tau\right)h_{\lambda}\left(\tau,\vec{k}\right)=\Pi^*_{lm,\lambda}\left(\hat{k}\right){\cal S}_{lm}\left(\tau,\vec{k}\right)\equiv {\cal S}_{\lambda}\left(\tau,\vec{k}\right)
\end{equation}

The particular solution, corresponding to the source term can be written as 

\begin{equation}
h_{\lambda}\left(\tau,\vec{k}\right)=\int^{\tau}_{\tau_{\rm in}} {\rm d}\tau' G_{k}\left(\tau,\tau'\right){\cal S}_{\lambda}\left(\tau',\vec{k}\right)
\label{eq:GWeom}
\end{equation}

During matter domination the scale factor scales with conformal time in the following way 

\begin{equation}
a\left(\tau\right)=\left(\frac{\tau}{\tau_{\rm in}}\right)^2
\label{scale}
\end{equation}

Where $\tau_{\rm in}$ denotes the starting point of our calculation in which the scale factor is normalized to $1$. During matter domination the differential operator corresponding to the homogeneous equation can be written as

\begin{equation}
{\cal D}=\left(\partial_\tau^2+k^2+\frac{4}{\tau}\partial_\tau\right)
\end{equation}

The corresponding retarded Green function is

\begin{eqnarray}
G_k\left(\tau,\tau'\right)&=&\frac{1}{k^5\tau'^{4}}\{-3 k \left(\tau-\tau'\right)\left(3+k^2 \tau \tau'\right)\cos k\left(\tau-\tau'\right)\nonumber\\
&&\quad\quad\quad\quad \quad\quad\quad\quad \quad\quad\quad+\left[9+k^4\tau^2\tau'^2-3 k^2\left(\tau^2-3\tau \tau' +\tau'^2\right)\right]\sin k\left(\tau- \tau'\right)\}\nonumber\\
\label{Green}
\end{eqnarray}

For $\tau>\tau'$ while $G_k\left(\tau<\tau'\right)=0$

Now that we have the Green function, we turn our attention to the source and compute it in terms of the gauge field. We will make no specific assumption about the gauge field or the evolution of the universe at this stage.

\begin{eqnarray}
{\cal S}_{ij}\left(\tau,\vec{k}\right)&=&-\frac{2 a^2}{M_p^2}\int \frac{d^3 x}{\left(2\pi\right)^{3/2}}{ \rm e}^{-i \vec{k}\cdot\vec{x}}\left[\frac{1}{a^4}\partial_\tau A_i(\tau,\vec{x}) \; \partial_\tau A_j(\tau,\vec{x})+\epsilon_{ifk}\partial_f A_k(\tau,\vec{x}) \;\epsilon_{jln} \partial_l A_n(\tau,\vec{x})\right]\nonumber\\
&=&-\frac{2}{a^2 M_p^2}\int \frac{d^3 k'}{\left(2\pi\right)^{3/2}}\Big[\partial_\tau A_i\left(\tau,\vec{k}'\right) \; \partial_\tau A_j\left(\tau,\vec{k}-\vec{k}'\right)\nonumber\\
&& \quad\quad\quad\quad \quad\quad\quad\quad \quad\quad\quad\quad \quad\quad- \epsilon_{ifk}k'_f A_k\left(\tau,\vec{k}'\right)  \;\epsilon_{jln} \left(k_l-k'_l\right) A_n\left(\tau,\vec{k}-\vec{k}'\right) \Big]\nonumber\\
\end{eqnarray}

We promote the gauge field to a quantum operator in the following way

\begin{equation}
A_i\left(\tau,\vec{k}\right)\rightarrow \hat{A}_i\left(\tau,\vec{k}\right)=\sum_{\lambda=+,-} \epsilon_{i,\lambda}\left(\vec{k}\right)\left[A_\lambda\left(\tau,k\right)\hat{a}^{\vec{k}}_\lambda+A^*_{\lambda}\left(\tau,k\right) \hat{a}^{\dagger-\vec{k}}_\lambda\right]
\end{equation}

The circular polarization vectors satisfy $\vec{k}\cdot\vec{\epsilon}_{\pm}\left(\vec{k}\right)=0$, $\vec{k}\times \vec{\epsilon}_{\pm}\left(\vec{k}\right)=\mp i k\vec{\epsilon}_{\pm}\left(\vec{k}\right)$, $\vec{\epsilon}_{\pm}\left(-\vec{k}\right)=\vec{\epsilon}_{\pm}\left(\vec{k}\right)^*$ and their normalization is $\vec{\epsilon}_\lambda\left(\vec{k}\right)^*\cdot \vec{\epsilon}_{\lambda'}\left(\vec{k}\right)=\delta_{\lambda \lambda'}$. The creation and annihilation operators satisfy canonical comutation relations $\left[\hat{a}_\lambda^{\vec{k}},\hat{a}^{\dagger\vec{k}'}_\sigma\right]=\delta_{\lambda\sigma}\delta^{(3)}\left(\vec{k}-\vec{k}'\right)$.

For simplicity we ignore the left handed polarization of the gauge field and maintain only the right handed one $A_+$. This can be done because without loss of generality we can choose which of the two polarizations is the unstable one that grows tachyonically. Any effects of the non-tachyonic polarization will be negligible contributions to our final result. We use the properties of the polarization vectors and write the source in terms of mode functions.

\begin{eqnarray}
{\cal S}_{ij}\left(\tau,\vec{k}\right)&=&
-\frac{2}{a^2 M_p^2}\int \frac{d^3 k'}{\left(2\pi\right)^{3/2}}\epsilon_{i,+}\left(\vec{k}'\right)\epsilon_{j,+}\left(\vec{k}-\vec{k}'\right)\nonumber\\
&&\!\!\!\!\!\!\!\! \!\!\!\!\!\!\!\! \!\!\!\!\!\!\!\! \!\!\!\!\!\!\!\!\times\Big\{\left[A'_+\left(\tau,k'\right)\hat{a}_+^{\vec{k}'}+{A'}^{*}_+\left(\tau,k'\right)\hat{a}_+^{\dagger-\vec{k}'} \right] \left[A'_+\left(\tau,\left\vert\vec{k}-\vec{k}'\right\vert\right)\hat{a}_+^{\vec{k}-\vec{k}'}+{A'}^{*}_+\left(\tau,\left\vert\vec{k}-\vec{k}'\right\vert\right)\hat{a}_+^{\dagger \vec{k}'-\vec{k}} \right]\nonumber\\
&&\!\!\!\!\!\!\!\! \!\!\!\!\!\!\!\! \!\!\!\!\!\!\!\! \!\!\!\!\!\!\!\!+ k' \left\vert \vec{k}-\vec{k}'\right\vert \left[A_+\left(\tau,k'\right)\hat{a}_+^{\vec{k}'}+{A}^{*}_+\left(\tau,k'\right)\hat{a}_+^{\dagger-\vec{k}'} \right] \left[A_+\left(\tau,\left\vert\vec{k}-\vec{k}'\right\vert\right)\hat{a}_+^{\vec{k}-\vec{k}'}+{A}^{*}_+\left(\tau,\left\vert\vec{k}-\vec{k}'\right\vert\right)\hat{a}_+^{\dagger \vec{k}'-\vec{k}} \right]\Big\}\nonumber\\
\end{eqnarray}

We next multiply the source with the polarization tensor to form the quantity ${\cal S}_\lambda$ defined earlier and compute the two point function of the source by performing Wick's theorem. The final answer is 

\begin{eqnarray}
\left\langle{\cal S}_{\lambda}\left(\tau',\vec{k}\right){\cal S}_{\lambda}\left(\tau'',\vec{p}\right)\right\rangle&=&
\frac{4\;\delta^{(3)}\left(\vec{k}+\vec{p}\right)}{a(\tau')^2 a(\tau'')^2 M_p^4}\int \frac{d^3 k'}{\left(2\pi\right)^{3}}\left\vert\vec{\epsilon}_{\lambda}\left(\vec{p}\right)\cdot\vec{\epsilon}_{+}\left(\vec{k}'+\vec{p}\right)\right\vert^2\left\vert\vec{\epsilon}_{\lambda}\left(\vec{p}\right)\cdot\vec{\epsilon}_{+}\left(-\vec{k}'\right)\right\vert^2\nonumber\\
&&\!\!\!\!\!\! \!\!\!\!\!\! \!\!\!\!\!\! \!\!\!\!\!\!\times\left[\left\vert\vec{k}'+\vec{p}\right\vert k' A_+\left(\tau',k'\right)A_+\left(\tau',\left\vert\vec{k}'+\vec{p}\right\vert\right)+A'_+\left(\tau',k'\right)A'_+\left(\tau',\left\vert\vec{k}'+\vec{p}\right\vert\right)\right]\nonumber\\
&&\!\!\!\!\!\! \!\!\!\!\!\!\!\!\!\!\!\! \!\!\!\!\!\!\times \left[\left\vert\vec{k}'+\vec{p}\right\vert k' A^*_+\left(\tau'',k'\right)A^*_+\left(\tau'',\left\vert\vec{k}'+\vec{p}\right\vert\right)+{A'}^{*}_+\left(\tau'',k'\right){A'}^{*}_+\left(\tau'',\left\vert\vec{k}'+\vec{p}\right\vert\right)\right]\nonumber\\
\label{eq:source}
\end{eqnarray}

Using (\ref{eq:source}) and (\ref{eq:GWeom}) it is possible to write the two point function of gravitational waves which, after some simplification,  becomes

\begin{eqnarray}
\left\langle h_\lambda' \left(\tau,\vec{k}\right)h_\lambda' \left(\tau,\vec{p}\right)\right\rangle&=&\frac{4\;\delta^{(3)}\left(\vec{k}+\vec{p}\right)}{M_p^4}\int \frac{d^3 k'}{\left(2\pi\right)^{3}}\left\vert\vec{\epsilon}_{\lambda}\left(\vec{p}\right)\cdot\vec{\epsilon}_{+}\left(\vec{k}'+\vec{p}\right)\right\vert^2\left\vert\vec{\epsilon}_{\lambda}\left(\vec{p}\right)\cdot\vec{\epsilon}_{+}\left(-\vec{k}'\right)\right\vert^2\nonumber\\
&\!\!\!\!\!\!\!\!\! \!\!\!\!\!\!\!\!\! \!\!\!\!\!\!\!\!\! \!\!\!\!\!\!\!\!\! \!\!\!\!\!\!\!\!\! \!\!\!\!\!\!\!\!\! \!\!\!\!\!\!\!\!\! \!\!\!\!\!\!\!\!\! \!\!\!\!\!\!\!\!\! \!\!\!\!\!\!\!\!\! \!\!\!\!\!\!\!\!\! \times&\!\!\!\!\!\!\!\!\! \!\!\!\!\!\!\!\!\! \!\!\!\!\!\!\!\!\! \!\!\!\!\!\!\!\!\! \!\!\!\!\!\!\!\!\! \!\!\!\!\!\!\!\!\!\left\vert \int_{\tau_{\rm in}}^{\tau}\frac{{\rm d}\tau'}{a(\tau')^2} \,\frac{d}{d\tau} G_p\left(\tau,\tau'\right)\left[\left\vert\vec{k}'+\vec{p}\right\vert k' A_+\left(\tau',k'\right)A_+\left(\tau',\left\vert\vec{k}'+\vec{p}\right\vert\right)+A'_+\left(\tau',k'\right)A'_+\left(\tau',\left\vert\vec{k}'+\vec{p}\right\vert\right)\right]\right\vert^2\nonumber\\
\end{eqnarray}

Using the polarization vector identity

\begin{equation}
\left\vert \vec{\epsilon}_\lambda\left(\vec{p}\right)\cdot \vec{\epsilon}_+\left(\vec{q}\right)\right\vert^2=\frac{1}{4}\left[1-\lambda\frac{\vec{p}\cdot\vec{q}}{\vert \vec{p}\vert\,\vert\vec{q}\vert}\right]^2
\end{equation}

We arrive at our final expression which is ready to be integrated. 

\begin{eqnarray}
\left\langle h_\lambda' \left(\tau,\vec{k}\right)h_\lambda' \left(\tau,\vec{p}\right)\right\rangle&=&\frac{\delta^{(3)}\left(\vec{k}+\vec{p}\right)}{4 M_p^4}\int \frac{d^3 k'}{\left(2\pi\right)^{3}}\left[1-\lambda\frac{\vec{p}\cdot\left(\vec{k}'+\vec{p}\right)}{p\,\vert\vec{k}'+\vec{p}\vert}\right]^2 \left[1+\lambda\frac{\vec{p}\cdot\vec{k}'}{p\,k}\right]^2\nonumber\\
&\!\!\!\!\!\!\!\!\! \!\!\!\!\!\!\!\!\! \!\!\!\!\!\!\!\!\! \!\!\!\!\!\!\!\!\! \!\!\!\!\!\!\!\!\! \!\!\!\!\!\!\!\!\! \!\!\!\!\!\!\!\!\! \!\!\!\!\!\!\!\!\! \!\!\!\!\!\!\!\!\! \!\!\!\!\!\!\!\!\! \!\!\!\!\!\!\!\!\! \times&\!\!\!\!\!\!\!\!\! \!\!\!\!\!\!\!\!\! \!\!\!\!\!\!\!\!\! \!\!\!\!\!\!\!\!\! \!\!\!\!\!\!\!\!\! \!\!\!\!\!\!\!\!\!\left\vert \int_{\tau_{\rm in}}^{\tau}\frac{{\rm d}\tau'}{a(\tau')^2} \,\frac{d}{d\tau} G_p\left(\tau,\tau'\right)\left[\left\vert\vec{k}'+\vec{p}\right\vert k' A_+\left(\tau',k'\right)A_+\left(\tau',\left\vert\vec{k}'+\vec{p}\right\vert\right)+A'_+\left(\tau',k'\right)A'_+\left(\tau',\left\vert\vec{k}'+\vec{p}\right\vert\right)\right]\right\vert^2\nonumber\\
\end{eqnarray}

All the necessary ingredients for computing the above integrals are available in exact analytic form. The expressions for the scale factor, the Green function and the gauge field are available in various parts of the present work so we rewrite them collectively here for convenience.

\begin{eqnarray}
a\left(\tau\right)&=&\left(\frac{\tau}{\tau_{\rm in}}\right)^2\nonumber\\
G_k\left(\tau,\tau'\right)&=&\frac{1}{k^5\tau'^{4}}\{-3 k \left(\tau-\tau'\right)\left(3+k^2 \tau \tau'\right)\cos k\left(\tau-\tau'\right)\nonumber\\
&&\quad\quad\quad\quad \quad\quad\quad\quad \quad\quad\quad+\left[9+k^4\tau^2\tau'^2-3 k^2\left(\tau^2-3\tau \tau' +\tau'^2\right)\right]\sin k\left(\tau- \tau'\right)\}\nonumber\\
A_{+,{\rm early}}\left(\tau,k\right)&=&\frac{1}{\sqrt{2 k}}{\rm e}^{\frac{2\tau\sqrt{f k M_p\left(V_0 \lambda \tau^5-9 f k M_p \tau_{\rm in}^4\right)}}{21 f M_p \tau_{\rm in}^2}+\frac{5 i 3^{2/5}\sqrt{\pi}\left(\frac{f k^6 M_p \tau_{\rm in}^4}{V_0 \lambda}\right)^{1/5}\Gamma\left(\frac{6}{5}\right)}{7 \Gamma\left(\frac{7}{10}\right)}-\frac{5}{7}i k \tau \,{}_2F_1\left[\frac{1}{5},\frac{1}{2},\frac{6}{5},\frac{V_0 \lambda \tau^5}{9 f k M_p \tau_{\rm in}^4}\right]}\nonumber\\
\end{eqnarray}

We would also like to point out for the sake of transparency, that we are integrating the time from $\tau_{\rm in}$ until $\tau_{\rm backreaction}$. The result is extremely sensitive with respect to the upper limit of the time integration and the approximate expression we give in the main text in (\ref{eq:timeback}) is not sensitive enough as it is only accurate to the $1\%$ level. Instead, in order to be more safe we evolved the numerical code and compared the quantity ${\cal B}_{EB}$ defined in \ref{back} evaluated numerically to the same quantity evaluated by our approximate solutions and we found the latest possible time for which the two were overlapping.

\end{document}